\documentclass[aps,prl,preprintnumbers,showpacs,nofootinbib]{revtex4}
\usepackage{graphicx}% Include figure files
\usepackage{dcolumn}% Align table columns on decimal point
\usepackage{bm}% bold math
\usepackage{subfig}
\usepackage{rotating} % Rotating tabl
\usepackage{amsmath}
\usepackage{amsfonts}
\usepackage{amssymb}

\usepackage[maxfloats=100]{morefloats}
\usepackage{young}
\begin{document}
\maxdeadcycles=1000
\newcommand{\newc}{\newcommand}
\newc{\ra}{\rightarrow}
\newc{\lra}{\leftrightarrow}
\newc{\beq}{\begin{equation}}
\newc{\eeq}{\end{equation}}
\newc{\barr}{\begin{eqnarray}}
\newc{\earr}{\end{eqnarray}}
%%%%%%%%%%%%%%%%%%%%%%%%%%%%%%%%%%%%%%%%%%%
\newcommand{\Od}{{\cal O}}
\newcommand{\lsim}   {\mathrel{\mathop{\kern 0pt \rlap
  {\raise.2ex\hbox{$<$}}}
  \lower.9ex\hbox{\kern-.190em $\sim$}}}
\newcommand{\gsim}   {\mathrel{\mathop{\kern 0pt \rlap
  {\raise.2ex\hbox{$>$}}}
  \lower.9ex\hbox{\kern-.190em $\sim$}}}
  \def\rpm{R_p \hspace{-0.8em}/\;\:}
%\preprint{APS/123-QED}
%\date{\today}
\title {Clebcsh-Gordan coefficients in the symmric group $S_n$.}
\author{ J. D. Vergados$^{1,2}$ }
\affiliation{1 KAIST University  and  IBS, Daejeon 305-701, Republic of Korea\footnote{ Permanent address: University of Ioannina, Ioannina, GR 45110, Greece.}}
\affiliation{2 University of S. Carolina, Columbia, SC 29208, USA.}
\vspace{0.5cm}

\begin{abstract}
 The coefficients of fractional parentage (CFP) or Clebcsh-Gordan coefficients of the outer product of representations of the symmetric group $S_n$ are  evaluated using an build up algorithm  defined in terms of the chain involving the chain $S_{n-1}\subset S_n$ and $S_{n-2}\subset S_n$ chains. Some applications in mathematical physics are considered by combining them with the  Clebcsh-Gordan (C-G) coefficients of the inner product of representations of the symmetric group $S_n$ 
\end{abstract}
%\pacs{95.35.+d, 12.60.Jv}
%\pacs{ 13.15.+g, 14.60Lm, 14.60Bq, 23.40.-s, 95.55.Vj, 12.15.-y}
%%%%%%%%%%%%%%%%%%%%%%%%%%%%%%%%%%%%%%%%%%%%%%%%%%%%%%%%%%%%%%%%%%%%%
%\date{\today}
\maketitle
%\end{keyword}
%\end{frontmatter}
%%%%%%%%%%%%%%%%%%%%%%%%%%%%%%%%%%%%%%%%%%%%%%%%%%%%%%%%%%%%%%%%%%%%%
%%%%%%%%%%%%%%%%%%%%%%%%%%%%%%%%%%%%%%%%%%%%%%%%%%%%%%%%%%%%%%%%%%%%%
\section{Introduction}
 It is well known that the set of permutations of $n$ symbols form a discreet group with $n!$ elements indicated by $S_n$. This simple group has sufficient structure, so that for a mathematician the task of finding the irreducible representations is one of the classic examples of involved algebraic structure. Furthermore the classification of tensors into irreducible sets of any linear group in $n$ dimensions is greatly facilitated once the irreducible representations of the symmetry groups are known.  Furthermore it is known that among the solutions of the many-body problem only those of a given symmetry are acceptable. Thus, e.g. the wave function of particles with integral spin must symmetric under the exchange of any two of them, while for particles with half integral spin it must be anti-symmetric.

The structure of the irreducible representations of the symmetric group is fairly well known \cite{SymBooks}, \cite{Hamermesh}. The irreducible representations are of $S_n$ described in terms of Young Tableaux [f], containing $f_1$ boxes in the first row, $f_2$ boxes in the second row,  $f_3$ etc subject to the condition $$f_1\geq f_2\geq f_3 \cdots f_r,\,f_1+f_2+f_3+\cdots f_r=n.$$
Then the number of basis states in a representation is obtained by putting on of the integers $1,2,\cdots n$ in each box so that they are increasing from left to right in a given row and from top to bottom in each column. Thus for $n=4$ we find: \\

$[4]$=\begin{Young}
      1 &2&3&4 \cr
     \end{Young}\\
		
		$\left[3,1\right]$=\begin{Young}
      1 &2&3\cr
			4 \cr
     \end{Young},\,\begin{Young}
      1 &2&4\cr
			3 \cr
     \end{Young},\,\begin{Young}
      1 &3&4\cr
			2 \cr
     \end{Young}\\
		
$\left[2,2\right]$=\begin{Young}
      1 &2\cr
			3&4 \cr
     \end{Young},\,\begin{Young}
      1 &3\cr
			2&4 \cr
			 \end{Young}\\
			
			$\left[2,1,1\right]\equiv\left[2,1^2\right]$=\begin{Young}
      1 &2\cr
			3\cr 4 \cr
     \end{Young},\,\begin{Young}
      1 &3\cr
			2\cr 4 \cr
			 \end{Young},\,
		\begin{Young}
      1 &4\cr
			2\cr 3 \cr
			 \end{Young}\\
			
				$\left[1^4\right]$=\begin{Young}
      1 \cr 2\cr
			3\cr 4 \cr
			 \end{Young}.\\
			
			We also see that the representations $ [4],\,[3,1],\,[2,2],\,[2,1^2]$ and $1^4$ are 1, 3, 2, 3-dimensional representations respectively. The representation obtained by interchanging the rows and columns of a Young tableaux is called its conjugate or adjoined and is indicated by putting the symbol tilde  over it, $[f]\rightarrow[\tilde{f}].$ If $[f]=[\tilde{f}]$ the representation is called self-adjoined. We thus see that the adjoined of [4] is $[1^4]$ , the adjoined of [3,1] is $[2,1^3]$, while [2,2] is self adjoined. Alternatively a given state of a representation is given by the Yamanouchi symbol, in which we indicate the row in which each symbol is starting from the highest. It is of the form $Y=r_n,r_{n-1}\cdots r_2,1$. Thus the three states of [3,1], correspond to Yamanouchi symbols 
			$$Y_1=2111,\,Y_2=1211,,\,Y_3=1121.$$
			Naturally the number of one in the Yamanouchi symbol is 3 ($f_1=3$), while the 2 appears only once ($f_2=1$).
			 
Let us consider the  fundamental representation $(\alpha)=\alpha^i_j$ of a group $G$  defined on a set of vectors $v^{i}$, i.e. $v^{'i}=\alpha^i_jv^{j}$. Then a tensor of rank $r$ is an object  $F^{i_1,i_2\cdots i_r}$ is an object that transforms  \cite{Hamermesh} as 
$$F^{'\,i_1,i_2\cdots i_r} =\alpha^{i_1}_{j_1}\alpha^{i_2}_{j_2}\cdots \alpha^{i_r}_{j_n} F_{j_1,j_2 \cdots j_r}\Leftrightarrow F^{'\,(i)}=[A^r]^{(i)}_{(j)}F^{(j)}.$$
Let us now consider a permutation 
$$P=\left ( \begin{array}{cccc}1&2&\cdots&r\\
p_1&p_2&\cdots&p_r\\  \end{array} \right ) ,$$
acting simultaneously on both the upper and the lower indices. Then it is easy to see  \cite{Hamermesh}that:
$$ [A^r]^{P(i)}_{P(j)}=[A^r]^{(i)}_{(j)}.$$
Then it is not difficult to show that:
$$PF^{'\,i}=[A^r]^{(i)}_{(j)}PF^{(j)}\Leftrightarrow P [A^r]^{(i)}_{(j)}= [A^r]^{(i)}_{(j)}P.$$
Since the operators $P$ of the symmetric group commute those of $G$ they can be simultaneously diagonalized. It is therefore convenient not to start with a an basis in the space of tensors, but select one that transforms according to the symmetric group $S_r$. In other words the transformations $ [A^r]^{(i)}_{(j)}$  preserve the symmetry $[f]$ of a representation of the symmetric group. 

The role of the symmetric group in specifying the representations of continuous groups has been explicitly demonstrated in the case of many symmetries which play a role in theoretical physics \cite{Wigner37}-\cite{Swart63} during the last hundred years (for more recent work see \cite{HechtAdler69} and  the  recent review  by Liang Meng and  Zhou \cite{HLM15}). The connection between discreet and continuous groups has also been recently reviewed \cite{RowCarRep12}.

			Through the Young tableaux one can find the number of times a given irreducible representation appears in the Kronecker product of two representations of $S_n$,  which refer to the same particle number. This is called inner product. This, e.g. appears in applications when one representation acts on the space of functions $\phi_a,\phi_b, \phi_c$, e.g.  $\phi_a(1),\phi_b(2), \phi_c(3)$etc, and the other $\chi_a,\chi_b, \chi_c$, e.g.  $\chi_a(1),\chi_b(2), \chi_c(3)$. E.g. the system $\phi_{\ell}(i)$ may refer to the orbital part $\ell$  of the particle $i$, 
while $\chi_s(j)$ refers to the spin s of particle $j$. 

			\section{The reduction of the inner product of representations of the symmetry group}
			\label{sec:InnerCG}
			
			As we have mentioned this arises in the construction of many particle  functions of a given symmetry, when they are defined in more than one space. This section is intended to motivate the reader about the importance of the next section, which deals with all the representations  discussed here.\\
The construction of the corresponding C-G coefficients has been obtained by Murganan \cite{Murganan} and summarized by Hamememesh \cite{Hamermesh}. Here are some interesting simple results:
$$[n-1,1] \times [n-1,1]=[n]+[n-1,1]+[n-2,2]+[n-2,1^2],$$
$$[n-1,1] \times [n-2,2]=[n-1,1]+[n-2,2]+[n-2,1^2]+[n-3,1^3]+[n-3,2,1],\,n>4,$$
$$[n-1,1] \times [n-2,1^2]=[n-1,1]+[n-2,2]+[n-2,1^2]+[n-3,1^3]+[n-3,2,1],[n-3,1^3],\,n>4,$$
$$[n-2,2] \times [n-2,2]=[n]+[n-1,1]+2[n-2,2]+[n-2,1^2]+[n-3,3]+[n-3,1^3]+2[n-3,2,1]+[n-3,1^3]+$$ $$[n-4,4]+[n-4,3,1]+[n-4,2^2],\,n>5.$$
Of special interest are the completely symmetric $[n]$ and the completely antisymmetric case $[1^n]$. Furthermore:
$$\mbox{ i)  } [f]\times[1^n]=[\tilde{f}],$$
$$\mbox{ ii)  }[f]\times[f] \mbox{ contains }[n],\,[f]\times[\tilde{f}] \mbox{ contains }[1^n].$$
Specializing the above formulas we get:
\begin{itemize}
\item [i)] In the case  of $S_3$
 $$[3]\otimes[3]=[3],\,[3]\otimes[1^3]=[1^3],\,[1^3]\otimes[1^3]=[3]$$
$$[2,1]\otimes[2,1]=[3]+2[2,1]+[1^3],$$
\item [ii)] In the case  of $S_4$
$$[3,1]\times[3,1]=[4]+[3,1]+[2^2+[2 ,1^2],$$
$$[3,1]\times[2^2]=[3,1]+[2,1^2],$$
$$[2^2]\times[2^2]=[4]+[2^2].$$
\end{itemize}
The above reduction formulae are useful but one needs the C-G coefficients. Let us consider the interesting possibility of the C-G coefficients in the case of $S_3$.  The non trivial C-G involve the [2,1]. We notice that
both [3] and $[1^3]$ appear in the product, since [2,1] is a self adjoined, two dimensional representation. One can show that  \cite{Hamermesh} that 
$$[3]=\frac{1}{2}\left( [2,1]_1 [2,1]_1+ [2,1]_2 [2,1]_2\right ),\,[1^3]=\frac{1}{2}\left( [2,1]_1 [2,1]_2- [2,1]_2 [2,1]_1 \right ) $$
and
$$[2,1]_1=\frac{1}{2}\left( [2,1]_1 [2,1]_1- [2,1]_2 [2,1]_2\right ),\,[2,1]2=\frac{1}{2}\left( [2,1]_1 [2,1]_2+ [2,1]_2 [2,1]_1.
\right )$$
This problem occurs in particle physics, when one considers a particle described in terms of three quarks appearing 
in three flavors \cite{Lightenberg}, \cite{DHPerkins}, in which case the symmetry is $SU(3)$:
\begin{itemize}
\item  Example 1: The quark content of the nucleon.\\
The quarks  at low energies appear in three flavors $u$, $d$ and $s$, which have spin 1/2. It turns out that onder SU(3)  the flavor and the spin symmetry functions each  transform like the [2,1]. The quarks also have a color degree of freedom, appearing in three colors, say red ($r$), green ($g$) and blue ($b$), transforming like the unitary group $U_c(3)$. By the rules the proton must be colorless transforming like the $1^3$ representation of $U(3)$color. So the flavor spin part of the wave function must be symmetric to yield a totally antisymmetric function for the proton. The   quarks.  We have seen above that  it is possible to combine the two [2,1] representations to yield this.\\
Were one to write the above wave function in components one would get for the proton with spin up the monster. 
 $$ p(\uparrow)=\frac{1}{9 \sqrt{2}}\left \{2\left [ d(\downarrow) u(\uparrow)u(\uparrow) + u(\uparrow)u(\uparrow)d(\downarrow)+u(\uparrow)d(\downarrow)u(\uparrow)\right ]\right .$$  
	$$-\left[  d(\uparrow)\left (u(\uparrow)u(\downarrow)+u(\downarrow)u(\uparrow)\right )+\left (u(\uparrow)u(\downarrow)+u(\downarrow)u(\uparrow)\right )d(\uparrow)\right ]$$
	$$\left .-\left[u(\downarrow)d(\uparrow)u(\uparrow)+u(\uparrow)d(\uparrow)u(\downarrow)\right ]\right \}.$$ 
	(the particle indices 1,2,3 labeling the quarks are understood). This function is symmetric. Thus, one obtains an anti-symmetric total wave function,  if one multiplies it with the component that includes  the color degree of freedom, i.e.  the colorless antisymmetric combination
	$$[1^3]=\frac{1}{\sqrt{6}}\left ((rg-gr)b+(gb-bg)r+(br-rb)g \right),$$
where again we have omitted the particle indices 1,2 3.
Fortunately one need not do this. All one needs is to know the CFP (coefficients of fractional parentage) of the representations involved. Such CFP's, which will be discussed in this work,   are also needed in extensions \cite{FujSuzNak07} of the flavor group $SU(6)$ for the six known quarks.

\item Example 2: Multiquark systems \cite{Strobag79}.\\
In this case  one has the following degrees of freedom: i) The orbital degree of freedom. The quarks are moving in a 3-dimensional  harmonic oscillator potential. They thus transform under $U(3)$ with a symmetry characterized  by a Young tableaux $[f]_L $of at most three rows. ii) they have flavor as above. If one considers two flavors the symmetry in question is $U_I(2)$ with irreducible representation $[f]_I$ with at most two rows. iii) They also have spin 1/2, i.e. a symmetry U(2),  iv) the quarks have color. As we have mentioned the group here is $U(3)$. For $n$ quarks we must a totally antisymmetric w.f. To achieve it is advantageous to enlarge the symmetry. It turns out that that the best way is to combine the spin and color into one symmetry $U_{sc}(6)$ \cite{IsoStr79}. This approach has recently been taken up  with the fancy name color-spin locking in a self-consistent Dyson-Schwinger approach \cite{Buballa05}. The irreducible representations  then transform like $[f]_{sc}$. So the   scheme is:
$$\left \{ \left \{ [f]_L\times [f]_{sc}\right\}_{{\tilde f}}\times[f] _I \right \}1^n,\, \mbox{ where } {\tilde f} \mbox{ is the adjoined of} [f]_I. $$
Additional quantum numbers are needed, depending on the subgroups of the groups indicated above. We will not address such issues in this work.  The allowed representations in the case of six quarks are presented in table \ref{table:QN}. Fortunately relatively simple representations were relevant in this problem \cite{StrVer15}. The orbital symmetries considered were restricted by considering orbitals with excitation energies less than $\leq 2 \hbar \omega$.
\end{itemize}
\begin{table}[h]
\caption{ The various symmetries appearing in the case of six quark cluster. The flavor symmetry is that consistent with the quantum numbers of two nucleons (isospin 1 and 0) i.e.  $[f]_I=[42] $ and [3,3]. Thus $[{\tilde f}]=[2^2,1^2]$ and $[2^3]$ respectively. 
The representations $[f]_{cs}$  selected are those that contain a colorless representation  of the subgroup $SU_c(3)$ (they contain 
the $[2^6]$ of $U_c(3)$).}  
%\resizebox{10cm}{!}{
\begin{tabular}{|c|c|c|ccc|}
\hline
$[f]_I$&$\tilde{f}$&$[f]_{cs}$&&$f_L$&\\
\hline
&&$[3,2,1]$& &[5,1]&[4,2]\\
&&$[4,1^2]$& & &[4,2]\\
$[4,2]$&$[2^2,1^2]$&$[2^3]$&& [5,1]&[4,2]\\
  &&$[3,1^3]$& &[5,1]&[4,2]\\  
	&&$[2^2]$&[6]&[5,1]&[4,2]\\
	&&$[2,1^4]$& &[5,1]&[4,2]\\
	&&$[3,3]$&&&[4,2]\\
	\hline
&&	$[3,1^3]$&& [5,1]&[4,2]\\
	$[3,3]$&$[2^6]$&[3,2,1]&&&[42]	\\	
	&&$[2^2,1^2]$&&[5,1]&\\
	&&$[2^3]$&[6]&&\\
	\hline								
\end{tabular}
\label{table:QN}
\end{table}
			
	\section{The outer product of representations of the symmetric group}
	In this case one defines the outer product of a representation defined in the space of $1,2,\cdots, n_1$ particles with  another  defined  in the space of  $n_1,n_1,\cdots,n$ particles. This product viewed under the symmetry group $S_n$ is in general reducible.
	Some very simple examples are:\\
	\\
	\begin{Young}
      & \cr
			 \end{Young} $\times$	\begin{Young}
       \cr
			 \end{Young} $=$	\begin{Young}
    & & \cr
			 \end{Young}+	\begin{Young}
      & \cr
			\cr
			 \end{Young},\\
			\\
			\begin{Young}
      & \cr
			 \end{Young}$\times$	\begin{Young}
       &\cr
			 \end{Young}=	\begin{Young}
    && & \cr
			 \end{Young}+	\begin{Young}
     && \cr
			\cr
			 \end{Young}+	\begin{Young}
     & \cr
			&\cr
			 \end{Young},\\
			\\
				\begin{Young}
      & \cr
			 \end{Young}$\times$	\begin{Young}
       \cr
			\cr
			 \end{Young}=	\begin{Young}
    & & \cr
			\cr \end{Young}+	
			\begin{Young}
     & \cr
			 \cr
			\cr
			 \end{Young},\\
			\\
					\begin{Young}
       \cr
			\cr
			 \end{Young}$\times$	\begin{Young}
       \cr
			\cr
			 \end{Young}=	\begin{Young}
     \cr
			\cr 
			\cr
			\cr\end{Young}+	
			\begin{Young}
     & \cr
			 \cr
			\cr
			 \end{Young}+	
			\begin{Young}
     & \cr
			 &\cr
			 \end{Young}.\\
These and more complicated products will be explicitly derived in the next section in a built up process, which will also allow the evaluation of C-G coefficients for the outer product  or coefficients of fractional parentage (CFP).
\section{A build up algorithm for construction of 1-particle CFP's}
We will develop an algorithm for constructing the reduction of the outer product of representations $S_n$ and at the same time will construct the corresponding expansion coefficients, using a recursion formula starting from $n=1$ and moving up to $n=6$. We will consider the most general case in which all the particles involved occupy different states. Various other possibilities may be considered as special cases.

We will exploit the fact that here exists an operator $P$ with eigenvalues which depend only on the symmetry [f]. Indeed the operator $P=\sum_{i<j} p(i,j)$ of all transposition permutations depends only on the  Young 	Tableaux. Its eigenvalues are $\Lambda([f])=\Lambda_{\alpha}-\Lambda_{\beta}$, where 
$$\Lambda_{\alpha} =\sum_{i<j} p(i,j),\,  p(i,j)=\mbox{a permutation  on the elements of the row } \alpha\mbox{  of the Young tableaux } [f],$$
$$\Lambda_{\beta} =\sum_{i<j} p(i,j),\,  p(i,j)=\mbox{a permutation acting on the elements of the column } \beta\mbox{  of the Young tableaux } [f].$$
Thus, e.g., $\Lambda([3])=3$, $\Lambda([2,1])=0$, $\Lambda([1^3])=-3$ etc.

Suppose that a symmetry $[f]$ of  $S_{n-1}$  has already been constructed. If $r$ such symmetries $[f]$  exit we indicate them by $[f]_i$, $i=1,2,\cdots,r$ . Construct now a basis for $S_n$ by adding one more particle, say $p$ , add taking the product 
$[i,p]=[f]_i Y\left (p\notin\sigma_n \right );p$ where $\sigma_n$ is the set of  integers up to and including $n$. In other words the state 
$ [f]_i Y\left (p\notin\sigma_n \right )$ contains all labels of $S_n$ with the exception of $p$. Let us now consider the matrix elements
$$m([i,p,j,q)=\prec[f]_i Y\left (p\notin\sigma_n \right );p|P|[f]_j Y\left (q\notin\sigma_n \right );q\succ.$$
Clearly for the diagonal matrix element $i=j,p=q$ the operator acts on [f] and its eigenvalue is known. For the off diagonal matrix element one gets a contribution only from the transposition  involving the elements $p$ and $q$. Thus for the off diagonal matrix element we get 
$$m([i,p,j,q)=\prec [f]_i Y\left (p\notin\sigma_n \right )|[f]_j Y\left (p\notin\sigma_n \right) \succ\mbox { off diagonal }. $$
These are known, if the reduction of $S_{n-2}\times [1]$ is already known.\\
%Let us start with $n=3$ and consider the case$ [2]\times[1]$. A convenient basis is ${1,2}3,\,{{1,3}2,\,{3,2}1$, where ${a,b}$ is the normalized symmetric combination of $a$ and  $b$. Then it is trivial 

For $S_3$ there are two one dimensional representations of  $S_2$, the symmetric and the antisymmetric. In terms of these we construct the representations of $S_3$ in a product basis:
\begin{itemize}
\item using the representation [2,0]=
\begin{Young}
       & \cr
     \end{Young}
		of $S_2$.\\
 In this we get the basis: \\
\\
		\begin{Young}
     1  &  2 \cr
     \end{Young}
	$\times$	\begin{Young}
     3\cr
     \end{Young}$\,^,$
		\begin{Young}
     1  &  3 \cr
     \end{Young}
		\begin{Young}
     2\cr
     \end{Young}$\,^,$
		\begin{Young}
     3&  2 \cr
     \end{Young}
		\begin{Young}
     1\cr
     \end{Young}.\\
		\\
 In this basis one can evaluate the matrix elements of the operator$ P$ by noting that \begin{Young}
       a& b\cr
     \end{Young}$^{=\mbox{(a b+ba)/}\sqrt{2}}$ . The resulting matrix is given below:
 \beq
\mbox{matrix P}=\left(
\begin{array}{ccc}
 1 & 1 & 1 \\
 1 & 1 & 1 \\
 1 & 1 & 1 \\
\end{array}
\right),
\mbox{ Matrix of eigenvectors of P}=\left(
\begin{array}{c|cc}
\hline
[3]&[2,1]k=1;&[2,1]k=2\\
\hline
 \frac{1}{\sqrt{3}} & \frac{1}{\sqrt{2}} & \frac{1}{\sqrt{6}} \\
 \frac{1}{\sqrt{3}} & -\frac{1}{\sqrt{2}} & \frac{1}{\sqrt{6}} \\
 \frac{1}{\sqrt{3}} & 0 & -\sqrt{\frac{2}{3}} \\
\end{array}
\right).
 \eeq
As expected the eigenvalues of $P$ are 3,0, and 0 identifying [3],[2,1]k=1 and [2,1] k=2 respectively. Since the eigenvalue 0 occurs more then once, the corresponding eigenvectors were chosen arbitrarily.
\item using the representation [1,1]=
  \begin{Young}
        \cr
		\cr
     \end{Young}  of $S_2$,
		\\
		we get the basis:\\
		\\
		\begin{Young}
    1   \cr
		2\cr
     \end{Young}
		\begin{Young}
     3\cr
     \end{Young}$\,^,$
		\begin{Young}
     1  \cr
		3\cr
     \end{Young}
		\begin{Young}
     2\cr
     \end{Young}$\,^,$
		 \begin{Young}
     3 \cr
		2\cr
     \end{Young}
		\begin{Young}
     1\cr
     \end{Young}.\\
The resulting $P$  matrix can be calculated by noting that  \begin{Young}
       a\cr
			b\cr
     \end{Young}$^{=\mbox{(a b-ba)/}\sqrt{2}}$   and it is given below:
 \beq
\mbox{matrix P}=\left(
\begin{array}{rrr}
 -1 & 1 & -1 \\
 1 & -1 & 1 \\
 -1 & 1 & -1 \\
\end{array}
\right),
\mbox{ Matrix of eigenvectors of P}=\left(
\begin{array}{c|cc}
\hline
[1^3]&[2,1]k=1;&[2,1]k=2\\
\hline
 \frac{1}{\sqrt{3}} & \frac{1}{\sqrt{2}} & -\frac{1}{\sqrt{6}} \\
 -\frac{1}{\sqrt{3}} & \frac{1}{\sqrt{2}} & \frac{1}{\sqrt{6}} \\
 \frac{1}{\sqrt{3}} & 0 & \sqrt{\frac{2}{3}} \\
\end{array}
\right).
 \eeq
Again as expected the eigenvalues of $P$ are -3,0, and 0 identifying [1,1,1],[2,1]k=1 and [2,1]k=2 respectively.\\
We should mention at this point that had we chosen the basis:\\
		\begin{Young}
    1   \cr
		2\cr
     \end{Young}
		\begin{Young}
     3\cr
     \end{Young}$\,^,$
		\begin{Young}
     3  \cr
		2\cr
     \end{Young}
		\begin{Young}
     1\cr
     \end{Young}$\,^,$
		 \begin{Young}
     3 \cr
		1\cr
     \end{Young}
		\begin{Young}
     2\cr
     \end{Young}.\\
		we would have obtained:
		\beq
\mbox{matrix P}=\left(
\begin{array}{ccc}
 -1 & -1 & -1 \\
 -1 & -1 & -1 \\
 -1 & -1 & -1 \\
\end{array}
\right),
\mbox{ Matrix of eigenvectors of P}=\left(
\begin{array}{c|cc}
[1^3]&[2,1]k=1;&[2,1]k=2\\
 \frac{1}{\sqrt{3}} & \frac{1}{\sqrt{2}} & \frac{1}{\sqrt{6}} \\
 \frac{1}{\sqrt{3}} & -\frac{1}{\sqrt{2}} & \frac{1}{\sqrt{6}} \\
 \frac{1}{\sqrt{3}} & 0 & -\sqrt{\frac{2}{3}} \\
\end{array}
\right).
 \eeq
The latter has the advantage that the expansion the coefficients involving the symmetry\footnote{We will write $\ell^k$  whenever the k rows  of the young tableaux have the same length $\ell$.}  [1,1,1,1]= $[1^3]$  are of the same sign, which convenient when discussing the color symmetry for 3-quark systems. We found it, however, more convenient to retain our rule of ordering the last particle as we moved on to higher $n$.
\end{itemize}
\section{Applications}
The $S_2\subset S_3$ (outer) G-G series has already been discussed.
\subsection{ The $S_3\subset S_4$ (outer) G-G series (1-particle CFP's).}
The possible representations of $S_4$ are [4], [3,1],[2,2], [2,1,1,1] and [1,1,1,1]. We will use a basis the $S_3$ representations  we  constructed above and $S_1$. Thus:
\begin{itemize}
\item The product $[3]\otimes 1\rightarrow [4]+3[3,1]$.\\
The starting basis is \\	
\\
	\begin{Young}
     1&2&3\cr
     \end{Young} $\times$
		\begin{Young}
     4\cr
     \end{Young} $^,$
		\begin{Young}
     1&2&4\cr
     \end{Young} $\times$
		\begin{Young}
     3\cr
     \end{Young} $^,$
\begin{Young}
     1&4&3\cr
     \end{Young} 
		\begin{Young}$\times$
     2\cr
     \end{Young}$^,$
		\begin{Young}
     4&2&3\cr
     \end{Young} 
		\begin{Young}$\times$
     1\cr
     \end{Young}$^{.}$\\
The obtained P matrix and the matrix of its eigenvectors are:
\beq
m=\left(
\begin{array}{cccc}
 3 & 1 & 1 & 1 \\
 1 & 3 & 1 & 1 \\
 1 & 1 & 3 & 1 \\
 1 & 1 & 1 & 3 \\
\end{array}
\right),
v=\left(
\begin{array}{c|ccc}
\hline
[4]&[3,1]_1&[3,1]_2&[3,1]_3\\
6&2&2& 2\\
\hline
 \frac{1}{2} & \frac{1}{\sqrt{2}} & 0 & \frac{1}{2} \\
 \frac{1}{2} & -\frac{1}{\sqrt{2}} & 0 & \frac{1}{2} \\
 \frac{1}{2} & 0 & \frac{1}{\sqrt{2}} & -\frac{1}{2} \\
 \frac{1}{2} & 0 & -\frac{1}{\sqrt{2}} & -\frac{1}{2} \\
\end{array}
\right),
\eeq
with the eigenvectors labeled by their  symmetry and their eigenvalues, 6,2,2 and 2. Since three of them are degenerate their selection was arbitrary an arbitrary linear combination subject  to the condition that they are orthogonal to each other  and properly normalized.
\item The product $[1,1,1]\otimes 1\rightarrow [1,1,1,1]+3[2,1,1]$.\\
The starting basis is \\	
\\
	\begin{Young}
     1\cr2\cr3\cr
     \end{Young} $\times$
		\begin{Young}
     4\cr
     \end{Young} $^,$
		\begin{Young}
     1\cr2\cr4\cr
     \end{Young} $\times$
		\begin{Young}
     3\cr
     \end{Young} $^,$
\begin{Young}
     1\cr4\cr3\cr
     \end{Young} 
		\begin{Young}$\times$
     2\cr
     \end{Young}$^,$
		\begin{Young}
     4\cr2\cr3\cr
     \end{Young} 
		\begin{Young}$\times$
     1\cr
     \end{Young}$^{.}$\\
The obtained P matrix and the matrix of its eigenvectors are:
\beq
m=\left(
\begin{array}{rrrr}
 -3 & 1 & 1 & 1 \\
 1 & -3 & -1 & -1 \\
 1 & -1 & -3 & -1 \\
 1 & -1 & -1 & -3 \\
\end{array}
\right),
v=\left(
\begin{array}{c|ccc}
\hline
[1^4]&[2,1^2_1&[2,1^2_2&[2,1^2_3]\\
-6&-2&-2&-2\\
\hline
 -\frac{1}{2} & \frac{1}{\sqrt{2}} & 0 & -\frac{1}{2} \\
 \frac{1}{2} & \frac{1}{\sqrt{2}} & 0 & \frac{1}{2} \\
 \frac{1}{2} & 0 & \frac{1}{\sqrt{2}} & -\frac{1}{2} \\
 \frac{1}{2} & 0 & -\frac{1}{\sqrt{2}} & -\frac{1}{2} \\
\end{array}
\right),
\eeq
with the eigenvectors labeled not only by their symmetry but  by their eigenvalues, i.e. -6,-2,-2 and- 2 as well. Since three of them are degenerate their selection was arbitrary subject to the condition that they are orthogonal to themselves (they are automatically orthogonal to the other one) and properly normalized.
\item The product $[2,1]\otimes 1\rightarrow 3[3,1]+2[2,2]+3[2,1^2]$.\\
A suitable basis is $[2,1]_1(123);4$, $[2,1]_1(124);3$, $[2,1]_1(143);2$, $[2,1]_1(423);1$, $[2,1]_2(123);4$, $[2,1]_2(124);3$, $[2,1]_2(143);2$, $[2,1]_2(423);1$
	The resulting P matrix is:
		$$
		m=\left(
\begin{array}{cccccccc}
 \frac{1}{\sqrt{3}} & 0 &
   -\frac{\sqrt{\frac{3}{2}}}{2} & 0 & 0 &
   \frac{1}{2 \sqrt{6}} & 0 & \frac{1}{2} \\
 -\frac{1}{\sqrt{3}} & 0 &
   -\frac{\sqrt{\frac{3}{2}}}{4} & -\frac{3}{4
   \sqrt{2}} & \frac{1}{4 \sqrt{2}} & \frac{1}{4
   \sqrt{6}} & \frac{\sqrt{3}}{4} & \frac{1}{4}
   \\
 -\frac{1}{\sqrt{3}} & 0 &
   -\frac{\sqrt{\frac{3}{2}}}{4} & \frac{3}{4
   \sqrt{2}} & -\frac{1}{4 \sqrt{2}} &
   \frac{1}{4 \sqrt{6}} & -\frac{\sqrt{3}}{4} &
   \frac{1}{4} \\
 0 & 0 & -\frac{\sqrt{\frac{3}{2}}}{2} & 0 & 0 &
   -\frac{\sqrt{\frac{3}{2}}}{2} & 0 &
   -\frac{1}{2} \\
 0 & \frac{1}{\sqrt{3}} & 0 & \frac{1}{2
   \sqrt{6}} & -\frac{\sqrt{\frac{3}{2}}}{2} & 0
   & \frac{1}{2} & 0 \\
 0 & \frac{1}{\sqrt{3}} & \frac{1}{4 \sqrt{2}} &
   -\frac{1}{4 \sqrt{6}} &
   \frac{\sqrt{\frac{3}{2}}}{4} & -\frac{3}{4
   \sqrt{2}} & -\frac{1}{4} & \frac{\sqrt{3}}{4}
   \\
 0 & \frac{1}{\sqrt{3}} & -\frac{1}{4 \sqrt{2}}
   & -\frac{1}{4 \sqrt{6}} &
   \frac{\sqrt{\frac{3}{2}}}{4} & \frac{3}{4
   \sqrt{2}} & -\frac{1}{4} &
   -\frac{\sqrt{3}}{4} \\
 0 & 0 & 0 & \frac{\sqrt{\frac{3}{2}}}{2} &
   \frac{\sqrt{\frac{3}{2}}}{2} & 0 &
   \frac{1}{2} & 0 \\
\end{array}
\right)
		$$
	with eigenvalues (-2,-2,-2,2,2,2,0,0). A set of eigenvectors   is:	
	\beq
	v=\left(
\begin{array}{|ccc|ccc|cc|}
\hline
[2,1^2]_1&,[2,1^2]_2&[2,1^2]_3&[3,1]_1&[3,1]_2&[3,1]_3&[2,2]_1&[2,2,_1]\\
\hline
-2&-2&-2&2&2&2&0&0\\
\hline
 \frac{1}{\sqrt{3}} & 0 &
   -\frac{\sqrt{\frac{3}{2}}}{2} & 0 & 0 &
   \frac{1}{2 \sqrt{6}} & 0 & \frac{1}{2} \\
 -\frac{1}{\sqrt{3}} & 0 &
   -\frac{\sqrt{\frac{3}{2}}}{4} & -\frac{3}{4
   \sqrt{2}} & \frac{1}{4 \sqrt{2}} & \frac{1}{4
   \sqrt{6}} & \frac{\sqrt{3}}{4} & \frac{1}{4}
   \\
 -\frac{1}{\sqrt{3}} & 0 &
   -\frac{\sqrt{\frac{3}{2}}}{4} & \frac{3}{4
   \sqrt{2}} & -\frac{1}{4 \sqrt{2}} &
   \frac{1}{4 \sqrt{6}} & -\frac{\sqrt{3}}{4} &
   \frac{1}{4} \\
 0 & 0 & -\frac{\sqrt{\frac{3}{2}}}{2} & 0 & 0 &
   -\frac{\sqrt{\frac{3}{2}}}{2} & 0 &
   -\frac{1}{2} \\
 0 & \frac{1}{\sqrt{3}} & 0 & \frac{1}{2
   \sqrt{6}} & -\frac{\sqrt{\frac{3}{2}}}{2} & 0
   & \frac{1}{2} & 0 \\
 0 & \frac{1}{\sqrt{3}} & \frac{1}{4 \sqrt{2}} &
   -\frac{1}{4 \sqrt{6}} &
   \frac{\sqrt{\frac{3}{2}}}{4} & -\frac{3}{4
   \sqrt{2}} & -\frac{1}{4} & \frac{\sqrt{3}}{4}
   \\
 0 & \frac{1}{\sqrt{3}} & -\frac{1}{4 \sqrt{2}}
   & -\frac{1}{4 \sqrt{6}} &
   \frac{\sqrt{\frac{3}{2}}}{4} & \frac{3}{4
   \sqrt{2}} & -\frac{1}{4} &
   -\frac{\sqrt{3}}{4} \\
 0 & 0 & 0 & \frac{\sqrt{\frac{3}{2}}}{2} &
   \frac{\sqrt{\frac{3}{2}}}{2} & 0 &
   \frac{1}{2} & 0 \\
\end{array}
\right).
	\eeq
	Again, since we encounter degeneracy,  any linear combination of the degenerate ones is also an accepted solution.
	\end{itemize}
	
\subsection{ The $S_4\subset S_5$ (outer) G-G series (1-particle CFP's).}
	%	\section{The Example $S_5$}
	We will begin with the reduction:\\ $[4]\otimes [1]\rightarrow[5]+4 [4,1]$ and  $[1^4]\otimes [1]\rightarrow[1^5]+4 [2,1^3]$.
	The obtained results are shown in the table \ref{t4x1.1111x1}.
	\begin{table}
	\caption{The reductions  $[4]\otimes [1]\rightarrow[5]+4 [4,1]$ and  $[1^4]\otimes [1]\rightarrow[1^5]+4 [2,1^3]$.}
	\label{t4x1.1111x1}
	$$ 
	\left(
\begin{array}{|c|ccccc}
\hline
&[5]&[4,1]_1&,[4,1]_2&[4,1]_3&[4,1]_4\\
\hline
(1234)5 &\frac{1}{\sqrt{5}} & \frac{1}{\sqrt{2}} & 0 &
   \frac{1}{2} & \frac{1}{2 \sqrt{5}} \\
 (1235)4&\frac{1}{\sqrt{5}} & -\frac{1}{\sqrt{2}} & 0 &
   \frac{1}{2} & \frac{1}{2 \sqrt{5}} \\
 (1254)3&\frac{1}{\sqrt{5}} & 0 & \frac{1}{\sqrt{2}} &
   -\frac{1}{2} & \frac{1}{2 \sqrt{5}} \\
(1534)2& \frac{1}{\sqrt{5}} & 0 & -\frac{1}{\sqrt{2}} &
   -\frac{1}{2} & \frac{1}{2 \sqrt{5}} \\
 (5234)1&\frac{1}{\sqrt{5}} & 0 & 0 & 0 & -\frac{2}{\sqrt{5}} \\
\end{array}
\right),
\left(
\begin{array}{|c|cccc}
\hline
[1^5]&[2,1^3]_1&,[2,1^3]_2&[2,1^3]_3&[2,1^3]_4\\
\hline
 -\frac{1}{\sqrt{5}} & \frac{1}{\sqrt{2}} & 0 &
   \frac{1}{2} & -\frac{1}{2 \sqrt{5}} \\
 \frac{1}{\sqrt{5}} & \frac{1}{\sqrt{2}} & 0 &
   -\frac{1}{2} & \frac{1}{2 \sqrt{5}} \\
 \frac{1}{\sqrt{5}} & 0 & \frac{1}{\sqrt{2}} &
   \frac{1}{2} & \frac{1}{2 \sqrt{5}} \\
 \frac{1}{\sqrt{5}} & 0 & -\frac{1}{\sqrt{2}} &
   \frac{1}{2} & \frac{1}{2 \sqrt{5}} \\
 \frac{1}{\sqrt{5}} & 0 & 0 & 0 & -\frac{2}{\sqrt{5}} \\
\end{array}
\right).
	$$
	\end{table}
	The labeling of the rows is essentially indicated by the label of the last particle as explicitly  indicated in the case $[4]\times 1$. The numbers (1234) etc indicate a completely symmetric combination of the indicated particles. In the case of $1^4$ the situation is analogous. Now (1234) etc indicate a completely antisymmetric combination.
	
We will continue with the reduction 
$$[3,1]\otimes[1]\rightarrow 4[4,1]+5[3,2]+6[3,1,1],$$
starting for simplicity  with the symmetric three particle function. The obtained eigenvalues were 
	5,2 and 0 respectively with multiplicities 5,5 and 2 respectively.
	\begin{table}
	\caption{The reduction  $[3,1]\otimes[1]\rightarrow 4[4,1]+5[3,2]+6[3,1,1]$.}
	\label{t31x1}
	$$\left(
\begin{array}{cccc}
\hline
[4,1]_1&[4,1]_2&[4,1]_3&[4,1]_4\\
\hline
 -\frac{2}{\sqrt{15}} & 0 & 0 & 0 \\
 -\frac{1}{\sqrt{15}} & 0 & \frac{1}{\sqrt{6}} &
   -\frac{1}{\sqrt{30}} \\
 -\frac{1}{\sqrt{15}} & 0 & -\frac{1}{\sqrt{6}} &
   \frac{1}{\sqrt{30}} \\
 -\frac{2}{\sqrt{15}} & 0 & 0 & 0 \\
 -\frac{2}{\sqrt{15}} & 0 & 0 & 0 \\
 0 & \frac{2}{\sqrt{15}} & 0 & 0 \\
 0 & \frac{2}{\sqrt{15}} & 0 & 0 \\
 0 & \frac{2}{\sqrt{15}} & 0 & 0 \\
 0 & \frac{1}{\sqrt{15}} & \frac{1}{\sqrt{6}} &
   \frac{1}{\sqrt{30}} \\
 0 & \frac{1}{\sqrt{15}} & -\frac{1}{\sqrt{6}} &
   -\frac{1}{\sqrt{30}} \\
 0 & 0 & 0 & \frac{2}{\sqrt{15}} \\
 \frac{1}{\sqrt{30}} & 0 & \frac{1}{2 \sqrt{3}} &
   \frac{\sqrt{\frac{3}{5}}}{2} \\
 -\frac{1}{\sqrt{30}} & 0 & \frac{1}{2 \sqrt{3}} &
   \frac{\sqrt{\frac{3}{5}}}{2} \\
 0 & \frac{1}{\sqrt{30}} & -\frac{1}{2 \sqrt{3}} &
   \frac{\sqrt{\frac{3}{5}}}{2} \\
 0 & -\frac{1}{\sqrt{30}} & -\frac{1}{2 \sqrt{3}} &
   \frac{\sqrt{\frac{3}{5}}}{2} \\
\end{array}
\right),
 \left(
\begin{array}{ccccc}
\hline
[3,2]_1&[3,2]_2&[3,2]_3&[3,2]_4&[3,2]_5\\
\hline
 0 & 0 & -\frac{1}{\sqrt{3}} & 0 & 0 \\
 \frac{1}{2 \sqrt{3}} & 0 & -\frac{1}{2 \sqrt{3}} & 0 &
   \frac{1}{\sqrt{6}} \\
 -\frac{1}{2 \sqrt{3}} & 0 & -\frac{1}{2 \sqrt{3}} & 0 &
   -\frac{1}{\sqrt{6}} \\
 0 & 0 & \frac{1}{2 \sqrt{3}} & -\frac{1}{2} & 0 \\
 0 & 0 & \frac{1}{2 \sqrt{3}} & \frac{1}{2} & 0 \\
 0 & \frac{1}{\sqrt{3}} & 0 & 0 & 0 \\
 0 & -\frac{1}{2 \sqrt{3}} & 0 & -\frac{1}{2} & 0 \\
 0 & -\frac{1}{2 \sqrt{3}} & 0 & \frac{1}{2} & 0 \\
 \frac{1}{2 \sqrt{3}} & \frac{1}{2 \sqrt{3}} & 0 & 0 &
   -\frac{1}{\sqrt{6}} \\
 -\frac{1}{2 \sqrt{3}} & \frac{1}{2 \sqrt{3}} & 0 & 0 &
   \frac{1}{\sqrt{6}} \\
 0 & 0 & 0 & 0 & \frac{1}{\sqrt{3}} \\
 -\frac{1}{\sqrt{6}} & 0 & -\frac{1}{\sqrt{6}} & 0 & 0
   \\
 -\frac{1}{\sqrt{6}} & 0 & \frac{1}{\sqrt{6}} & 0 & 0 \\
 \frac{1}{\sqrt{6}} & -\frac{1}{\sqrt{6}} & 0 & 0 & 0 \\
 \frac{1}{\sqrt{6}} & \frac{1}{\sqrt{6}} & 0 & 0 & 0 \\
\end{array}
\right)$$
\end{table}
	\begin{table}
	\caption{The reduction  $[3,1]\otimes[1]\rightarrow 4[4,1]+5[3,2]+6[3,1,1]$ (continued).}
	\label{t31x1a}
$$
\left(
\begin{array}{cccccc}
\hline
[3,1^2]_1&[3,1^2]_2&[3,1^2]_3&[3,1^2]_4&[3,1^2]_5&[3,1^2]_6\\
\hline
 0 & 0 & -\frac{1}{\sqrt{3}} & -\frac{1}{5 \sqrt{3}} &
   -\frac{1}{10 \sqrt{3}} & \frac{1}{2 \sqrt{5}} \\
 0 & 0 & \frac{1}{\sqrt{3}} & \frac{1}{5 \sqrt{3}} &
   \frac{1}{10 \sqrt{3}} & \frac{1}{2 \sqrt{5}} \\
 0 & 0 & \frac{1}{\sqrt{3}} & -\frac{2}{5 \sqrt{3}} &
   -\frac{1}{5 \sqrt{3}} & 0 \\
 0 & \frac{1}{2} & 0 & \frac{\sqrt{3}}{10} &
   \frac{\sqrt{3}}{20} & -\frac{3}{4 \sqrt{5}} \\
 0 & -\frac{1}{2} & 0 & \frac{\sqrt{3}}{10} &
   \frac{\sqrt{3}}{20} & -\frac{3}{4 \sqrt{5}} \\
 -\frac{1}{\sqrt{3}} & 0 & 0 & -\frac{1}{5 \sqrt{3}} &
   \frac{2}{5 \sqrt{3}} & 0 \\
 0 & -\frac{1}{2} & 0 & \frac{\sqrt{3}}{10} &
   -\frac{\sqrt{3}}{5} & 0 \\
 0 & \frac{1}{2} & 0 & \frac{\sqrt{3}}{10} &
   -\frac{\sqrt{3}}{5} & 0 \\
 \frac{1}{\sqrt{3}} & 0 & 0 & -\frac{2}{5 \sqrt{3}} &
   \frac{1}{20 \sqrt{3}} & -\frac{1}{4 \sqrt{5}} \\
 \frac{1}{\sqrt{3}} & 0 & 0 & \frac{1}{5 \sqrt{3}} &
   \frac{7}{20 \sqrt{3}} & \frac{1}{4 \sqrt{5}} \\
 0 & 0 & 0 & -\frac{\sqrt{6}}{5} &
   -\frac{\sqrt{\frac{3}{2}}}{5} & -\frac{1}{\sqrt{10}}
   \\
 0 & 0 & 0 & \frac{\sqrt{6}}{5} &
   \frac{\sqrt{\frac{3}{2}}}{5} & -\frac{1}{\sqrt{10}}
   \\
 0 & 0 & 0 & 0 & 0 & \sqrt{\frac{2}{5}} \\
 0 & 0 & 0 & 0 & \frac{\sqrt{\frac{3}{2}}}{2} &
   \frac{1}{2 \sqrt{10}} \\
 0 & 0 & 0 & \frac{\sqrt{6}}{5} & -\frac{3
   \sqrt{\frac{3}{2}}}{10} & \frac{1}{2 \sqrt{10}} \\
\end{array}
\right).
$$
\end{table}
$$[2,1^2]\otimes[1]\rightarrow 4[2,1^2]+5[2^2,1]+6[3,1,1].$$
Eigenvalues:(-5,-2,0). The obtained eigenvectors  are given in table \ref{211x1}.\\
	\begin{table}
	\caption{The reduction  $[2,1^2]\otimes[1]\rightarrow 4[4,1]+5[3,2]+6[3,1,1]$.}
	\label{211x1}
$$
\left(
\begin{array}{cccc}
\hline
[2,1^3]_1&[2,1^3]_2&[2,1^3]_3&[2,1^3]_4\\
\hline
 -\frac{2}{\sqrt{15}} & 0 & 0 & 0 \\
 \frac{1}{\sqrt{15}} & 0 & -\frac{1}{\sqrt{6}} &
   \frac{1}{\sqrt{30}} \\
 \frac{1}{\sqrt{15}} & 0 & \frac{1}{\sqrt{6}} &
   -\frac{1}{\sqrt{30}} \\
 \frac{2}{\sqrt{15}} & 0 & 0 & 0 \\
 \frac{2}{\sqrt{15}} & 0 & 0 & 0 \\
 0 & -\frac{2}{\sqrt{15}} & 0 & 0 \\
 0 & \frac{2}{\sqrt{15}} & 0 & 0 \\
 0 & \frac{2}{\sqrt{15}} & 0 & 0 \\
 0 & \frac{1}{\sqrt{15}} & \frac{1}{\sqrt{6}} &
   \frac{1}{\sqrt{30}} \\
 0 & \frac{1}{\sqrt{15}} & -\frac{1}{\sqrt{6}} &
   -\frac{1}{\sqrt{30}} \\
 0 & 0 & 0 & -\frac{2}{\sqrt{15}} \\
 \frac{1}{\sqrt{30}} & 0 & \frac{1}{2 \sqrt{3}} &
   \frac{\sqrt{\frac{3}{5}}}{2} \\
 -\frac{1}{\sqrt{30}} & 0 & \frac{1}{2 \sqrt{3}} &
   \frac{\sqrt{\frac{3}{5}}}{2} \\
 0 & \frac{1}{\sqrt{30}} & -\frac{1}{2 \sqrt{3}} &
   \frac{\sqrt{\frac{3}{5}}}{2} \\
 0 & -\frac{1}{\sqrt{30}} & -\frac{1}{2 \sqrt{3}} &
   \frac{\sqrt{\frac{3}{5}}}{2} \\
\end{array}
\right),
\left(
\begin{array}{ccccc}
\hline
[2^2,1]_1&[2^2,1]_2&[2^2,1]_3&[2^2,1]_4&[2^2,1]_5\\
\hline
 -\frac{1}{\sqrt{6}} & 0 & 0 & 0 &
   -\frac{1}{\sqrt{6}} \\
 0 & 0 & \frac{1}{\sqrt{6}} & 0 & \frac{1}{\sqrt{6}}
   \\
 \frac{1}{\sqrt{6}} & 0 & -\frac{1}{\sqrt{6}} & 0 &
   0 \\
 -\frac{1}{2 \sqrt{6}} & \frac{1}{4} & 0 &
   \frac{\sqrt{3}}{4} & -\frac{1}{2 \sqrt{6}} \\
 -\frac{1}{2 \sqrt{6}} & -\frac{1}{4} & 0 &
   -\frac{\sqrt{3}}{4} & -\frac{1}{2 \sqrt{6}} \\
 0 & -\frac{1}{2} & 0 & \frac{1}{2 \sqrt{3}} & 0 \\
 0 & -\frac{1}{2} & 0 & -\frac{1}{2 \sqrt{3}} & 0 \\
 0 & 0 & 0 & \frac{1}{\sqrt{3}} & 0 \\
 \frac{1}{2 \sqrt{6}} & \frac{1}{4} &
   \frac{1}{\sqrt{6}} & -\frac{1}{4 \sqrt{3}} &
   -\frac{1}{2 \sqrt{6}} \\
 -\frac{1}{2 \sqrt{6}} & \frac{1}{4} &
   -\frac{1}{\sqrt{6}} & -\frac{1}{4 \sqrt{3}} &
   \frac{1}{2 \sqrt{6}} \\
 0 & 0 & \frac{1}{\sqrt{3}} & 0 & 0 \\
 -\frac{1}{\sqrt{3}} & 0 & 0 & 0 & 0 \\
 0 & 0 & 0 & 0 & \frac{1}{\sqrt{3}} \\
 \frac{1}{2 \sqrt{3}} & -\frac{1}{2 \sqrt{2}} & 0 &
   \frac{1}{2 \sqrt{6}} & -\frac{1}{2 \sqrt{3}} \\
 \frac{1}{2 \sqrt{3}} & \frac{1}{2 \sqrt{2}} & 0 &
   -\frac{1}{2 \sqrt{6}} & -\frac{1}{2 \sqrt{3}} \\
\end{array}
\right),
$$
\end{table}
	\begin{table}
	\caption{The reduction  $[3,1]\otimes[1]\rightarrow 4[4,1]+5[3,2]+6[3,1,1]$ (continued).}
	\label{t31x1b}
$$
\left(
\begin{array}{cccccc}
\hline
[3,1^2]_2&[3,1^2]_2&[3,1^2]_3&[3,1^2]_4&[3,1^2]_5&[3,1^2]_6\\
\hline
 \frac{1}{\sqrt{3}} & 0 & 0 & \frac{1}{\sqrt{15}} &
   0 & 0 \\
 \frac{1}{\sqrt{3}} & 0 & 0 & -\frac{1}{2 \sqrt{15}}
   & 0 & -\frac{1}{2 \sqrt{5}} \\
 \frac{1}{\sqrt{3}} & 0 & 0 & -\frac{1}{2 \sqrt{15}}
   & 0 & \frac{1}{2 \sqrt{5}} \\
 0 & -\frac{1}{2} & 0 & \frac{\sqrt{\frac{3}{5}}}{2}
   & 0 & 0 \\
 0 & \frac{1}{2} & 0 & \frac{\sqrt{\frac{3}{5}}}{2}
   & 0 & 0 \\
 0 & 0 & \frac{1}{\sqrt{3}} & 0 &
   \frac{1}{\sqrt{15}} & 0 \\
 0 & -\frac{1}{2} & 0 & 0 &
   \frac{\sqrt{\frac{3}{5}}}{2} & 0 \\
 0 & \frac{1}{2} & 0 & 0 &
   \frac{\sqrt{\frac{3}{5}}}{2} & 0 \\
 0 & 0 & \frac{1}{\sqrt{3}} & 0 & -\frac{1}{2
   \sqrt{15}} & -\frac{1}{2 \sqrt{5}} \\
 0 & 0 & \frac{1}{\sqrt{3}} & 0 & -\frac{1}{2
   \sqrt{15}} & \frac{1}{2 \sqrt{5}} \\
 0 & 0 & 0 & 0 & 0 & \sqrt{\frac{2}{5}} \\
 0 & 0 & 0 & -\sqrt{\frac{3}{10}} & 0 &
   \frac{1}{\sqrt{10}} \\
 0 & 0 & 0 & \sqrt{\frac{3}{10}} & 0 &
   \frac{1}{\sqrt{10}} \\
 0 & 0 & 0 & 0 & -\sqrt{\frac{3}{10}} &
   \frac{1}{\sqrt{10}} \\
 0 & 0 & 0 & 0 & \sqrt{\frac{3}{10}} &
   \frac{1}{\sqrt{10}} \\
\end{array}
\right).
$$
\end{table}
Finally we consider the chain:
$$[2^2]\otimes[1]\rightarrow 5[3,2]+5[2^2,1].$$
The obtained eigenvalues are (2,-2) quadruply degenerate. The corresponding eigenvectors shown in table \ref{t22x1}:
\begin{table}
	\caption{The reduction  $[2^2]\otimes[1]\rightarrow 5[3,2]+5[2^2,1]$.}
	\label{t22x1}
$$
\left(
\begin{array}{ccccc}
\hline
[3,2]_1&[3,2]_2&[3,2]_3&[3,2]_4&[3,2]_5\\
\hline
 0 & -\frac{\sqrt{\frac{11}{2}}}{4} & \frac{1}{4
   \sqrt{2}} & 0 & -\frac{1}{2 \sqrt{2}} \\
 0 & \frac{\sqrt{\frac{11}{2}}}{4} & -\frac{1}{4
   \sqrt{2}} & 0 & -\frac{1}{2 \sqrt{2}} \\
 \sqrt{\frac{3}{11}} & \frac{5}{4 \sqrt{22}} &
   -\frac{1}{4 \sqrt{2}} &
   -\frac{\sqrt{\frac{3}{2}}}{4} & \frac{1}{4
   \sqrt{2}} \\
 -\sqrt{\frac{3}{11}} & \frac{3}{2 \sqrt{22}} & 0 &
   \frac{\sqrt{\frac{3}{2}}}{4} & \frac{1}{4
   \sqrt{2}} \\
 0 & 0 & 0 & 0 & \frac{1}{\sqrt{2}} \\
 -\frac{1}{\sqrt{11}} & -\frac{5}{4 \sqrt{66}} &
   -\frac{5}{4 \sqrt{6}} & -\frac{1}{2 \sqrt{2}} & 0
   \\
 0 & 0 & 0 & \frac{1}{\sqrt{2}} & 0 \\
 0 & 0 & \frac{\sqrt{\frac{3}{2}}}{2} & -\frac{1}{4
   \sqrt{2}} & \frac{\sqrt{\frac{3}{2}}}{4} \\
 0 & \frac{\sqrt{\frac{11}{6}}}{4} & \frac{5}{4
   \sqrt{6}} & -\frac{1}{4 \sqrt{2}} &
   -\frac{\sqrt{\frac{3}{2}}}{4} \\
 \frac{2}{\sqrt{11}} & -\frac{1}{4 \sqrt{66}} &
   -\frac{1}{4 \sqrt{6}} & \frac{1}{2 \sqrt{2}} & 0
   \\
\end{array}
\right),
\left(
\begin{array}{ccccc}
\hline
[2^2,1]_1&[2^2,1]_2&[2^2,1]_3&[2^2,1]_4&[2^2,1]_5\\
\hline
 -\frac{1}{\sqrt{11}} & -\frac{5}{4 \sqrt{66}} &
   \frac{\sqrt{\frac{5}{6}}}{4} &
   \frac{\sqrt{\frac{5}{6}}}{2} & \frac{1}{2
   \sqrt{2}} \\
 0 & -\frac{\sqrt{\frac{11}{6}}}{4} & \frac{11}{4
   \sqrt{30}} & \frac{7}{4 \sqrt{30}} & -\frac{1}{4
   \sqrt{2}} \\
 -\frac{2}{\sqrt{11}} & \frac{1}{4 \sqrt{66}} &
   -\frac{1}{4 \sqrt{30}} & -\frac{1}{2 \sqrt{30}} &
   \frac{1}{2 \sqrt{2}} \\
 0 & 0 & 0 & 0 & \frac{1}{\sqrt{2}} \\
 0 & 0 & 0 & \frac{\sqrt{\frac{15}{2}}}{4} &
   -\frac{1}{4 \sqrt{2}} \\
 0 & \frac{\sqrt{\frac{11}{2}}}{4} &
   \frac{\sqrt{\frac{5}{2}}}{4} & 0 & 0 \\
 -\sqrt{\frac{3}{11}} & \frac{3}{2 \sqrt{22}} &
   \frac{1}{2 \sqrt{10}} & -\frac{1}{4 \sqrt{10}} &
   -\frac{\sqrt{\frac{3}{2}}}{4} \\
 0 & 0 & \sqrt{\frac{2}{5}} & -\frac{1}{\sqrt{10}} &
   0 \\
 0 & \frac{\sqrt{\frac{11}{2}}}{4} & -\frac{3}{4
   \sqrt{10}} & \frac{1}{\sqrt{10}} & 0 \\
 \sqrt{\frac{3}{11}} & \frac{5}{4 \sqrt{22}} &
   \frac{3}{4 \sqrt{10}} & \frac{1}{4 \sqrt{10}} &
   \frac{\sqrt{\frac{3}{2}}}{4} \\
\end{array}
\right).
$$
\end{table}
At this point it should be mentioned that the above  procedure  suffers in this case from a bit of complication. The two states [2,2] are expressed in terms of basis of $[2,1]\times [1]$, which is   8-dimensional and is given by 
$$[2,2]_1=\left\{\frac{1}{2},\frac{1}{4},\frac{1}{4},-\frac{1}
   {2},0,\frac{\sqrt{3}}{4},-\frac{\sqrt{3}}{4},0\right\},$$
	$$[2,2]_2=\left\{\frac{1}{2},\frac{1}{4},\frac{1}{4},-\frac{1}
   {2},0,\frac{\sqrt{3}}{4},-\frac{\sqrt{3}}{4},0\right\}.$$
	To proceed further they should be expressed in terms of the 12 4-particle arrangements:
	$$\left(
\begin{array}{ccc}
 \{1,2\} & 3 & 4 \\
 \{1,2\} & 4 & 3 \\
 \{1,4\} & 3 & 2 \\
 \{4,2\} & 3 & 1 \\
 \{1,3\} & 2 & 4 \\
 \{1,4\} & 2 & 3 \\
 \{1,3\} & 4 & 2 \\
 \{4,3\} & 2 & 1 \\
 \{3,2\} & 1 & 4 \\
 \{4,2\} & 1 & 3 \\
 \{3,4\} & 1 & 2 \\
 \{3,2\} & 4 & 1 \\
\end{array}
\right). $$ One finds:
	$$[2,2]_1=\left\{\frac{1}{2 \sqrt{6}},\frac{1}{2
   \sqrt{6}},-\frac{1}{\sqrt{6}},\frac{1}{2
   \sqrt{6}},\frac{1}{2
   \sqrt{6}},-\frac{1}{\sqrt{6}},\frac{1}{2
   \sqrt{6}},\frac{1}{2
   \sqrt{6}},-\frac{1}{\sqrt{6}},\frac{1}{2
   \sqrt{6}},\frac{1}{2
   \sqrt{6}},-\frac{1}{\sqrt{6}}\right\},$$
		$$[2,2]_2=\left\{\frac{1}{2 \sqrt{2}},\frac{1}{2
   \sqrt{2}},0,-\frac{1}{2 \sqrt{2}},-\frac{1}{2
   \sqrt{2}},0,-\frac{1}{2 \sqrt{2}},\frac{1}{2
   \sqrt{2}},0,-\frac{1}{2 \sqrt{2}},\frac{1}{2
   \sqrt{2}},0\right\}.$$
	We will frequently encounter this problem in the construction of the C-G series in the case of $S_6$.
	%\end{document}
	\subsection{ The $S_5\subset S_6$ (outer) G-G series (1particle CFP's).}
	%\section{Example $S_6$}
Since this a bit more complicated problem we remind the reader of some of the needed 
%We will try to find the C-G coeffiecients up to $S_6$. We summarize  the 
ingredients  in Table \ref{table.S6}.
\begin{table}
\caption{Some facts about $S_6$}
\begin{center}
\begin{tabular}{|c|c|c|}
\hline
\hline
representation&dimension&P-eigenvalue\\
$\left[ 6,0\right ]$&1&15\\
$\left[ 5,1\right ]$&5&9\\
$\left [ 4,2\right]$&9&5\\
$\left [ 4,1^2\right ]$&10&3\\
$\left [ 3,3\right ]$&5&3\\
$\left [ 3,2,1\right ]$&16&0\\
$\left [ 3,1^3\right ]$&10&-3\\
$\left [ 2^3\right ]$&5&-3\\
$\left [ 2^2,1^2\right]$&9&-5\\
$\left [ 2,1^4\right]$&5&-9\\
$\left [ 1^6\right]$&1&-15\\
\hline
\hline
\end{tabular}
\end{center}
\label{table.S6}
\end{table}
%The C-G coefficients regarding $S_5$ and $S_{6}$ are more complicated and will be discussed below. Here we examine two simple cases:
	%\begin{itemize}
	%\item 
%	\item The reduction $[5]\otimes [1]\rightarrow[6]+5 [5,1]$ and  $[1^5]\otimes [1]\rightarrow[1^6]+5 [2,1^4]$.\\
	The obtained results are shown in the table \ref{t51x1-21111.x1}. :
	\begin{table}
	\caption{The reduction  $[5]\otimes[1]\rightarrow [6]+5[5,1]$ (left) and$[1^5]\otimes[1]\rightarrow [1^6]+5[2,1^4]$ (right).}
	\label{t51x1-21111.x1}
	$$
	\left(
\begin{array}{|c|ccccc}
\hline
[6]&[5,1]_1&,[5,1]_2&[5,1]_3&[5,1]_4&[5,1]_5\\
\hline
 \frac{1}{\sqrt{6}} & \frac{1}{\sqrt{2}} & 0 &
   \frac{1}{2 \sqrt{5}} & \frac{1}{2} &
   \frac{1}{\sqrt{30}} \\
 \frac{1}{\sqrt{6}} & -\frac{1}{\sqrt{2}} & 0 &
   \frac{1}{2 \sqrt{5}} & \frac{1}{2} &
   \frac{1}{\sqrt{30}} \\
 \frac{1}{\sqrt{6}} & 0 & \frac{1}{\sqrt{2}} &
   \frac{1}{2 \sqrt{5}} & -\frac{1}{2} &
   \frac{1}{\sqrt{30}} \\
 \frac{1}{\sqrt{6}} & 0 & -\frac{1}{\sqrt{2}} &
   \frac{1}{2 \sqrt{5}} & -\frac{1}{2} &
   \frac{1}{\sqrt{30}} \\
 \frac{1}{\sqrt{6}} & 0 & 0 & -\frac{2}{\sqrt{5}} & 0 &
   \frac{1}{\sqrt{30}} \\
 \frac{1}{\sqrt{6}} & 0 & 0 & 0 & 0 &
   -\sqrt{\frac{5}{6}} \\
\end{array}
\right),
\left(
\begin{array}{|c|ccccc}
\hline
[1^6]&[2,1^4]_1&[2,1^4]_2&[2,1^4]_3&[2,1^4]_4&[2,1^4]_5\\
\hline
 -\frac{1}{\sqrt{6}} & \frac{1}{\sqrt{2}} & 0 &
   \frac{1}{2 \sqrt{5}} & \frac{1}{2} &
   -\frac{1}{\sqrt{30}} \\
 \frac{1}{\sqrt{6}} & \frac{1}{\sqrt{2}} & 0 &
   -\frac{1}{2 \sqrt{5}} & -\frac{1}{2} &
   \frac{1}{\sqrt{30}} \\
 \frac{1}{\sqrt{6}} & 0 & \frac{1}{\sqrt{2}} &
   -\frac{1}{2 \sqrt{5}} & \frac{1}{2} &
   \frac{1}{\sqrt{30}} \\
 \frac{1}{\sqrt{6}} & 0 & -\frac{1}{\sqrt{2}} &
   -\frac{1}{2 \sqrt{5}} & \frac{1}{2} &
   \frac{1}{\sqrt{30}} \\
 \frac{1}{\sqrt{6}} & 0 & 0 & \frac{2}{\sqrt{5}} & 0 &
   \frac{1}{\sqrt{30}} \\
 \frac{1}{\sqrt{6}} & 0 & 0 & 0 & 0 &
   -\sqrt{\frac{5}{6}} \\
\end{array}
\right).
	$$
	\end{table}
	The labeling of the rows is analogous to the previous case for $S_5$. Clearly, due to degeneracy, the selection of the columns, other than the first, is arbitrary.
	
	In many applications the CFP's (outer C-G coefficients) needed are those involving the most symmetric or the most antisymmetric representations of $S_6$ given above. We will compute the CFP's involving the more complicated representations of $S_6$ and present them in the appendices. We will only discuss some technical difficulties involved by considering, e.g.,  the series:
%We will try to eventually  get the C-G coefficients for the series  
$$[3,2]\otimes[1]\rightarrow 9[4,2]+5[3,3]+16[3,2,1],$$
with eigenvalues 5,3 and 0 respectively and multiplicities 9, 5 and 16 respectively\\. 
We have already discussed the reduction:
$$[3,1]\otimes[1]\rightarrow 4[4,1]+5[3,2]+6[3,1,1],$$
	with eigenvalues 5,3 and 0 respectively. We have seen that the corresponding eigenstates for the [3,2] are:
	$$[3,2]_1= \left\{0,\frac{1}{2 \sqrt{3}},-\frac{1}{2
   \sqrt{3}},0,0,0,0,0,\frac{1}{2
   \sqrt{3}},-\frac{1}{2
   \sqrt{3}},0,-\frac{1}{\sqrt{6}},-\frac{1}{
   \sqrt{6}},\frac{1}{\sqrt{6}},\frac{1}{\sqrt{6}}\right\},$$
	$$[3,2]_2=\left\{0,0,0,0,0,\frac{1}{\sqrt{3}},-\frac{1}{2 \sqrt{3}},-\frac{1}{2
   \sqrt{3}},\frac{1}{2 \sqrt{3}},\frac{1}{2
   \sqrt{3}},0,0,0,-\frac{1}{\sqrt{6}},\frac{1}{\sqrt{6}}\right\},$$
		$$[3,2]_3=\left\{-\frac{1}{\sqrt{3}},-\frac{1}{2 \sqrt{3}},-\frac{1}{2 \sqrt{3}},\frac{1}{2
   \sqrt{3}},\frac{1}{2
   \sqrt{3}},0,0,0,0,0,0,-\frac{1}{\sqrt{6}},\frac{1}{\sqrt{6}},0,0\right\},$$
		$$[3,2]_4=\left\{0,0,0,-\frac{1}{2},\frac{1}{2},0,-\frac{1}{2},\frac{1}{2},0,0,0,0,0,0,0 \right\},$$
		$$[3,2]_5=\left\{0,\frac{1}{\sqrt{6}},-\frac{1}{\sqrt{6}},0,0,0,0,0,-\frac{1}{\sqrt{6}},
		\frac{1}{\sqrt{6}},\frac{1}{\sqrt{3}},0,0,0,0\right\}.$$
		These functions given in the 15-dimensional basis $[3,1]\otimes[1]$ must be expressed in terms of the left set of 5-particle states (see table \ref{tab:set5}).
	\begin{table}
	\caption{A convenient basis for labeling the 5-particle states}
	\label{tab:set5}
		$$\left(
\begin{array}{ccc}
 \{1,2,3\} & 4 & 5 \\
 \{1,2,4\} & 3 & 5 \\
 \{1,4,3\} & 2 & 5 \\
 \{4,2,3\} & 1 & 5 \\
 \{1,2,3\} & 5 & 4 \\
 \{1,2,5\} & 3 & 4 \\
 \{1,5,3\} & 2 & 4 \\
 \{5,2,3\} & 1 & 4 \\
 \{1,2,5\} & 4 & 3 \\
 \{1,2,4\} & 5 & 3 \\
 \{1,4,5\} & 2 & 3 \\
 \{4,2,5\} & 1 & 3 \\
 \{1,5,3\} & 4 & 2 \\
 \{1,5,4\} & 3 & 2 \\
 \{1,4,3\} & 5 & 2 \\
 \{4,5,3\} & 1 & 2 \\
 \{5,2,3\} & 4 & 1 \\
 \{5,2,4\} & 3 & 1 \\
 \{5,4,3\} & 2 & 1 \\
 \{4,2,3\} & 5 & 1 \\
\end{array}
\right), \left(
\begin{array}{cc}
 \{1,2,3,4\} & 5 \\
 \{1,2,3,5\} & 4 \\
 \{1,2,5,4\} & 3 \\
 \{1,5,3,4\} & 2 \\
 \{5,2,3,4\} & 1 \\
 \{1,2,4,3\} & 5 \\
 \{1,2,5,3\} & 4 \\
 \{1,2,4,5\} & 3 \\
 \{1,5,4,3\} & 2 \\
 \{5,2,4,3\} & 1 \\
 \{1,4,3,2\} & 5 \\
 \{1,5,3,2\} & 4 \\
 \{1,4,5,2\} & 3 \\
 \{1,4,3,5\} & 2 \\
 \{5,4,3,2\} & 1 \\
 \{4,2,3,1\} & 5 \\
 \{5,2,3,1\} & 4 \\
 \{4,2,5,1\} & 3 \\
 \{4,5,3,1\} & 2 \\
 \{4,2,3,5\} & 1 \\
\end{array}
\right)$$
\end{table}
Where the curly bracket indicates a symmetric state of the indicated three particles (left pattern) or four particles right.\\
The next step is to express the 5 [3,2] states  in the above basis by expanding the obtained eigenvectors  in terms of the [31] states. The result is:
$$ [3,2]_1=\left\{0,0,0,0,0,-\frac{1}{\sqrt{6}},\frac{1}{2 \sqrt{6}},\frac{1}{2
   \sqrt{6}},-\frac{1}{\sqrt{6}},0,\frac{1}{2 \sqrt{6}},\frac{1}{2
   \sqrt{6}},\frac{1}{2 \sqrt{6}},\frac{1}{2
   \sqrt{6}},0,-\frac{1}{\sqrt{6}},\frac{1}{2 \sqrt{6}},\frac{1}{2
   \sqrt{6}},-\frac{1}{\sqrt{6}},0\right\},$$
	$$ [3,2]_2=\left\{0,0,\frac{1}{\sqrt{6}},-\frac{1}{\sqrt{6}},0,0,-\frac{1}{2
   \sqrt{6}},\frac{1}{2 \sqrt{6}},0,0,-\frac{1}{2 \sqrt{6}},\frac{1}{2
   \sqrt{6}},-\frac{1}{2 \sqrt{6}},-\frac{1}{2
   \sqrt{6}},\frac{1}{\sqrt{6}},0,\frac{1}{2 \sqrt{6}},\frac{1}{2
   \sqrt{6}},0,-\frac{1}{\sqrt{6}}\right\},$$
		$$ [3,2]_3=\left\{-\frac{1}{\sqrt{6}},\frac{1}{\sqrt{6}},0,0,-\frac{1}{\sqrt{6}},0,
		\frac{1}{2 \sqrt{6}},\frac{1}{2
   \sqrt{6}},0,\frac{1}{\sqrt{6}},-\frac{1}{2 \sqrt{6}},-\frac{1}{2
   \sqrt{6}},\frac{1}{2 \sqrt{6}},-\frac{1}{2 \sqrt{6}},0,0,\frac{1}{2
   \sqrt{6}},-\frac{1}{2 \sqrt{6}},0,0\right\},
	$$
		$$ [3,2]_4=\left\{0,0,0,0,0,0,-\frac{1}{2 \sqrt{2}},\frac{1}{2
   \sqrt{2}},0,0,\frac{1}{2 \sqrt{2}},-\frac{1}{2 \sqrt{2}},-\frac{1}{2
   \sqrt{2}},\frac{1}{2 \sqrt{2}},0,0,\frac{1}{2 \sqrt{2}},-\frac{1}{2
   \sqrt{2}},0,0\right\},$$
		$$ [3,2]_5=\left\{\frac{1}{2 \sqrt{3}},\frac{1}{2 \sqrt{3}},-\frac{1}{2
   \sqrt{3}},-\frac{1}{2 \sqrt{3}},\frac{1}{2 \sqrt{3}},-\frac{1}{2
   \sqrt{3}},0,0,-\frac{1}{2 \sqrt{3}},\frac{1}{2
   \sqrt{3}},0,0,0,0,-\frac{1}{2 \sqrt{3}},\frac{1}{2
   \sqrt{3}},0,0,\frac{1}{2 \sqrt{3}},-\frac{1}{2 \sqrt{3}}\right\}.$$
%This is essentially the matrix  $\Lambda^{j}_{\alpha(p,q),p,q}$ .
%(the index $r$ is not really needed  at this point). 
The product space is defined by the set $([3,2]_j,r)$. We first very the index $j$ for fixed  $r$, 
e.g. ($([3,2]_1$,6), ($([3,2]_2,6),\cdots$,($([3,2]_5$,6),($([3,2]_1,5)\cdots$ etc.\\
The obtained results are included in the appendix A.
\section{A build up algorithm for construction of 2-particle CFP's}
$$[f]_1\times [f_2]\rightarrow[f], \mbox{ with }[f_2]=[2],\,[1,1].$$
These can be constructed from the 1-particle CFP's obtained above, a practice, which  is  standard 
 in group theory. In the case of $S_n$ , however, they can be obtained directly by extending  the  method developed above.
		
 Again we need worry about the off-diagonal matrix elements. So only permutations of the type $P_{\alpha,i}$ with $\alpha$ one of the indices appearing in $ [f]_1$, while $j$ appearing in $f_2$. The resulting overlaps can be calculated starting from  a number of particles $n=4$ in the combined space.

Before proceeding further we should mention that there appears a  problem in the case that there exist two representations, which may happen to be degenerate, i.e. they happen to have the same eigenvalue. This, in fact, happens in our calculation in the case of the reductions: 
$$[3,1]\otimes[2]\rightarrow [f],\,[f]=[4,1^2],[3,3], \mbox{ and } [2,1^2]\otimes[1^2]\rightarrow [f],\,[f]=[3,1^3],[2^3]$$
Then one can obtain the states that have the symmetry [f] after projection with  the Young operator \cite{Hamermesh} associated with the corresponding Young Tableaux  defined by:
\beq
Y[f]=\sum_c Q_c  \sum_r P_r,\,Q_c=\sum_q \delta_q q \mbox{ (for column c) },P_r=\sum_p  p \mbox{ (for row r)}
\eeq
where $q$ and $p$ are the allowed permutations of the indices of the  relevant columns and rows respectively and $\delta_q$ is one, if the permutation is even, and $-1$, if it is odd. The operators $Q$ and $P$ are sometimes called anti-symmetrizer and symmetrizer respectively. This operator acting on any basis state gives a state with definite symmetry $[f]$.  We found it convenient to choose as a basis the degenerate eigenvectors in each case. Any allowed pattern of the Young tableau can be chosen, provided that it does not yield a zero projection. The resulting states must be properly orthonormalized. The whole procedure is quite tedious but straightforward. 
\\In our case we found it convenient to use:\\
$Y([3,3])$=\begin{Young}
     1&2&3\cr
		4&5&6\cr
     \end{Young},\, $Y([4,1^2])$=\begin{Young}
     1&2&3&6\cr
		4\cr
		5\cr
     \end{Young},\,$Y([2^3)$=\begin{Young}
     1&3\cr
		2&5\cr
		4&6\cr
     \end{Young},\, $Y([3,1^3])$=\begin{Young}
     1&2&6\cr
		3\cr
		4\cr
		5\cr
     \end{Young}
		\\Thus, e.g., for the pattern $Y[2^3]$ we find:
		$$P_1=\frac{1}{\sqrt{2}}(1+(1,3)),\,P_2=\frac{1}{\sqrt{2}}(1+(2,5)),\,P_3=\frac{1}{\sqrt{2}}(1+(4,6))$$
				$$Q_1=\frac{1}{\sqrt{6}}(1-(1,2)-(1,4)-(2,4)+(1,2,4)+(2,1,4)),\,Q_2=\frac{1}{\sqrt{6}}(1-(3,5)-(3,6)-(5,6)+(3,5,6)+(5,3,6)) \mbox{ etc}$$
\subsection{ The $S_2\subset S_4$ two-particle CFP's.}
	The same technique can be applied in evaluating the two particle CFP's. For $n=3$ they coincide with the one particle CFP's obtained above.
	For n=4 we have the reductions:
	\begin{itemize}
	\item $ [2]\times[2]=[4]+3[3,1]+2[2,2].$
	A convenient basis is the order pairs  of particles:
	$$(\{1,2\},\{3,4\}),(\{1,3\},\{2,4\}),(\{1,4\},\{2,3\}),(\{2,3\},\{1,4\}),(\{2,4\},\{1,3\}),(\{3,4\},\{1,2\}),$$
	where $\{a,b\}$ represents the symmetric combination of $a$ and $b$. 
	The resulting eigenvalues are $6,2,2,2,0,0$. 
	The eigenvectors are given in table \ref{t2x2.11x11}.
	\item $ [2]\times[2]=[4]+3[3,1]+2[2,2].$
	The basis is a above except that $\{a,b\}$ represents the anti-symmetric combination of $a$ and $b$. The Eigenvalues are:$-6,-2,-2,-2,0,0$ and the eigenvectors are also given in table\ref{t2x2.11x11}.
	\item $ [2]\times[1,1]=3 [3,1]+3[2,1^2].$
	The basis is given by the same notation except that the first pair is symmetrically coupled , while the second is coupled anti-symmetrically.
	The eigenvalues are: $-2,-2,-2,2,2,2$ and the eigenvectors as given table \ref{t2x11.11x2}.
	\item $ [1,1]\times[2]=3 [3,1]+3[2,1^2].$
	This reduction can be made to be the same results as in the previous case, but we will not bother to accomplish this. The obtained results are given in table \ref{t2x11.11x2}.
	\end{itemize}
	\begin{table}
	\caption{The reductions  $ [2]\times[2]=[4]+3[3,1]+2[2,2]$ (left) and  $ [2]\times[2]=[4]+3[3,1]+2[2,2].$ (right)}
	\label{t2x2.11x11}
	$$
	\left(
\begin{array}{cccccc}
\hline
[4]&[3,1]_1&[3,1]_2&[3,1]_3&[2,2]_1&[2,2]_2\\
\hline
 \frac{1}{\sqrt{6}} & -\frac{1}{\sqrt{2}} &
   0 & 0 & \frac{1}{2} & -\frac{1}{2
   \sqrt{3}} \\
 \frac{1}{\sqrt{6}} & 0 &
   -\frac{1}{\sqrt{2}} & 0 & 0 &
   \frac{1}{\sqrt{3}} \\
 \frac{1}{\sqrt{6}} & 0 & 0 &
   -\frac{1}{\sqrt{2}} & -\frac{1}{2} &
   -\frac{1}{2 \sqrt{3}} \\
 \frac{1}{\sqrt{6}} & 0 & 0 &
   \frac{1}{\sqrt{2}} & -\frac{1}{2} &
   -\frac{1}{2 \sqrt{3}} \\
 \frac{1}{\sqrt{6}} & 0 & \frac{1}{\sqrt{2}}
   & 0 & 0 & \frac{1}{\sqrt{3}} \\
 \frac{1}{\sqrt{6}} & \frac{1}{\sqrt{2}} & 0
   & 0 & \frac{1}{2} & -\frac{1}{2 \sqrt{3}}
   \\
\end{array}
\right)
	\left(
\begin{array}{cccccc}
\hline
[1^4]&[2,1^2]_1&[2,1^2]_2&[2,1^2]_3&[2,2]_1&[2,2]_2\\
\hline
 \frac{1}{\sqrt{6}} & -\frac{1}{\sqrt{2}} &
   0 & 0 & \frac{1}{2} & \frac{1}{2
   \sqrt{3}} \\
 -\frac{1}{\sqrt{6}} & 0 &
   -\frac{1}{\sqrt{2}} & 0 & 0 &
   \frac{1}{\sqrt{3}} \\
 \frac{1}{\sqrt{6}} & 0 & 0 &
   -\frac{1}{\sqrt{2}} & -\frac{1}{2} &
   \frac{1}{2 \sqrt{3}} \\
 \frac{1}{\sqrt{6}} & 0 & 0 &
   \frac{1}{\sqrt{2}} & -\frac{1}{2} &
   \frac{1}{2 \sqrt{3}} \\
 -\frac{1}{\sqrt{6}} & 0 &
   \frac{1}{\sqrt{2}} & 0 & 0 &
   \frac{1}{\sqrt{3}} \\
 \frac{1}{\sqrt{6}} & \frac{1}{\sqrt{2}} & 0
   & 0 & \frac{1}{2} & \frac{1}{2 \sqrt{3}}
   \\
\end{array}
\right)
	$$
		\end{table}
		\begin{table}
	\caption{The reductions  $ [2]\times[1^2]=3[2,1^2]+3[3,1]$ (left) and the  $ [1^2]\times[2]=3[2,1^2]+3[3,1]$ (right)}
	\label{t2x11.11x2}
	$$
	\left(
\begin{array}{cccccc}
\hline
[2,1^2]_1&[2,1^2]_2&[2,1^2]_3&[3,1]_1&[3,1]_2&[3,1]_3\\
\hline
 0 & \frac{1}{2} & 0 & -\frac{1}{\sqrt{3}} &
   -\frac{1}{\sqrt{15}} & \frac{3}{2
   \sqrt{10}} \\
 \frac{1}{2} & 0 & 0 & -\frac{1}{\sqrt{3}} &
   \frac{2}{\sqrt{15}} & -\frac{1}{2
   \sqrt{10}} \\
 \frac{1}{2} & -\frac{1}{2} &
   \frac{1}{\sqrt{3}} & 0 & 0 & 0 \\
 0 & -\frac{1}{2} & -\frac{1}{\sqrt{3}} & 0
   & 0 & \frac{\sqrt{\frac{5}{2}}}{2} \\
 -\frac{1}{2} & 0 & \frac{1}{\sqrt{3}} & 0 &
   \sqrt{\frac{3}{5}} & \frac{1}{2
   \sqrt{10}} \\
 \frac{1}{2} & \frac{1}{2} & 0 &
   \frac{1}{\sqrt{3}} & \frac{1}{\sqrt{15}}
   & \frac{1}{\sqrt{10}} \\
\end{array}
\right)
\left(
\begin{array}{cccccc}
\hline
[2,1^2]_1&[2,1^2]_2&[2,1^2]_3&[3,1]_1&[3,1]_2&[3,1]_3\\
\hline
 0 & \frac{1}{2}\sqrt{\frac{3}{2}} &
   \frac{1}{\sqrt{3}} & \frac{1}{\sqrt{3}} &
   0 & -\frac{1}{2}\sqrt{\frac{3}{2}} \\
 \frac{1}{\sqrt{3}} & -\frac{1}{2 \sqrt{6}}
   & \frac{1}{\sqrt{3}} &
   -\frac{1}{\sqrt{3}} & -\frac{1}{\sqrt{3}}
   & -\frac{1}{2 \sqrt{6}} \\
 -\frac{1}{\sqrt{3}} & -\frac{1}{\sqrt{6}} &
   \frac{1}{\sqrt{3}} & 0 & 0 & 0 \\
 0 & 0 & 0 & \frac{1}{\sqrt{3}} &
   -\frac{1}{\sqrt{3}} & \frac{1}{\sqrt{6}}
   \\
 0 & \frac{1}{2}\sqrt{\frac{3}{2}} & 0 & 0 &
   0 & \frac{1}{2}\sqrt{\frac{3}{2}} \\
 \frac{1}{\sqrt{3}} & -\frac{1}{2 \sqrt{6}}
   & 0 & 0 & \frac{1}{\sqrt{3}} & \frac{1}{2
   \sqrt{6}} \\
\end{array}
\right)
	$$
	\end{table}
	%\clearpage
	\subsection{ The $S_3\subset S_5$ two-particle CFP's.}
	The reductions:
	\begin{itemize}
	\item[i)]  
	$$[3]\times[2]\rightarrow[5,0]+4[4,1]+5[3,2],\quad[1^3]\times[1^2]\rightarrow[1^5]+4[2,1^3]+5[2^2,1]]$$
	\end{itemize}
	and
	\begin{itemize}
	\item[ii)]  
	$$[1^3]\times[1^2]\rightarrow[1^5]+4[2,1^3]+5[2^2,1]]\, \quad [1^3]\times[2]\rightarrow 4[2,1^3]+6[2^2,1],$$
	\end{itemize}
	are easy to obtain (see tables \ref{30x2}-\ref{111x2}. One needs a basis for the two sets, the set of three  particles and the set of two particles, e.g:
$$	\left(
\begin{array}{cc}
 \{1,2,3\} & \{4,5\} \\
 \{1,2,4\} & \{3,5\} \\
 \{1,4,3\} & \{2,5\} \\
 \{4,2,3\} & \{1,5\} \\
 \{1,2,5\} & \{3,4\} \\
 \{1,3,5\} & \{2,4\} \\
 \{2,3,5\} & \{1,4\} \\
 \{1,4,5\} & \{2,3\} \\
 \{4,2,5\} & \{1,3\} \\
 \{5,4,3\} & \{1,2\} \\
\end{array}
\right),
$$
with appropriate symmetry of each group understood.
\begin{table}
\caption{The reductions $[3]\times[2]\rightarrow[5,0]+4[4,1]+5[3,2]$.}
\label{30x2}
$$
\left(
\begin{array}{c|cccc}
\hline
[5]&[4,1]_1&[4,1]_2&[4,1]_3&[4,1]_4\\
%[\tilde{5}]&[2,1^3]_1&[2,1^3]_2&[2,1^3]_3&[2,1^3]_4\\
%[5]&[4,1]_1&[4,1]_2&[4,1]_3&[4,1]_4\\
%&&or&&\\
%[1^5]&[2,1^3]_1&[2,1^3]_2&[2,1^3]_3&[2,1^3]_4\\
\hline
 \frac{1}{\sqrt{10}}&-\frac{1}{2 \sqrt{3}} & -\frac{1}{2 \sqrt{3}} & -\frac{\sqrt{2}}{3} & \frac{1}{3 \sqrt{10}} \\
 \frac{1}{\sqrt{10}}&-\frac{1}{2 \sqrt{3}} & \frac{1}{2 \sqrt{3}} & -\frac{1}{3 \sqrt{2}} & -\frac{2 \sqrt{\frac{2}{5}}}{3} \\
 \frac{1}{\sqrt{10}}&\frac{1}{2 \sqrt{3}} & -\frac{1}{2 \sqrt{3}} & -\frac{1}{3 \sqrt{2}} & -\frac{2 \sqrt{\frac{2}{5}}}{3} \\
 \frac{1}{\sqrt{10}}&\frac{1}{2 \sqrt{3}} & \frac{1}{2 \sqrt{3}} & -\frac{\sqrt{2}}{3} & \frac{1}{3 \sqrt{10}} \\
\frac{1}{\sqrt{10}}& -\frac{1}{\sqrt{3}} & 0 & \frac{1}{3 \sqrt{2}} & \frac{1}{3 \sqrt{10}} \\
\frac{1}{\sqrt{10}}& 0 & 0 & 0 & \sqrt{\frac{2}{5}} \\
 \frac{1}{\sqrt{10}}&0 & 0 & \frac{\sqrt{2}}{3} & -\frac{2 \sqrt{\frac{2}{5}}}{3} \\
 \frac{1}{\sqrt{10}}&0 & \frac{1}{\sqrt{3}} & \frac{1}{3 \sqrt{2}} & \frac{1}{3 \sqrt{10}} \\
 \frac{1}{\sqrt{10}}&\frac{1}{\sqrt{3}} & 0 & \frac{1}{3 \sqrt{2}} & \frac{1}{3 \sqrt{10}} \\
\end{array}
\right),
\left(
\begin{array}{ccccc}
\hline
[3,2]_1&[3,2]_2&[3,2]_3&[3,2]_4&[3,2]_5\\
 \frac{1}{\sqrt{6}} & \frac{1}{\sqrt{10}} & \frac{1}{\sqrt{15}} & -\frac{1}{3} & -\frac{1}{3 \sqrt{2}} \\
 \frac{1}{\sqrt{6}} & -\frac{1}{\sqrt{10}} & -\frac{1}{\sqrt{15}} & \frac{1}{3} & \frac{1}{3 \sqrt{2}} \\
 -\frac{1}{\sqrt{6}} & \frac{1}{\sqrt{10}} & -\frac{\sqrt{\frac{3}{5}}}{2} & \frac{1}{6} & -\frac{1}{3
   \sqrt{2}} \\
 -\frac{1}{\sqrt{6}} & -\frac{1}{\sqrt{10}} & \frac{\sqrt{\frac{3}{5}}}{2} & -\frac{1}{6} & \frac{1}{3
   \sqrt{2}} \\
 -\frac{1}{\sqrt{6}} & -\frac{1}{\sqrt{10}} & -\frac{1}{\sqrt{15}} & -\frac{1}{3} & -\frac{1}{3 \sqrt{2}} \\
 0 & 0 & 0 & 0 & \frac{1}{\sqrt{2}} \\
 0 & 0 & 0 & \frac{2}{3} & -\frac{1}{3 \sqrt{2}} \\
 0 & 0 & \frac{\sqrt{\frac{5}{3}}}{2} & \frac{1}{6} & -\frac{1}{3 \sqrt{2}} \\
 0 & \sqrt{\frac{2}{5}} & -\frac{1}{2 \sqrt{15}} & -\frac{1}{6} & \frac{1}{3 \sqrt{2}} \\
 \frac{1}{\sqrt{6}} & -\frac{1}{\sqrt{10}} & -\frac{1}{\sqrt{15}} & -\frac{1}{3} & -\frac{1}{3 \sqrt{2}} \\
\end{array}
\right).
$$
\end{table}
 %Similarly  in ii):
\begin{table}
\caption{The reductions $[3]\times[1^2]\rightarrow 4[4,1]+6[3,1^2]$.}
\label{30x11}
$$
\left(
\begin{array}{cccc}
\hline
[4,1]_1&[4,1]_2&[4,1]_3&[4,1]_4\\
\hline
 0 & 0 & -\sqrt{\frac{3}{10}} & -\frac{1}{\sqrt{10}} \\
 0 & -\frac{2}{\sqrt{15}} & \frac{1}{\sqrt{30}} & \frac{1}{\sqrt{10}} \\
 \frac{1}{2} & -\frac{1}{2 \sqrt{15}} & \frac{1}{\sqrt{30}} & \frac{1}{\sqrt{10}} \\
 0 & 0 & 0 & \sqrt{\frac{2}{5}} \\
 0 & \frac{2}{\sqrt{15}} & \sqrt{\frac{2}{15}} & 0 \\
 \frac{1}{2} & -\frac{1}{2 \sqrt{15}} & -\sqrt{\frac{2}{15}} & 0 \\
 0 & 0 & \sqrt{\frac{3}{10}} & -\frac{1}{\sqrt{10}} \\
 -\frac{1}{2} & -\frac{\sqrt{\frac{3}{5}}}{2} & 0 & 0 \\
 0 & \frac{2}{\sqrt{15}} & -\frac{1}{\sqrt{30}} & \frac{1}{\sqrt{10}} \\
 \frac{1}{2} & -\frac{1}{2 \sqrt{15}} & \frac{1}{\sqrt{30}} & -\frac{1}{\sqrt{10}} \\
\end{array}
\right),
\left(
\begin{array}{cccccc}
\hline
[3,1^2]_1&[3,1^2]_2&[3,1^2]_3&[3,1^2]_4&[3,1^2]_5&[3,1^2]_6\\
\hline
 0 & 0 & 0 & \sqrt{\frac{2}{5}} & -\sqrt{\frac{2}{15}} & \frac{1}{\sqrt{15}} \\
 0 & \frac{\sqrt{\frac{3}{2}}}{2} & -\frac{1}{2 \sqrt{2}} & \frac{1}{2 \sqrt{10}} & -\frac{1}{2
   \sqrt{30}} & \frac{1}{\sqrt{15}} \\
 -\frac{1}{\sqrt{3}} & -\frac{1}{2 \sqrt{6}} & \frac{1}{2 \sqrt{2}} & \frac{1}{2 \sqrt{10}} &
   -\frac{\sqrt{\frac{3}{10}}}{2} & 0 \\
 \frac{1}{\sqrt{3}} & -\frac{1}{\sqrt{6}} & 0 & \frac{1}{\sqrt{10}} & 0 & 0 \\
 0 & 0 & 0 & 0 & 0 & \sqrt{\frac{3}{5}} \\
 0 & 0 & 0 & 0 & 2 \sqrt{\frac{2}{15}} & \frac{1}{\sqrt{15}} \\
 0 & 0 & 0 & \sqrt{\frac{2}{5}} & \sqrt{\frac{2}{15}} & -\frac{1}{\sqrt{15}} \\
 0 & 0 & \frac{1}{\sqrt{2}} & 0 & \frac{1}{\sqrt{30}} & \frac{1}{\sqrt{15}} \\
 0 & \frac{\sqrt{\frac{3}{2}}}{2} & \frac{1}{2 \sqrt{2}} & \frac{1}{2 \sqrt{10}} & \frac{1}{2
   \sqrt{30}} & -\frac{1}{\sqrt{15}} \\
 \frac{1}{\sqrt{3}} & \frac{1}{2 \sqrt{6}} & \frac{1}{2 \sqrt{2}} & -\frac{1}{2 \sqrt{10}} &
   -\frac{\sqrt{\frac{3}{10}}}{2} & 0 \\
\end{array}
\right).
$$
\end{table}
\begin{table}
\caption{The reductions $[1^3]\times[1^2]\rightarrow [1^5]+4[2,1^3]+5[2^2,1]$.}
\label{111x11}
$$
\left(
\begin{array}{c|cccc}
\hline
[1^5]&[2,1^3]_1&[2,1^3]_2&[2,1^3]_3&[2,1^3]_4\\
\hline
\frac{1}{\sqrt{10}}& \frac{1}{2 \sqrt{2}} & \frac{3}{2 \sqrt{14}} &
   \frac{1}{\sqrt{119}} & -38 \sqrt{\frac{2}{16711}} \\
 -\frac{1}{\sqrt{10}}&-\frac{1}{2 \sqrt{2}} & \frac{\sqrt{\frac{7}{2}}}{6} &
   \frac{\sqrt{\frac{7}{17}}}{3} & \frac{311}{3
   \sqrt{33422}} \\
 -\frac{1}{\sqrt{10}}&-\frac{1}{2 \sqrt{2}} & \frac{\sqrt{\frac{7}{2}}}{6} &
   -\frac{2 \sqrt{\frac{7}{17}}}{3} & \frac{46
   \sqrt{\frac{2}{16711}}}{3} \\
 \frac{1}{\sqrt{10}}&\frac{1}{2 \sqrt{2}} & \frac{3}{2 \sqrt{14}} &
   \frac{6}{\sqrt{119}} & 24 \sqrt{\frac{2}{16711}} \\
 \frac{1}{\sqrt{10}}&\frac{1}{2 \sqrt{2}} & \frac{3}{2 \sqrt{14}} &
   -\frac{1}{\sqrt{119}} & 13 \sqrt{\frac{2}{16711}} \\
0& \frac{1}{2 \sqrt{2}} & -\frac{\sqrt{\frac{7}{2}}}{6} &
   -\frac{\sqrt{\frac{7}{17}}}{3} & \frac{74
   \sqrt{\frac{2}{16711}}}{3} \\
-\frac{1}{\sqrt{10}}& 0 & 0 & \frac{\sqrt{\frac{7}{17}}}{2} & -\frac{131}{2
   \sqrt{33422}} \\
-\frac{1}{\sqrt{10}}& 0 & \frac{2 \sqrt{\frac{2}{7}}}{3} &
   -\frac{\sqrt{\frac{17}{7}}}{6} & -\frac{17
   \sqrt{\frac{17}{1966}}}{6} \\
0& -\frac{1}{2 \sqrt{2}} & -\frac{1}{6 \sqrt{14}} &
   \frac{31}{6 \sqrt{119}} & -\frac{313}{6 \sqrt{33422}}
   \\
0& \frac{1}{2 \sqrt{2}} & -\frac{\sqrt{\frac{7}{2}}}{6} &
   \frac{\sqrt{\frac{7}{17}}}{6} & \frac{209}{6
   \sqrt{33422}} \\
\end{array}
\right)
\left(
\begin{array}{ccccc}
\hline
[2,1^2]_1&[2,1^2]_2&[2,1^2]_3&[2,1^2]_4&[2,1^2]_5\\
\hline
 \frac{1}{\sqrt{11}} & -\frac{9}{\sqrt{319}} &
   -\frac{5}{2 \sqrt{87}} & -\frac{17}{2 \sqrt{921}} &
   \frac{31 \sqrt{\frac{5}{614}}}{13} \\
 0 & 0 & 0 & 0 & \frac{\sqrt{\frac{614}{5}}}{13} \\
 \frac{2}{\sqrt{11}} & \frac{4}{\sqrt{319}} &
   \frac{13}{3 \sqrt{87}} & \frac{37}{3 \sqrt{921}} & 2
   \sqrt{\frac{2}{1535}} \\
 -\frac{2}{\sqrt{11}} & -\frac{4}{\sqrt{319}} &
   \frac{16}{3 \sqrt{87}} & \frac{4}{3 \sqrt{921}} &
   \frac{36 \sqrt{\frac{2}{1535}}}{13} \\
 -\frac{1}{\sqrt{11}} & \frac{9}{\sqrt{319}} &
   -\frac{7}{3 \sqrt{87}} & \frac{5}{3 \sqrt{921}} &
   \frac{9 \sqrt{\frac{10}{307}}}{13} \\
 0 & 0 & 0 & 0 & \frac{\sqrt{\frac{307}{10}}}{13} \\
 0 & 0 & 0 & 6 \sqrt{\frac{3}{307}} & \frac{51}{13
   \sqrt{3070}} \\
 0 & 0 & \frac{\sqrt{\frac{29}{3}}}{6} & -\frac{67}{6
   \sqrt{921}} & \frac{11}{13 \sqrt{3070}} \\
 0 & \sqrt{\frac{11}{29}} & -\frac{1}{6 \sqrt{87}} &
   -\frac{61}{6 \sqrt{921}} & \sqrt{\frac{5}{614}} \\
 \frac{1}{\sqrt{11}} & \frac{2}{\sqrt{319}} &
   \frac{13}{6 \sqrt{87}} & -\frac{71}{6 \sqrt{921}} &
   -\frac{5 \sqrt{\frac{5}{614}}}{13} \\
\end{array}
\right)
$$
\end{table}
\begin{table}
\caption{The reductions $[1^3]\times[2]\rightarrow +[2,1^3]+6[3,1^2]$.}
\label{111x2}
$$
\left(
\begin{array}{cccc}
\hline
[2,1^3]_1&[2,1^3]_2&[2,1^3]_3&[2,1^3]_4\\
\hline
 0 & 0 & -\sqrt{\frac{3}{10}} & \frac{1}{\sqrt{10}} \\
 0 & -\frac{2}{\sqrt{15}} & -\frac{1}{\sqrt{30}} &
   \frac{1}{\sqrt{10}} \\
 -\frac{1}{2} & -\frac{1}{2 \sqrt{15}} &
   -\frac{1}{\sqrt{30}} & \frac{1}{\sqrt{10}} \\
 0 & 0 & 0 & \sqrt{\frac{2}{5}} \\
 0 & -\frac{2}{\sqrt{15}} & \sqrt{\frac{2}{15}} & 0 \\
 -\frac{1}{2} & -\frac{1}{2 \sqrt{15}} &
   \sqrt{\frac{2}{15}} & 0 \\
 0 & 0 & \sqrt{\frac{3}{10}} & \frac{1}{\sqrt{10}} \\
 -\frac{1}{2} & \frac{\sqrt{\frac{3}{5}}}{2} & 0 & 0 \\
 0 & \frac{2}{\sqrt{15}} & \frac{1}{\sqrt{30}} &
   \frac{1}{\sqrt{10}} \\
 \frac{1}{2} & \frac{1}{2 \sqrt{15}} &
   \frac{1}{\sqrt{30}} & \frac{1}{\sqrt{10}} \\
\end{array}
\right)
\left(
\begin{array}{cccccc}
\hline
[3,1^3]_1&[3,1^2]_2&[3,1^2]_3&[3,1^4]_3&[3,1^2]_5&[3,1^2]_6\\
\hline
 0 & 0 & 0 & \sqrt{\frac{2}{5}} & -\sqrt{\frac{2}{15}} &
   \frac{1}{\sqrt{15}} \\
 0 & -\frac{\sqrt{\frac{3}{2}}}{2} & -\frac{1}{2
   \sqrt{2}} & \frac{1}{2 \sqrt{10}} & -\frac{1}{2
   \sqrt{30}} & \frac{1}{\sqrt{15}} \\
 \frac{1}{\sqrt{3}} & \frac{1}{2 \sqrt{6}} & \frac{1}{2
   \sqrt{2}} & \frac{1}{2 \sqrt{10}} &
   -\frac{\sqrt{\frac{3}{10}}}{2} & 0 \\
 -\frac{1}{\sqrt{3}} & \frac{1}{\sqrt{6}} & 0 &
   \frac{1}{\sqrt{10}} & 0 & 0 \\
 0 & 0 & 0 & 0 & 0 & \sqrt{\frac{3}{5}} \\
 0 & 0 & 0 & 0 & 2 \sqrt{\frac{2}{15}} &
   \frac{1}{\sqrt{15}} \\
 0 & 0 & 0 & \sqrt{\frac{2}{5}} & \sqrt{\frac{2}{15}} &
   -\frac{1}{\sqrt{15}} \\
 0 & 0 & \frac{1}{\sqrt{2}} & 0 & \frac{1}{\sqrt{30}} &
   \frac{1}{\sqrt{15}} \\
 0 & \frac{\sqrt{\frac{3}{2}}}{2} & -\frac{1}{2
   \sqrt{2}} & -\frac{1}{2 \sqrt{10}} & -\frac{1}{2
   \sqrt{30}} & \frac{1}{\sqrt{15}} \\
 \frac{1}{\sqrt{3}} & \frac{1}{2 \sqrt{6}} & -\frac{1}{2
   \sqrt{2}} & \frac{1}{2 \sqrt{10}} &
   \frac{\sqrt{\frac{3}{10}}}{2} & 0 \\
\end{array}
\right)
$$
\end{table}
\clearpage
A bit more complicated are the reductions:
$$[2,1]\times[2]\rightarrow 4[4,1]+5[3,2]+5[2,2,1]+6[3,1^2],$$
$$ [2,1]\times[1^2]\rightarrow 4[2,1^3]+5[2,2,1]+5[3,2]+6[3,1^2]$$
The appropriate basis is analogous to the  one  above with the obvious substitution $$\{1,2,3\}\{4,5\},\cdots \{5,4,3\}\{1,2\}\rightarrow [2,1]_1(1,2,3)\{4,5\},\cdots [2,1]_1(5,4,3)\{1,2\},[2,1]_2(1,2,3)\{4,5\},\cdots [2,1]_2(5,4,3)\{1,2\}$$
(the [2,1] is two dimensional and the space is 20-dimensional).
Let us begin with the first. We get the results below in tables \ref{21x2} and  \ref{21x11}:
\begin{table}
\caption{The reduction $[2,1]\times[2]\rightarrow 4[4,1]+5[3,2]+5[2,2,1]+6[3,1^2]$}
\label{21x2}
$$
\left(
\begin{array}{cccc}
\hline
[4,1]_1&[4,1]_2&[4,1]_3&[4,1]_4\\
\hline
 \sqrt{\frac{2}{11}} & -\frac{1}{\sqrt{55}} & 0 & 0 \\
 \frac{1}{2 \sqrt{22}} & -\frac{3}{\sqrt{55}} & \frac{1}{2 \sqrt{30}} & -\frac{1}{2 \sqrt{15}} \\
 \frac{3}{2 \sqrt{22}} & \frac{2}{\sqrt{55}} & -\frac{1}{2 \sqrt{30}} & \frac{1}{2 \sqrt{15}} \\
 \sqrt{\frac{2}{11}} & -\frac{1}{\sqrt{55}} & 0 & 0 \\
 \frac{1}{\sqrt{22}} & -\frac{1}{2 \sqrt{55}} & 0 & -\frac{\sqrt{\frac{3}{5}}}{2} \\
 \frac{1}{\sqrt{22}} & -\frac{1}{2 \sqrt{55}} & 0 & \frac{\sqrt{\frac{3}{5}}}{2} \\
 \sqrt{\frac{2}{11}} & -\frac{1}{\sqrt{55}} & 0 & 0 \\
 \frac{1}{2 \sqrt{22}} & \frac{\sqrt{\frac{5}{11}}}{2} & -\frac{1}{2 \sqrt{30}} &
   -\frac{1}{\sqrt{15}} \\
 \frac{1}{\sqrt{22}} & -\frac{1}{2 \sqrt{55}} & 0 & -\frac{\sqrt{\frac{3}{5}}}{2} \\
 \frac{1}{\sqrt{22}} & -\frac{1}{2 \sqrt{55}} & 0 & \frac{\sqrt{\frac{3}{5}}}{2} \\
 \frac{1}{2 \sqrt{66}} & \frac{\sqrt{\frac{5}{33}}}{2} & \frac{7}{6 \sqrt{10}} & \frac{1}{3
   \sqrt{5}} \\
 -\frac{1}{\sqrt{66}} & \frac{1}{2 \sqrt{165}} & \frac{2 \sqrt{\frac{2}{5}}}{3} & \frac{1}{6
   \sqrt{5}} \\
 \frac{1}{\sqrt{66}} & -\frac{1}{2 \sqrt{165}} & \frac{2 \sqrt{\frac{2}{5}}}{3} & \frac{1}{6
   \sqrt{5}} \\
 \frac{1}{\sqrt{66}} & \sqrt{\frac{5}{33}} & -\frac{1}{3 \sqrt{10}} & \frac{1}{3 \sqrt{5}} \\
 -\frac{1}{2 \sqrt{66}} & \sqrt{\frac{3}{55}} & \frac{7}{6 \sqrt{10}} & -\frac{1}{6 \sqrt{5}} \\
 \frac{\sqrt{\frac{3}{22}}}{2} & \frac{2}{\sqrt{165}} & \frac{7}{6 \sqrt{10}} & -\frac{1}{6
   \sqrt{5}} \\
 0 & 0 & 0 & \frac{1}{\sqrt{5}} \\
 0 & 0 & \frac{2 \sqrt{\frac{2}{5}}}{3} & -\frac{1}{3 \sqrt{5}} \\
 0 & \frac{\sqrt{\frac{11}{15}}}{2} & -\frac{1}{3 \sqrt{10}} & -\frac{1}{6 \sqrt{5}} \\
 \sqrt{\frac{2}{33}} & \frac{3 \sqrt{\frac{3}{55}}}{2} & -\frac{1}{3 \sqrt{10}} & -\frac{1}{6
   \sqrt{5}} \\
\end{array}
\right),
\left(
\begin{array}{cccccc}
\hline
[3,1^2]_2&[3,1^2]_2&[3,1^2]_3&[3,1^2]_4&[3,1^2]_5&[3,1^2]_6\\
\hline
 -\frac{1}{\sqrt{6}} & \frac{1}{\sqrt{22}} & -\frac{\sqrt{\frac{3}{11}}}{4} & -\frac{\sqrt{3}}{8}
   & 0 & \frac{3}{8} \\
 0 & -\sqrt{\frac{2}{11}} & \frac{\sqrt{\frac{3}{11}}}{2} & 0 & 0 & 0 \\
 0 & 0 & \frac{\sqrt{\frac{11}{3}}}{4} & \frac{1}{8 \sqrt{3}} & 0 & -\frac{1}{8} \\
 -\frac{1}{\sqrt{6}} & 0 & -\frac{\sqrt{\frac{11}{3}}}{8} & \frac{1}{2 \sqrt{3}} &
   \frac{\sqrt{3}}{8} & -\frac{1}{4} \\
 \frac{1}{\sqrt{6}} & \frac{1}{\sqrt{22}} & -\frac{\sqrt{\frac{3}{11}}}{4} & \frac{\sqrt{3}}{8} &
   0 & -\frac{3}{8} \\
 \frac{1}{\sqrt{6}} & -\frac{1}{\sqrt{22}} & -\frac{2}{\sqrt{33}} & \frac{1}{4 \sqrt{3}} & 0 &
   -\frac{1}{4} \\
 0 & -\frac{1}{\sqrt{22}} & \frac{17}{8 \sqrt{33}} & -\frac{1}{8 \sqrt{3}} & -\frac{\sqrt{3}}{8} &
   -\frac{1}{8} \\
 0 & -\sqrt{\frac{2}{11}} & -\frac{5}{4 \sqrt{33}} & -\frac{1}{8 \sqrt{3}} & 0 & \frac{1}{8} \\
 \frac{1}{\sqrt{6}} & \frac{1}{\sqrt{22}} & \frac{5}{8 \sqrt{33}} & -\frac{1}{2 \sqrt{3}} &
   \frac{\sqrt{3}}{8} & \frac{1}{4} \\
 \frac{1}{\sqrt{6}} & 0 & 0 & 0 & 0 & \frac{1}{2} \\
 0 & 0 & 0 & \frac{3}{8} & -\frac{1}{4} & \frac{\sqrt{3}}{8} \\
 0 & 0 & 0 & 0 & \frac{1}{2} & 0 \\
 0 & 0 & 0 & -\frac{3}{8} & -\frac{1}{4} & -\frac{\sqrt{3}}{8} \\
 0 & -\sqrt{\frac{3}{22}} & -\frac{5}{8 \sqrt{11}} & -\frac{1}{4} & \frac{1}{8} & 0 \\
 0 & 0 & 0 & -\frac{3}{8} & -\frac{1}{4} & -\frac{\sqrt{3}}{8} \\
 0 & 0 & 0 & 0 & \frac{1}{2} & 0 \\
 0 & \sqrt{\frac{3}{22}} & \frac{5}{8 \sqrt{11}} & -\frac{1}{8} & \frac{1}{8} &
   -\frac{\sqrt{3}}{8} \\
 0 & 0 & 0 & \frac{3}{8} & -\frac{1}{4} & \frac{\sqrt{3}}{8} \\
 0 & 0 & \frac{\sqrt{11}}{8} & \frac{1}{4} & \frac{1}{8} & 0 \\
 0 & \sqrt{\frac{3}{22}} & -\frac{3}{4 \sqrt{11}} & 0 & -\frac{1}{4} & 0 \\
\end{array}
\right)
 $$
\end{table}
\begin{table}
\caption{The reduction $[2,1]\times[2]\rightarrow 4[4,1]+5[3,2]+5[2,2,1]+6[3,1^2]$ (table \ref{21x2} continued)}
\label{21x2a}
$$
\left(
\begin{array}{ccccc}
\hline
[3,2]_1&[3,2]_2&[3,2]_3&[3,2]_4&[3,2]_5\\
\hline
 \frac{31}{7 \sqrt{134}} & -\frac{\sqrt{\frac{3}{469}}}{2} & -\frac{1}{7 \sqrt{17}} & -\frac{5}{2
   \sqrt{357}} & -\frac{1}{2 \sqrt{3}} \\
 2 \sqrt{\frac{2}{67}} & \frac{9 \sqrt{\frac{3}{469}}}{4} & -\frac{5}{4 \sqrt{17}} &
   \frac{\sqrt{\frac{3}{119}}}{2} & 0 \\
 \frac{3}{7 \sqrt{134}} & -\frac{11 \sqrt{\frac{3}{469}}}{4} & \frac{31}{28 \sqrt{17}} &
   -\frac{4}{\sqrt{357}} & -\frac{1}{2 \sqrt{3}} \\
 -\frac{29}{7 \sqrt{134}} & \frac{17 \sqrt{\frac{3}{469}}}{4} & -\frac{1}{28 \sqrt{17}} &
   -\frac{5}{8 \sqrt{357}} & -\frac{1}{8 \sqrt{3}} \\
 \frac{4 \sqrt{\frac{2}{67}}}{7} & -\frac{\sqrt{\frac{21}{67}}}{4} & -\frac{41}{28 \sqrt{17}} &
   \frac{2}{\sqrt{357}} & -\frac{1}{2 \sqrt{3}} \\
 \frac{23}{7 \sqrt{134}} & \frac{5 \sqrt{\frac{3}{469}}}{4} & \frac{37}{28 \sqrt{17}} & -\frac{3
   \sqrt{\frac{3}{119}}}{2} & 0 \\
 -\frac{\sqrt{\frac{2}{67}}}{7} & -\frac{15 \sqrt{\frac{3}{469}}}{4} & \frac{5}{28 \sqrt{17}} &
   \frac{25}{8 \sqrt{357}} & \frac{5}{8 \sqrt{3}} \\
 -\frac{10 \sqrt{\frac{2}{67}}}{7} & -4 \sqrt{\frac{3}{469}} & -\frac{3}{14 \sqrt{17}} &
   \frac{1}{2 \sqrt{357}} & -\frac{1}{2 \sqrt{3}} \\
 -\frac{1}{\sqrt{134}} & \frac{29 \sqrt{\frac{3}{469}}}{8} & \frac{11}{8 \sqrt{17}} & \frac{31}{8
   \sqrt{357}} & -\frac{1}{8 \sqrt{3}} \\
 -\frac{11 \sqrt{\frac{2}{67}}}{7} & \frac{5 \sqrt{\frac{3}{469}}}{8} & -\frac{79}{56 \sqrt{17}} &
   -\frac{3 \sqrt{\frac{3}{119}}}{2} & 0 \\
 \frac{5 \sqrt{\frac{2}{201}}}{7} & \frac{2}{\sqrt{469}} & -\frac{23}{14 \sqrt{51}} & \frac{55}{12
   \sqrt{119}} & -\frac{1}{12} \\
 \frac{11 \sqrt{\frac{2}{201}}}{7} & -\frac{9}{\sqrt{469}} & -\frac{5}{7 \sqrt{51}} & -\frac{4}{3
   \sqrt{119}} & -\frac{1}{6} \\
 -\frac{\sqrt{\frac{67}{6}}}{7} & 0 & -\frac{3 \sqrt{\frac{3}{17}}}{14} & -\frac{11}{12
   \sqrt{119}} & -\frac{1}{12} \\
 \frac{10 \sqrt{\frac{2}{201}}}{7} & \frac{4}{\sqrt{469}} & \frac{\sqrt{\frac{3}{17}}}{14} &
   \frac{115}{24 \sqrt{119}} & -\frac{1}{24} \\
 \frac{5}{\sqrt{402}} & \sqrt{\frac{7}{67}} & -\frac{1}{2 \sqrt{51}} & -\frac{5
   \sqrt{\frac{7}{17}}}{12} & -\frac{1}{12} \\
 0 & 0 & 0 & 0 & \frac{1}{2} \\
 0 & 0 & 0 & \frac{\sqrt{119}}{24} & -\frac{5}{24} \\
 0 & 0 & \frac{7}{2 \sqrt{51}} & \frac{\sqrt{\frac{7}{17}}}{12} & -\frac{1}{12} \\
 0 & \frac{\sqrt{\frac{67}{7}}}{8} & -\frac{11}{8 \sqrt{51}} & -\frac{65}{24 \sqrt{119}} &
   -\frac{1}{24} \\
 \frac{25}{7 \sqrt{402}} & -\frac{27}{8 \sqrt{469}} & -\frac{55}{56 \sqrt{51}} & -\frac{5}{12
   \sqrt{119}} & \frac{5}{12} \\
\end{array}
\right),
\left(
\begin{array}{ccccc}
\hline
[2^2,1]_1&[2^2,1]_2&[2^2,1]_3&[2^2,1]_4&[2^2,1]_5\\
\hline
 -\frac{1}{\sqrt{22}} & \sqrt{\frac{5}{374}} & \frac{15 \sqrt{\frac{5}{374}}}{8} &
   \frac{\sqrt{\frac{3}{110}}}{8} & \frac{1}{2 \sqrt{10}} \\
 \sqrt{\frac{2}{11}} & -\sqrt{\frac{10}{187}} & \frac{\sqrt{\frac{5}{374}}}{2} & \frac{3
   \sqrt{\frac{3}{110}}}{2} & 0 \\
 \frac{1}{\sqrt{22}} & -8 \sqrt{\frac{2}{935}} & -\frac{19}{8 \sqrt{1870}} & -\frac{43}{8
   \sqrt{330}} & \frac{1}{2 \sqrt{10}} \\
 -\frac{1}{\sqrt{22}} & -\sqrt{\frac{17}{110}} & \frac{\sqrt{\frac{17}{110}}}{8} & \frac{13}{8
   \sqrt{330}} & -\frac{3}{4 \sqrt{10}} \\
 0 & \sqrt{\frac{11}{170}} & \frac{29}{8 \sqrt{1870}} & -\frac{9 \sqrt{\frac{3}{110}}}{8} &
   \frac{1}{2 \sqrt{10}} \\
 \frac{1}{\sqrt{22}} & \sqrt{\frac{17}{110}} & -\frac{3 \sqrt{\frac{17}{110}}}{4} &
   \frac{\sqrt{\frac{5}{66}}}{4} & 0 \\
 -\sqrt{\frac{2}{11}} & \sqrt{\frac{10}{187}} & -\frac{\sqrt{\frac{5}{374}}}{2} &
   \frac{1}{\sqrt{330}} & -\frac{3}{4 \sqrt{10}} \\
 0 & 0 & -\frac{\sqrt{\frac{85}{22}}}{8} & \frac{67}{8 \sqrt{330}} & \frac{1}{2 \sqrt{10}} \\
 \frac{1}{\sqrt{22}} & 3 \sqrt{\frac{2}{935}} & -\frac{131}{8 \sqrt{1870}} & -\frac{7}{8
   \sqrt{330}} & -\frac{3}{4 \sqrt{10}} \\
 0 & 0 & 0 & 0 & 0 \\
 0 & 0 & -\frac{\sqrt{\frac{255}{22}}}{8} & -\frac{21}{8 \sqrt{110}} & 0 \\
 0 & 0 & 0 & 0 & -\sqrt{\frac{3}{10}} \\
 \sqrt{\frac{3}{22}} & 3 \sqrt{\frac{6}{935}} & \frac{39 \sqrt{\frac{3}{1870}}}{8} & -\frac{9}{8
   \sqrt{110}} & 0 \\
 0 & 0 & \frac{\sqrt{\frac{255}{22}}}{8} & -\frac{23}{8 \sqrt{110}} &
   -\frac{\sqrt{\frac{3}{10}}}{4} \\
 -\sqrt{\frac{3}{22}} & -3 \sqrt{\frac{6}{935}} & -\frac{39 \sqrt{\frac{3}{1870}}}{8} & \frac{9}{8
   \sqrt{110}} & 0 \\
 0 & 0 & 0 & 0 & \sqrt{\frac{3}{10}} \\
 0 & 0 & 0 & \frac{\sqrt{\frac{11}{10}}}{2} & \frac{\sqrt{\frac{3}{10}}}{4} \\
 0 & 0 & \frac{\sqrt{\frac{255}{22}}}{8} & \frac{21}{8 \sqrt{110}} & 0 \\
 0 & \sqrt{\frac{33}{170}} & \frac{29 \sqrt{\frac{3}{1870}}}{8} & \frac{17}{8 \sqrt{110}} &
   -\frac{\sqrt{\frac{3}{10}}}{4} \\
 \sqrt{\frac{3}{22}} & -\sqrt{\frac{15}{374}} & \frac{\sqrt{\frac{15}{374}}}{4} & \frac{9}{4
   \sqrt{110}} & -\frac{\sqrt{\frac{3}{10}}}{2} \\
\end{array}
\right).
$$
\end{table}
%In the reduction:
%$$ [2,1]\times[1^2]\rightarrow 4[2,1^3]+5[2,2,1]+5[3,2]+6[3,1^2],$$
%the obtained results are:
\begin{table}
\caption{The reduction $ [2,1]\times[1^2]\rightarrow 4[2,1^3]+5[2,2,1]+5[3,2]+6[3,1^2]$}
\label{21x11}
$$
\left(
\begin{array}{cccc}
\hline
[2,1^3]_1&[2,1^3]_2&[2,1^3]_3&[2,1^3]_4\\
\hline
 -\frac{1}{2 \sqrt{6}} &
   \frac{\sqrt{\frac{5}{2}}}{6} & 0 & \frac{2}{3
   \sqrt{5}} \\
 0 & -\frac{\sqrt{\frac{2}{5}}}{3} & -\frac{1}{2
   \sqrt{15}} & -\frac{\sqrt{5}}{6} \\
 \frac{1}{2 \sqrt{6}} & -\frac{1}{6 \sqrt{10}} &
   -\frac{1}{2 \sqrt{15}} & -\frac{\sqrt{5}}{6} \\
 \frac{1}{2 \sqrt{6}} &
   -\frac{\sqrt{\frac{5}{2}}}{6} &
   \frac{1}{\sqrt{15}} & \frac{1}{3 \sqrt{5}} \\
 0 & \frac{\sqrt{\frac{2}{5}}}{3} & -\frac{1}{2
   \sqrt{15}} & \frac{\sqrt{5}}{6} \\
 -\frac{1}{2 \sqrt{6}} & \frac{1}{6 \sqrt{10}} &
   -\frac{1}{2 \sqrt{15}} & \frac{\sqrt{5}}{6} \\
 -\frac{1}{2 \sqrt{6}} &
   \frac{\sqrt{\frac{5}{2}}}{6} &
   \frac{1}{\sqrt{15}} & -\frac{1}{3 \sqrt{5}} \\
 0 & 0 & 0 & \frac{1}{\sqrt{5}} \\
 0 & \frac{\sqrt{\frac{2}{5}}}{3} &
   -\frac{\sqrt{\frac{3}{5}}}{2} & -\frac{1}{6
   \sqrt{5}} \\
 -\frac{1}{2 \sqrt{6}} & \frac{1}{6 \sqrt{10}} &
   -\frac{\sqrt{\frac{3}{5}}}{2} & -\frac{1}{6
   \sqrt{5}} \\
 \frac{1}{2 \sqrt{2}} &
   \frac{\sqrt{\frac{3}{10}}}{2} & 0 & 0 \\
 0 & -\sqrt{\frac{2}{15}} & \frac{1}{2 \sqrt{5}} &
   \frac{1}{2 \sqrt{15}} \\
 -\frac{1}{2 \sqrt{2}} & \frac{1}{2 \sqrt{30}} &
   -\frac{1}{2 \sqrt{5}} & -\frac{1}{2 \sqrt{15}}
   \\
 -\frac{1}{2 \sqrt{2}} &
   -\frac{\sqrt{\frac{3}{10}}}{2} & 0 & 0 \\
 0 & \sqrt{\frac{2}{15}} & \frac{1}{2 \sqrt{5}} &
   -\frac{1}{2 \sqrt{15}} \\
 \frac{1}{2 \sqrt{2}} & -\frac{1}{2 \sqrt{30}} &
   -\frac{1}{2 \sqrt{5}} & \frac{1}{2 \sqrt{15}} \\
 \frac{1}{2 \sqrt{2}} &
   \frac{\sqrt{\frac{3}{10}}}{2} & 0 & 0 \\
 0 & 0 & \frac{1}{\sqrt{5}} & 0 \\
 0 & \sqrt{\frac{2}{15}} & \frac{1}{2 \sqrt{5}} &
   -\frac{1}{2 \sqrt{15}} \\
 \frac{1}{2 \sqrt{2}} & -\frac{1}{2 \sqrt{30}} &
   -\frac{1}{2 \sqrt{5}} & \frac{1}{2 \sqrt{15}} \\
\end{array}
\right),
\left(
\begin{array}{cccccc}
\hline
[3,1^2]_1&3,1^2]_2&[3,1^2]_3&[3,1^2]_4&[3,1^2]_5&[3,1^2]_6\\
\hline
 4 \sqrt{\frac{2}{177}} &
   -\frac{\sqrt{\frac{19}{177}}}{4} &
   \frac{\sqrt{\frac{3}{5}}}{4} & 0 & -\frac{1}{2
   \sqrt{5}} & \frac{1}{2 \sqrt{10}} \\
 \frac{5}{\sqrt{354}} & \frac{109}{8 \sqrt{3363}} &
   -\frac{3 \sqrt{\frac{3}{5}}}{8} & \frac{5
   \sqrt{\frac{3}{19}}}{8} & \frac{1}{8 \sqrt{5}} &
   -\frac{1}{2 \sqrt{10}} \\
 \sqrt{\frac{3}{118}} & -\frac{49
   \sqrt{\frac{3}{1121}}}{8} & -\frac{3
   \sqrt{\frac{3}{5}}}{8} & -\frac{5
   \sqrt{\frac{3}{19}}}{8} & \frac{3}{8 \sqrt{5}} &
   0 \\
 2 \sqrt{\frac{2}{177}} &
   -\frac{\sqrt{\frac{19}{177}}}{8} & -\frac{3
   \sqrt{\frac{3}{5}}}{8} & 0 & -\frac{3}{4
   \sqrt{5}} & \frac{3}{4 \sqrt{10}} \\
 \sqrt{\frac{3}{118}} & -\frac{49
   \sqrt{\frac{3}{1121}}}{8} & -\frac{3
   \sqrt{\frac{3}{5}}}{8} & -\frac{5
   \sqrt{\frac{3}{19}}}{8} & \frac{3}{8 \sqrt{5}} &
   0 \\
 \frac{5}{\sqrt{354}} & \frac{109}{8 \sqrt{3363}} &
   -\frac{3 \sqrt{\frac{3}{5}}}{8} & \frac{5
   \sqrt{\frac{3}{19}}}{8} & \frac{1}{8 \sqrt{5}} &
   -\frac{1}{2 \sqrt{10}} \\
 -2 \sqrt{\frac{2}{177}} &
   \frac{\sqrt{\frac{19}{177}}}{8} &
   -\frac{\sqrt{15}}{8} & 0 & -\frac{1}{4 \sqrt{5}}
   & \frac{1}{4 \sqrt{10}} \\
 -4 \sqrt{\frac{2}{177}} &
   \frac{\sqrt{\frac{19}{177}}}{4} &
   -\frac{\sqrt{\frac{3}{5}}}{4} & 0 & \frac{1}{2
   \sqrt{5}} & -\frac{1}{2 \sqrt{10}} \\
 -\frac{5}{\sqrt{354}} & \frac{313}{16 \sqrt{3363}}
   & \frac{\sqrt{\frac{3}{5}}}{16} & -\frac{9
   \sqrt{\frac{3}{19}}}{16} & -\frac{9}{16
   \sqrt{5}} & 0 \\
 -\frac{7}{\sqrt{354}} & -\frac{199}{16
   \sqrt{3363}} & \frac{\sqrt{\frac{3}{5}}}{16} &
   \frac{9 \sqrt{\frac{3}{19}}}{16} & -\frac{3}{16
   \sqrt{5}} & \frac{3}{4 \sqrt{10}} \\
 -\frac{1}{\sqrt{118}} & -\frac{16}{\sqrt{1121}} &
   0 & \frac{1}{2 \sqrt{19}} & -\frac{1}{2
   \sqrt{15}} & -\frac{1}{\sqrt{30}} \\
 -\frac{3}{\sqrt{118}} & -\frac{15}{4 \sqrt{1121}}
   & -\frac{1}{4 \sqrt{5}} & \frac{7}{4 \sqrt{19}}
   & -\frac{1}{4 \sqrt{15}} & \frac{1}{\sqrt{30}}
   \\
 \frac{7}{3 \sqrt{118}} & -\frac{83}{12
   \sqrt{1121}} & \frac{1}{4 \sqrt{5}} & \frac{7}{4
   \sqrt{19}} & \frac{\sqrt{\frac{3}{5}}}{4} & 0 \\
 -\frac{1}{3 \sqrt{118}} & -\frac{16}{3
   \sqrt{1121}} & 0 & -\frac{5}{8 \sqrt{19}} &
   \frac{3 \sqrt{\frac{3}{5}}}{8} & \frac{3
   \sqrt{\frac{3}{10}}}{4} \\
 0 & 0 & 0 & 0 & 0 & \sqrt{\frac{3}{10}} \\
 0 & 0 & 0 & 0 & \frac{2}{\sqrt{15}} &
   \frac{1}{\sqrt{30}} \\
 0 & 0 & 0 & \frac{\sqrt{19}}{8} & \frac{1}{8
   \sqrt{15}} & \frac{1}{4 \sqrt{30}} \\
 0 & 0 & \frac{1}{\sqrt{5}} & 0 &
   \frac{1}{\sqrt{15}} & -\frac{1}{\sqrt{30}} \\
 0 & \frac{3 \sqrt{\frac{59}{19}}}{16} &
   -\frac{\sqrt{5}}{16} & \frac{1}{16 \sqrt{19}} &
   \frac{5 \sqrt{\frac{5}{3}}}{16} & \frac{1}{2
   \sqrt{30}} \\
 \frac{2 \sqrt{\frac{2}{59}}}{3} & \frac{493}{48
   \sqrt{1121}} & \frac{\sqrt{5}}{16} & \frac{1}{16
   \sqrt{19}} & -\frac{\sqrt{\frac{3}{5}}}{16} &
   \frac{3 \sqrt{\frac{3}{10}}}{4} \\
\end{array}
\right)
$$
\end{table}
\begin{table}
\caption{The reduction $ [2,1]\times[1^2]\rightarrow 4[2,1^3]+5[2,2,1]+5[3,2]+6[3,1^2]$ (table \ref{21x11} continued).}
\label{21x11a}
%$ [2,1]\times[1^2]\rightarrow 4[2,1^3]+5[2,2,1]+5[3,2]+6[3,1^2] (able \ref{21x11} contiued).$
$$
\left(
\begin{array}{ccccc}
\hline
[2^2,1]_1&[2^2,1]_2&[2^2,1]_3&[2^2,1]_4&[2^2,1]_5\\
\hline
 \sqrt{\frac{2}{17}} & -\frac{5}{2 \sqrt{561}} &
   \frac{4}{3 \sqrt{55}} & \frac{1}{6 \sqrt{5}} &
   \frac{1}{2 \sqrt{3}} \\
 -\frac{1}{\sqrt{34}} & -\frac{29}{4 \sqrt{561}} &
   \frac{31}{12 \sqrt{55}} & \frac{1}{6 \sqrt{5}} &
   0 \\
 \sqrt{\frac{2}{17}} & \frac{7}{4 \sqrt{561}} &
   \frac{13}{12 \sqrt{55}} & -\frac{1}{3 \sqrt{5}}
   & -\frac{1}{2 \sqrt{3}} \\
 -\sqrt{\frac{2}{17}} & \frac{9
   \sqrt{\frac{3}{187}}}{4} & -\frac{19}{12
   \sqrt{55}} & -\frac{1}{24 \sqrt{5}} &
   -\frac{1}{8 \sqrt{3}} \\
 0 & \frac{\sqrt{\frac{17}{33}}}{4} & \frac{19}{12
   \sqrt{55}} & \frac{2}{3 \sqrt{5}} & -\frac{1}{2
   \sqrt{3}} \\
 -\frac{1}{\sqrt{34}} & \frac{5}{4 \sqrt{561}} &
   \frac{5 \sqrt{\frac{5}{11}}}{12} &
   -\frac{\sqrt{5}}{6} & 0 \\
 \sqrt{\frac{2}{17}} & -\frac{9
   \sqrt{\frac{3}{187}}}{4} & -\frac{1}{4
   \sqrt{55}} & \frac{1}{8 \sqrt{5}} &
   \frac{\sqrt{3}}{8} \\
 0 & -\frac{\sqrt{\frac{17}{33}}}{2} & -\frac{4}{3
   \sqrt{55}} & -\frac{1}{6 \sqrt{5}} & -\frac{1}{2
   \sqrt{3}} \\
 -\frac{1}{\sqrt{34}} & \frac{9
   \sqrt{\frac{3}{187}}}{8} & \frac{47}{24
   \sqrt{55}} & \frac{19}{24 \sqrt{5}} &
   -\frac{1}{8 \sqrt{3}} \\
 0 & \frac{\sqrt{\frac{51}{11}}}{8} & \frac{7
   \sqrt{\frac{5}{11}}}{24} & -\frac{\sqrt{5}}{6} &
   0 \\
 -\sqrt{\frac{3}{34}} & -\frac{3}{\sqrt{187}} &
   \frac{\sqrt{\frac{3}{55}}}{2} &
   -\frac{\sqrt{\frac{3}{5}}}{4} & \frac{1}{4} \\
 \sqrt{\frac{3}{34}} & \frac{3}{\sqrt{187}} &
   \frac{4}{\sqrt{165}} & \frac{1}{2 \sqrt{15}} & 0
   \\
 0 & 0 & -\frac{\sqrt{\frac{11}{15}}}{2} &
   \frac{1}{4 \sqrt{15}} & -\frac{1}{4} \\
 -\sqrt{\frac{3}{34}} & -\frac{3}{\sqrt{187}} &
   \frac{\sqrt{\frac{3}{55}}}{2} & \frac{3
   \sqrt{\frac{3}{5}}}{8} & \frac{1}{8} \\
 -\sqrt{\frac{3}{34}} & -\frac{3}{\sqrt{187}} &
   \frac{\sqrt{\frac{3}{55}}}{2} &
   -\frac{\sqrt{\frac{3}{5}}}{4} & -\frac{1}{4} \\
 0 & 0 & 0 & 0 & \frac{1}{2} \\
 0 & 0 & 0 & \frac{\sqrt{15}}{8} & -\frac{1}{8} \\
 0 & 0 & \frac{\sqrt{\frac{11}{15}}}{2} &
   -\frac{1}{4 \sqrt{15}} & -\frac{1}{4} \\
 0 & \frac{3 \sqrt{\frac{17}{11}}}{8} & -\frac{3
   \sqrt{\frac{3}{55}}}{8} &
   -\frac{\sqrt{\frac{3}{5}}}{8} & \frac{1}{8} \\
 \sqrt{\frac{3}{34}} & -\frac{27}{8 \sqrt{187}} &
   -\frac{\sqrt{\frac{3}{55}}}{8} &
   -\frac{\sqrt{\frac{3}{5}}}{4} & -\frac{1}{4} \\
\end{array}
\right),
\left(
\begin{array}{ccccc}
\hline
[3,2]_1&[3,2]_2&[3,2]_3&[3,2]_4&[3,2]_5\\
\hline
 -\sqrt{\frac{2}{17}} & \frac{3}{2 \sqrt{17}} & 0 &
   0 & 0 \\
 \frac{1}{2 \sqrt{34}} & \frac{7}{4 \sqrt{17}} & 0
   & -\frac{1}{8} & -\frac{\sqrt{3}}{8} \\
 -\frac{5}{2 \sqrt{34}} & -\frac{1}{4 \sqrt{17}} &
   0 & \frac{1}{8} & \frac{\sqrt{3}}{8} \\
 -\frac{1}{\sqrt{34}} & \frac{3}{4 \sqrt{17}} &
   \frac{\sqrt{3}}{4} & 0 & 0 \\
 \frac{5}{2 \sqrt{34}} & \frac{1}{4 \sqrt{17}} & 0
   & -\frac{1}{8} & -\frac{\sqrt{3}}{8} \\
 -\frac{1}{2 \sqrt{34}} & -\frac{7}{4 \sqrt{17}} &
   0 & \frac{1}{8} & \frac{\sqrt{3}}{8} \\
 \frac{1}{\sqrt{34}} & -\frac{3}{4 \sqrt{17}} &
   \frac{\sqrt{3}}{4} & 0 & 0 \\
 -\sqrt{\frac{2}{17}} & \frac{3}{2 \sqrt{17}} & 0 &
   0 & 0 \\
 -\sqrt{\frac{2}{17}} & -\frac{5}{8 \sqrt{17}} &
   \frac{\sqrt{3}}{8} & -\frac{1}{4} & 0 \\
 \frac{1}{\sqrt{34}} & \frac{11}{8 \sqrt{17}} &
   \frac{\sqrt{3}}{8} & \frac{1}{4} & 0 \\
 0 & 0 & 0 & -\frac{\sqrt{3}}{4} & \frac{1}{4} \\
 -\frac{\sqrt{\frac{3}{34}}}{2} &
   -\frac{1}{\sqrt{51}} & -\frac{1}{4} &
   -\frac{5}{8 \sqrt{3}} & -\frac{1}{8} \\
 -\frac{\sqrt{\frac{3}{34}}}{2} &
   -\frac{1}{\sqrt{51}} & \frac{1}{4} & -\frac{5}{8
   \sqrt{3}} & -\frac{1}{8} \\
 0 & 0 & 0 & 0 & \frac{1}{2} \\
 -\frac{\sqrt{\frac{3}{34}}}{2} &
   -\frac{1}{\sqrt{51}} & -\frac{1}{4} & \frac{1}{8
   \sqrt{3}} & -\frac{3}{8} \\
 -\frac{\sqrt{\frac{3}{34}}}{2} &
   -\frac{1}{\sqrt{51}} & \frac{1}{4} & \frac{1}{8
   \sqrt{3}} & -\frac{3}{8} \\
 0 & 0 & 0 & \frac{\sqrt{3}}{4} & \frac{1}{4} \\
 0 & 0 & \frac{1}{2} & 0 & 0 \\
 0 & \frac{\sqrt{\frac{17}{3}}}{8} & -\frac{1}{8} &
   -\frac{1}{2 \sqrt{3}} & \frac{1}{4} \\
 \sqrt{\frac{3}{34}} & -\frac{1}{8 \sqrt{51}} &
   \frac{1}{8} & -\frac{1}{2 \sqrt{3}} &
   \frac{1}{4} \\
\end{array}
\right).
$$
\end{table}
\subsection{ The $S_4\subset S_6$ two-particle CFP's.}
 The two particle basis is taken to be the ordered particle labels:
$$ (5,6),(4,6),(3,6),(2,6),(1,6),(4,5),(3,5),(2,5),(1,5),(3,4),(2,4),(1,4),(2,3),(1,3),(1,2),$$
which are symmetrically or antisymmetrically coupled.
 The four particle basis contains the remaining 4 particle labels as was define above. Then proceeding as above one finds the appropriate 2 particle CFP's. The simplest and perhaps the most interesting for applications is the case  when the four particle representation is one dimensional, i.e. specified by the Young tableaux f=[4] and $[1^4] $
\begin{itemize}
\item  [4]$\otimes$[2]$\rightarrow$ [6]+5[5,1]+9[4,2].\\
The obtained results are given in the tables that follow.
\end{itemize}
\begin{table}
\caption{The reduction [4]$\otimes$[2]$\rightarrow$ [6]+5[5,1]+9[4,2].}
\label{42x6}
$$
\left(
\begin{array}{|c|ccccc}
\hline
[6]&[5,1]_1&[5,1]_2&[5,1]_3&[5,1]_4&[5,1]_5\\
\hline
\hline
 \frac{1}{\sqrt{15}}&0 & 0 & -\frac{1}{\sqrt{5}} & -\frac{3}{4 \sqrt{5}} & \frac{1}{4
   \sqrt{3}} \\
 \frac{1}{\sqrt{15}}& 0 & 0 & -\frac{1}{\sqrt{5}} & \frac{1}{2 \sqrt{5}} & -\frac{1}{2
   \sqrt{3}} \\
  \frac{1}{\sqrt{15}}&0 & \frac{\sqrt{3}}{4} & -\frac{1}{4 \sqrt{5}} & -\frac{1}{2 \sqrt{5}} &
   -\frac{1}{2 \sqrt{3}} \\
  \frac{1}{\sqrt{15}}&\frac{1}{\sqrt{6}} & -\frac{1}{4 \sqrt{3}} & -\frac{1}{4 \sqrt{5}} &
   -\frac{1}{2 \sqrt{5}} & -\frac{1}{2 \sqrt{3}} \\
  \frac{1}{\sqrt{15}}&\frac{1}{\sqrt{6}} & \frac{1}{2 \sqrt{3}} & -\frac{1}{2 \sqrt{5}} &
   \frac{1}{4 \sqrt{5}} & \frac{1}{4 \sqrt{3}} \\
  \frac{1}{\sqrt{15}}&-\frac{1}{\sqrt{6}} & -\frac{1}{2 \sqrt{3}} & -\frac{1}{2 \sqrt{5}} &
   \frac{1}{4 \sqrt{5}} & \frac{1}{4 \sqrt{3}} \\
  \frac{1}{\sqrt{15}}&-\frac{1}{\sqrt{6}} & \frac{1}{4 \sqrt{3}} & \frac{1}{4 \sqrt{5}} &
   -\frac{3}{4 \sqrt{5}} & \frac{1}{4 \sqrt{3}} \\
  \frac{1}{\sqrt{15}}&0 & -\frac{\sqrt{3}}{4} & \frac{1}{4 \sqrt{5}} & -\frac{3}{4 \sqrt{5}} &
   \frac{1}{4 \sqrt{3}} \\
  \frac{1}{\sqrt{15}}&0 & 0 & 0 & 0 & \frac{1}{\sqrt{3}} \\
 \frac{1}{15}&-\frac{1}{\sqrt{6}} & \frac{1}{4 \sqrt{3}} & \frac{1}{4 \sqrt{5}} &
   \frac{1}{2 \sqrt{5}} & -\frac{1}{2 \sqrt{3}} \\
 \frac{1}{\sqrt{15}}& 0 & -\frac{\sqrt{3}}{4} & \frac{1}{4 \sqrt{5}} & \frac{1}{2 \sqrt{5}} &
   -\frac{1}{2 \sqrt{3}} \\
  \frac{1}{\sqrt{15}}&0 & 0 & 0 & \frac{\sqrt{5}}{4} & \frac{1}{4 \sqrt{3}} \\
  \frac{1}{\sqrt{15}}&0 & 0 & \frac{1}{\sqrt{5}} & -\frac{1}{2 \sqrt{5}} & -\frac{1}{2
   \sqrt{3}} \\
  \frac{1}{\sqrt{15}}&0 & \frac{\sqrt{3}}{4} & \frac{3}{4 \sqrt{5}} & \frac{1}{4 \sqrt{5}} &
   \frac{1}{4 \sqrt{3}} \\
  \frac{1}{\sqrt{15}}&\frac{1}{\sqrt{6}} & -\frac{1}{4 \sqrt{3}} & \frac{3}{4 \sqrt{5}} &
   \frac{1}{4 \sqrt{5}} & \frac{1}{4 \sqrt{3}} \\
\end{array}
\right),
\left(
\begin{array}{ccccccccc}
\hline
[4,2]_1&[4,2]_2&[4,2]_3&[4,2]_4&[4,2]_5&[4,2]_6&[4,2]_7&[4,2]_8&[4,2]_9\\
\hline
 \frac{1}{\sqrt{6}} & \frac{1}{\sqrt{30}} & \frac{1}{\sqrt{70}} &
   \sqrt{\frac{3}{35}} & \sqrt{\frac{3}{65}} &
   \frac{\sqrt{\frac{3}{26}}}{2} & -\frac{1}{2 \sqrt{2}} & -\frac{1}{4} &
   -\frac{\sqrt{\frac{3}{5}}}{4} \\
 \frac{1}{\sqrt{6}} & \frac{1}{\sqrt{30}} & \frac{1}{\sqrt{70}} &
   -\frac{4}{\sqrt{105}} & -\frac{4}{\sqrt{195}} & -\sqrt{\frac{2}{39}} &
   \frac{1}{3 \sqrt{2}} & \frac{1}{6} & \frac{1}{2 \sqrt{15}} \\
 0 & -\sqrt{\frac{3}{10}} & -\frac{3}{\sqrt{70}} &
   \frac{\sqrt{\frac{3}{35}}}{2} & \frac{\sqrt{\frac{3}{65}}}{2} &
   -\frac{5}{2 \sqrt{78}} & \frac{1}{6 \sqrt{2}} & \frac{1}{12} &
   -\frac{\sqrt{\frac{3}{5}}}{4} \\
 -\frac{1}{\sqrt{6}} & \sqrt{\frac{2}{15}} & -\frac{3}{\sqrt{70}} &
   \frac{\sqrt{\frac{3}{35}}}{2} & -\frac{7}{2 \sqrt{195}} &
   \frac{\sqrt{\frac{3}{26}}}{2} & \frac{1}{6 \sqrt{2}} & -\frac{1}{6} &
   \frac{1}{2 \sqrt{15}} \\
 -\frac{1}{\sqrt{6}} & -\frac{1}{\sqrt{30}} & 2 \sqrt{\frac{2}{35}} &
   -\frac{2}{\sqrt{105}} & \sqrt{\frac{3}{65}} &
   \frac{\sqrt{\frac{3}{26}}}{2} & -\frac{1}{6 \sqrt{2}} & \frac{1}{6} &
   \frac{1}{2 \sqrt{15}} \\
 -\frac{1}{\sqrt{6}} & -\frac{1}{\sqrt{30}} & -\frac{1}{\sqrt{70}} &
   -\sqrt{\frac{3}{35}} & -\sqrt{\frac{3}{65}} &
   -\frac{\sqrt{\frac{3}{26}}}{2} & -\frac{1}{2 \sqrt{2}} & -\frac{1}{4}
   & -\frac{\sqrt{\frac{3}{5}}}{4} \\
 0 & 0 & 0 & 0 & 0 & 0 & 0 & 0 & \sqrt{\frac{3}{5}} \\
 0 & 0 & 0 & 0 & 0 & 0 & 0 & \frac{3}{4} & -\frac{\sqrt{\frac{3}{5}}}{4}
   \\
 0 & 0 & 0 & 0 & 0 & 0 & \frac{1}{\sqrt{2}} & -\frac{1}{4} &
   -\frac{\sqrt{\frac{3}{5}}}{4} \\
 0 & 0 & 0 & 0 & 0 & \frac{\sqrt{\frac{13}{6}}}{2} & \frac{1}{6 \sqrt{2}}
   & \frac{1}{12} & -\frac{\sqrt{\frac{3}{5}}}{4} \\
 0 & 0 & 0 & 0 & 2 \sqrt{\frac{5}{39}} & -\frac{\sqrt{\frac{3}{26}}}{2} &
   \frac{1}{6 \sqrt{2}} & -\frac{1}{6} & \frac{1}{2 \sqrt{15}} \\
 0 & 0 & 0 & \sqrt{\frac{7}{15}} & -\sqrt{\frac{3}{65}} &
   -\frac{\sqrt{\frac{3}{26}}}{2} & -\frac{1}{6 \sqrt{2}} & \frac{1}{6} &
   \frac{1}{2 \sqrt{15}} \\
 0 & 0 & \sqrt{\frac{5}{14}} & \frac{1}{\sqrt{105}} &
   -\frac{4}{\sqrt{195}} & -\sqrt{\frac{2}{39}} & 0 & -\frac{1}{4} &
   -\frac{\sqrt{\frac{3}{5}}}{4} \\
 0 & \sqrt{\frac{3}{10}} & -\sqrt{\frac{2}{35}} &
   -\frac{\sqrt{\frac{5}{21}}}{2} & \frac{\sqrt{\frac{5}{39}}}{2} &
   -\sqrt{\frac{2}{39}} & -\frac{1}{3 \sqrt{2}} & \frac{1}{12} &
   -\frac{\sqrt{\frac{3}{5}}}{4} \\
 \frac{1}{\sqrt{6}} & -\sqrt{\frac{2}{15}} & -\sqrt{\frac{2}{35}} &
   -\frac{\sqrt{\frac{5}{21}}}{2} & -\frac{\sqrt{\frac{5}{39}}}{2} &
   \sqrt{\frac{2}{39}} & -\frac{1}{3 \sqrt{2}} & -\frac{1}{6} &
   \frac{1}{2 \sqrt{15}} \\
\end{array}
\right).
$$
\end{table}
\begin{itemize}
\item  $[4]\otimes [1^2]\rightarrow 5[5,1]+10[4,1^2]$. 
\end{itemize}
\begin{table}
\caption{The reduction $[4]\otimes [1^2]\rightarrow 5[5,1]+10[4,1^2]$.}
\label{42x11}
$$\left(
\begin{array}{ccccc}
\hline
[5,1]_1&[5,1]_2&[5,1]_3&[5,1]_4&[5,1]_5\\
\hline
 0 & 0 & 0 & -\frac{1}{2} & \frac{1}{2 \sqrt{3}} \\
 0 & 0 & -\frac{\sqrt{2}}{3} & -\frac{1}{6} & \frac{1}{2 \sqrt{3}} \\
 0 & -\frac{\sqrt{\frac{5}{6}}}{2} & -\frac{1}{6 \sqrt{2}} & -\frac{1}{6}
   & \frac{1}{2 \sqrt{3}} \\
 -\frac{1}{\sqrt{5}} & -\frac{1}{2 \sqrt{30}} & -\frac{1}{6 \sqrt{2}} &
   -\frac{1}{6} & \frac{1}{2 \sqrt{3}} \\
 0 & 0 & 0 & 0 & \frac{1}{\sqrt{3}} \\
 0 & 0 & -\frac{\sqrt{2}}{3} & \frac{1}{3} & 0 \\
 0 & -\frac{\sqrt{\frac{5}{6}}}{2} & -\frac{1}{6 \sqrt{2}} & \frac{1}{3}
   & 0 \\
 -\frac{1}{\sqrt{5}} & -\frac{1}{2 \sqrt{30}} & -\frac{1}{6 \sqrt{2}} &
   \frac{1}{3} & 0 \\
 0 & 0 & 0 & \frac{1}{2} & \frac{1}{2 \sqrt{3}} \\
 0 & -\frac{\sqrt{\frac{5}{6}}}{2} & \frac{1}{2 \sqrt{2}} & 0 & 0 \\
 -\frac{1}{\sqrt{5}} & -\frac{1}{2 \sqrt{30}} & \frac{1}{2 \sqrt{2}} & 0
   & 0 \\
 0 & 0 & \frac{\sqrt{2}}{3} & \frac{1}{6} & \frac{1}{2 \sqrt{3}} \\
 -\frac{1}{\sqrt{5}} & \sqrt{\frac{2}{15}} & 0 & 0 & 0 \\
 0 & \frac{\sqrt{\frac{5}{6}}}{2} & \frac{1}{6 \sqrt{2}} & \frac{1}{6} &
   \frac{1}{2 \sqrt{3}} \\
 \frac{1}{\sqrt{5}} & \frac{1}{2 \sqrt{30}} & \frac{1}{6 \sqrt{2}} &
   \frac{1}{6} & \frac{1}{2 \sqrt{3}} \\
\end{array}
\right ),
\left (
\begin{array}{cccccccccc}
\hline
&&&&[4,1^2]_i)&&&&\\
i=1&2&3&4&5&6&7&8&9&\\
%[4,1^2]_1&[4,1^2]_2&[4,1^2]_3&[4,1^2]_4&[4,1^2]_5&[4,1^2]_6&[4,1^2]_7&[4,1^2]_8&[4,1^2]_9&[4,1^2]_{10}\\
\hline
 0 & 0 & 0 & 0 & 0 & 0 & \frac{\sqrt{\frac{5}{3}}}{2} &
   \frac{\sqrt{5}}{6} & \frac{\sqrt{\frac{5}{2}}}{6} & \frac{1}{2
   \sqrt{6}} \\
 0 & 0 & 0 & \sqrt{\frac{2}{5}} & \sqrt{\frac{2}{15}} &
   \frac{1}{\sqrt{15}} & -\frac{1}{2 \sqrt{15}} & -\frac{1}{6 \sqrt{5}} &
   -\frac{1}{6 \sqrt{10}} & -\frac{1}{2 \sqrt{6}} \\
 0 & \frac{\sqrt{\frac{3}{2}}}{2} & \frac{1}{2 \sqrt{2}} & -\frac{1}{2
   \sqrt{10}} & -\frac{1}{2 \sqrt{30}} & -\frac{1}{\sqrt{15}} &
   -\frac{1}{2 \sqrt{15}} & -\frac{1}{6 \sqrt{5}} &
   -\frac{\sqrt{\frac{2}{5}}}{3} & 0 \\
 \frac{1}{\sqrt{3}} & -\frac{1}{2 \sqrt{6}} & -\frac{1}{2 \sqrt{2}} &
   -\frac{1}{2 \sqrt{10}} & -\frac{\sqrt{\frac{3}{10}}}{2} & 0 &
   -\frac{1}{2 \sqrt{15}} & -\frac{1}{2 \sqrt{5}} & 0 & 0 \\
 -\frac{1}{\sqrt{3}} & -\frac{1}{\sqrt{6}} & 0 & -\frac{1}{\sqrt{10}} & 0
   & 0 & -\frac{1}{\sqrt{15}} & 0 & 0 & 0 \\
 0 & 0 & 0 & 0 & 0 & 0 & 0 & 0 & 0 & \sqrt{\frac{2}{3}} \\
 0 & 0 & 0 & 0 & 0 & 0 & 0 & 0 & \frac{\sqrt{\frac{5}{2}}}{2} &
   -\frac{1}{2 \sqrt{6}} \\
 0 & 0 & 0 & 0 & 0 & 0 & 0 & \frac{\sqrt{5}}{3} &
   -\frac{\sqrt{\frac{5}{2}}}{6} & -\frac{1}{2 \sqrt{6}} \\
 0 & 0 & 0 & 0 & 0 & 0 & \frac{\sqrt{\frac{5}{3}}}{2} &
   -\frac{\sqrt{5}}{6} & -\frac{\sqrt{\frac{5}{2}}}{6} & -\frac{1}{2
   \sqrt{6}} \\
 0 & 0 & 0 & 0 & 0 & \sqrt{\frac{3}{5}} & 0 & 0 & -\frac{1}{2 \sqrt{10}}
   & \frac{1}{2 \sqrt{6}} \\
 0 & 0 & 0 & 0 & 2 \sqrt{\frac{2}{15}} & -\frac{1}{\sqrt{15}} & 0 &
   -\frac{1}{3 \sqrt{5}} & \frac{1}{6 \sqrt{10}} & \frac{1}{2 \sqrt{6}}
   \\
 0 & 0 & 0 & \sqrt{\frac{2}{5}} & -\sqrt{\frac{2}{15}} &
   -\frac{1}{\sqrt{15}} & -\frac{1}{2 \sqrt{15}} & \frac{1}{6 \sqrt{5}} &
   \frac{1}{6 \sqrt{10}} & \frac{1}{2 \sqrt{6}} \\
 0 & 0 & \frac{1}{\sqrt{2}} & 0 & -\frac{1}{\sqrt{30}} &
   \frac{1}{\sqrt{15}} & 0 & -\frac{1}{3 \sqrt{5}} &
   \frac{\sqrt{\frac{2}{5}}}{3} & 0 \\
 0 & \frac{\sqrt{\frac{3}{2}}}{2} & -\frac{1}{2 \sqrt{2}} & -\frac{1}{2
   \sqrt{10}} & \frac{1}{2 \sqrt{30}} & \frac{1}{\sqrt{15}} & -\frac{1}{2
   \sqrt{15}} & \frac{1}{6 \sqrt{5}} & \frac{\sqrt{\frac{2}{5}}}{3} & 0
   \\
 \frac{1}{\sqrt{3}} & -\frac{1}{2 \sqrt{6}} & \frac{1}{2 \sqrt{2}} &
   -\frac{1}{2 \sqrt{10}} & \frac{\sqrt{\frac{3}{10}}}{2} & 0 &
   -\frac{1}{2 \sqrt{15}} & \frac{1}{2 \sqrt{5}} & 0 & 0 \\
\end{array}
\right ).
$$
\end{table}
\begin{itemize}
\item  $[1^4]\otimes [2] \rightarrow 5[2,1^4]+10[3,1^3]$. 
\end{itemize}
\begin{table}
\caption{The reduction $[1^4]\otimes [2] \rightarrow 5[2,1^4]+10[3,1^3]$.}
\label{1111x11}
$$
\left (
\begin{array}{ccccc}
\hline
[2,1^4]_1&[2,1^4]_2&[2,1^4]_3&[2,1^4]_4&[2,1^4]_5\\
\hline
 0 & 0 & 0 & -\frac{1}{2} & -\frac{1}{2 \sqrt{3}} \\
 0 & 0 & \frac{\sqrt{2}}{3} & \frac{1}{6} & \frac{1}{2 \sqrt{3}} \\
 0 & \frac{\sqrt{\frac{5}{6}}}{2} & \frac{1}{6 \sqrt{2}} & \frac{1}{6} &
   \frac{1}{2 \sqrt{3}} \\
 \frac{1}{\sqrt{5}} & \frac{1}{2 \sqrt{30}} & \frac{1}{6 \sqrt{2}} &
   \frac{1}{6} & \frac{1}{2 \sqrt{3}} \\
 0 & 0 & 0 & 0 & \frac{1}{\sqrt{3}} \\
 0 & 0 & -\frac{\sqrt{2}}{3} & \frac{1}{3} & 0 \\
 0 & -\frac{\sqrt{\frac{5}{6}}}{2} & -\frac{1}{6 \sqrt{2}} & \frac{1}{3}
   & 0 \\
 -\frac{1}{\sqrt{5}} & -\frac{1}{2 \sqrt{30}} & -\frac{1}{6 \sqrt{2}} &
   \frac{1}{3} & 0 \\
 0 & 0 & 0 & \frac{1}{2} & -\frac{1}{2 \sqrt{3}} \\
 0 & \frac{\sqrt{\frac{5}{6}}}{2} & -\frac{1}{2 \sqrt{2}} & 0 & 0 \\
 -\frac{1}{\sqrt{5}} & -\frac{1}{2 \sqrt{30}} & \frac{1}{2 \sqrt{2}} & 0
   & 0 \\
 0 & 0 & \frac{\sqrt{2}}{3} & \frac{1}{6} & -\frac{1}{2 \sqrt{3}} \\
 -\frac{1}{\sqrt{5}} & \sqrt{\frac{2}{15}} & 0 & 0 & 0 \\
 0 & \frac{\sqrt{\frac{5}{6}}}{2} & \frac{1}{6 \sqrt{2}} & \frac{1}{6} &
   -\frac{1}{2 \sqrt{3}} \\
 \frac{1}{\sqrt{5}} & \frac{1}{2 \sqrt{30}} & \frac{1}{6 \sqrt{2}} &
   \frac{1}{6} & -\frac{1}{2 \sqrt{3}} \\
\end{array}
\right ),
\left .
\begin{array}{|cccccccccc|}
\hline
&&&&&[3,1^3]_i&&&&\\
1&2&3&4&5&6&7&8&9&10\\
%[3,1^3]_1&[3,1^3]_2&[3,1^3]_3&[3,1^3]_4&[3,1^3]_5&[3,1^3]_6&[3,1^3]_7&[3,1^3]_8&[3,1^3]_9&[3,1^3]_{10}\\
\hline
 0 & 0 & 0 & 0 & 0 & 0 & \frac{\sqrt{\frac{5}{3}}}{2} &
   \frac{\sqrt{5}}{6} & \frac{\sqrt{\frac{5}{2}}}{6} & \frac{1}{2
   \sqrt{6}} \\
 0 & 0 & 0 & -\sqrt{\frac{2}{5}} & -\sqrt{\frac{2}{15}} &
   \frac{1}{\sqrt{15}} & \frac{1}{2 \sqrt{15}} & \frac{1}{6 \sqrt{5}} &
   \frac{1}{6 \sqrt{10}} & \frac{1}{2 \sqrt{6}} \\
 0 & -\frac{\sqrt{\frac{3}{2}}}{2} & -\frac{1}{2 \sqrt{2}} & \frac{1}{2
   \sqrt{10}} & \frac{1}{2 \sqrt{30}} & -\frac{1}{\sqrt{15}} & \frac{1}{2
   \sqrt{15}} & \frac{1}{6 \sqrt{5}} & \frac{\sqrt{\frac{2}{5}}}{3} & 0
   \\
 -\frac{1}{\sqrt{3}} & \frac{1}{2 \sqrt{6}} & \frac{1}{2 \sqrt{2}} &
   \frac{1}{2 \sqrt{10}} & \frac{\sqrt{\frac{3}{10}}}{2} & 0 & \frac{1}{2
   \sqrt{15}} & \frac{1}{2 \sqrt{5}} & 0 & 0 \\
 \frac{1}{\sqrt{3}} & \frac{1}{\sqrt{6}} & 0 & \frac{1}{\sqrt{10}} & 0 &
   0 & \frac{1}{\sqrt{15}} & 0 & 0 & 0 \\
 0 & 0 & 0 & 0 & 0 & 0 & 0 & 0 & 0 & \sqrt{\frac{2}{3}} \\
 0 & 0 & 0 & 0 & 0 & 0 & 0 & 0 & \frac{\sqrt{\frac{5}{2}}}{2} &
   -\frac{1}{2 \sqrt{6}} \\
 0 & 0 & 0 & 0 & 0 & 0 & 0 & \frac{\sqrt{5}}{3} &
   -\frac{\sqrt{\frac{5}{2}}}{6} & -\frac{1}{2 \sqrt{6}} \\
 0 & 0 & 0 & 0 & 0 & 0 & \frac{\sqrt{\frac{5}{3}}}{2} &
   -\frac{\sqrt{5}}{6} & -\frac{\sqrt{\frac{5}{2}}}{6} & -\frac{1}{2
   \sqrt{6}} \\
 0 & 0 & 0 & 0 & 0 & \sqrt{\frac{3}{5}} & 0 & 0 & \frac{1}{2 \sqrt{10}} &
   -\frac{1}{2 \sqrt{6}} \\
 0 & 0 & 0 & 0 & 2 \sqrt{\frac{2}{15}} & \frac{1}{\sqrt{15}} & 0 &
   -\frac{1}{3 \sqrt{5}} & \frac{1}{6 \sqrt{10}} & \frac{1}{2 \sqrt{6}}
   \\
 0 & 0 & 0 & \sqrt{\frac{2}{5}} & -\sqrt{\frac{2}{15}} &
   \frac{1}{\sqrt{15}} & -\frac{1}{2 \sqrt{15}} & \frac{1}{6 \sqrt{5}} &
   \frac{1}{6 \sqrt{10}} & \frac{1}{2 \sqrt{6}} \\
 0 & 0 & \frac{1}{\sqrt{2}} & 0 & -\frac{1}{\sqrt{30}} &
   -\frac{1}{\sqrt{15}} & 0 & -\frac{1}{3 \sqrt{5}} &
   \frac{\sqrt{\frac{2}{5}}}{3} & 0 \\
 0 & \frac{\sqrt{\frac{3}{2}}}{2} & -\frac{1}{2 \sqrt{2}} & -\frac{1}{2
   \sqrt{10}} & \frac{1}{2 \sqrt{30}} & -\frac{1}{\sqrt{15}} &
   -\frac{1}{2 \sqrt{15}} & \frac{1}{6 \sqrt{5}} &
   \frac{\sqrt{\frac{2}{5}}}{3} & 0 \\
 \frac{1}{\sqrt{3}} & -\frac{1}{2 \sqrt{6}} & \frac{1}{2 \sqrt{2}} &
   -\frac{1}{2 \sqrt{10}} & \frac{\sqrt{\frac{3}{10}}}{2} & 0 &
   -\frac{1}{2 \sqrt{15}} & \frac{1}{2 \sqrt{5}} & 0 & 0 \\
	\hline
\end{array}
\right . .
$$
\end{table}
\begin{itemize}
\item  $[1^4]\otimes [1^2] \rightarrow [1^6]+5[2,1^4]+9[2^2,1^2]$.\\
\end{itemize}
\begin{table}
\caption{$[1^4]\otimes [1^2] \rightarrow [1^6]+5[2,1^4]+9[2^2,1^2]$.}
\label{1111x11a}
$$
\left(
\begin{array}{cccccc}
\hline
[1^6]&[2,1^5]_1&[2,1^5]_2&[2,1^5]_3&[2,1^5]_4&[2,1^5]_5\\
\hline
  \frac{1}{\sqrt{15}}&0 & 0 & -\frac{1}{\sqrt{5}} & -\frac{3}{4 \sqrt{5}} & \frac{1}{4
   \sqrt{3}} \\
 - \frac{1}{\sqrt{15}}&0 & 0 & \frac{1}{\sqrt{5}} & -\frac{1}{2 \sqrt{5}} & \frac{1}{2
   \sqrt{3}} \\
  -\frac{1}{\sqrt{15}}&0 & -\frac{\sqrt{3}}{4} & \frac{1}{4 \sqrt{5}} & \frac{1}{2 \sqrt{5}} &
   \frac{1}{2 \sqrt{3}} \\
 -\frac{1}{\sqrt{15}}&-\frac{1}{\sqrt{6}} & \frac{1}{4 \sqrt{3}} & \frac{1}{4 \sqrt{5}} &
   \frac{1}{2 \sqrt{5}} & \frac{1}{2 \sqrt{3}} \\
  -\frac{1}{\sqrt{15}}&-\frac{1}{\sqrt{6}} & -\frac{1}{2 \sqrt{3}} & \frac{1}{2 \sqrt{5}} &
   -\frac{1}{4 \sqrt{5}} & -\frac{1}{4 \sqrt{3}} \\
  \frac{1}{\sqrt{15}}&-\frac{1}{\sqrt{6}} & -\frac{1}{2 \sqrt{3}} & -\frac{1}{2 \sqrt{5}} &
   \frac{1}{4 \sqrt{5}} & \frac{1}{4 \sqrt{3}} \\
  \frac{1}{\sqrt{15}}&-\frac{1}{\sqrt{6}} & \frac{1}{4 \sqrt{3}} & \frac{1}{4 \sqrt{5}} &
   -\frac{3}{4 \sqrt{5}} & \frac{1}{4 \sqrt{3}} \\
 \frac{1}{\sqrt{15}}&0 & -\frac{\sqrt{3}}{4} & \frac{1}{4 \sqrt{5}} & -\frac{3}{4 \sqrt{5}} &
   \frac{1}{4 \sqrt{3}} \\
 \frac{1}{\sqrt{15}}&0 & 0 & 0 & 0 & \frac{1}{\sqrt{3}} \\
  -\frac{1}{\sqrt{15}}&\frac{1}{\sqrt{6}} & -\frac{1}{4 \sqrt{3}} & -\frac{1}{4 \sqrt{5}} &
   -\frac{1}{2 \sqrt{5}} & \frac{1}{2 \sqrt{3}} \\
 \frac{1}{\sqrt{15}}& 0 & -\frac{\sqrt{3}}{4} & \frac{1}{4 \sqrt{5}} & \frac{1}{2 \sqrt{5}} &
   -\frac{1}{2 \sqrt{3}} \\
 \frac{1}{\sqrt{15}}& 0 & 0 & 0 & \frac{\sqrt{5}}{4} & \frac{1}{4 \sqrt{3}} \\
  \frac{1}{\sqrt{15}}&0 & 0 & \frac{1}{\sqrt{5}} & -\frac{1}{2 \sqrt{5}} & -\frac{1}{2
   \sqrt{3}} \\
  \frac{1}{\sqrt{15}}&0 & \frac{\sqrt{3}}{4} & \frac{3}{4 \sqrt{5}} & \frac{1}{4 \sqrt{5}} &
   \frac{1}{4 \sqrt{3}} \\
  \frac{1}{\sqrt{15}}&\frac{1}{\sqrt{6}} & -\frac{1}{4 \sqrt{3}} & \frac{3}{4 \sqrt{5}} &
   \frac{1}{4 \sqrt{5}} & \frac{1}{4 \sqrt{3}} \\
\end{array}
\right),
\left(
\begin{array}{cccccccccc}
\hline
&&&&[2^2,1^2]_i&&&&\\
1&2&3&4&5&6&7&8&9\\
\hline
 \frac{1}{\sqrt{6}} & \frac{1}{\sqrt{30}} & \frac{1}{\sqrt{70}} &
   \sqrt{\frac{3}{35}} & \sqrt{\frac{3}{65}} &
   -\frac{\sqrt{\frac{3}{26}}}{2} & -\frac{1}{2 \sqrt{2}} & -\frac{1}{4}
   & -\frac{\sqrt{\frac{3}{5}}}{4} \\
 -\frac{1}{\sqrt{6}} & -\frac{1}{\sqrt{30}} & -\frac{1}{\sqrt{70}} &
   \frac{4}{\sqrt{105}} & \frac{4}{\sqrt{195}} & -\sqrt{\frac{2}{39}} &
   -\frac{1}{3 \sqrt{2}} & -\frac{1}{6} & -\frac{1}{2 \sqrt{15}} \\
 0 & \sqrt{\frac{3}{10}} & \frac{3}{\sqrt{70}} &
   -\frac{\sqrt{\frac{3}{35}}}{2} & -\frac{\sqrt{\frac{3}{65}}}{2} &
   -\frac{5}{2 \sqrt{78}} & -\frac{1}{6 \sqrt{2}} & -\frac{1}{12} &
   \frac{\sqrt{\frac{3}{5}}}{4} \\
 \frac{1}{\sqrt{6}} & -\sqrt{\frac{2}{15}} & \frac{3}{\sqrt{70}} &
   -\frac{\sqrt{\frac{3}{35}}}{2} & \frac{7}{2 \sqrt{195}} &
   \frac{\sqrt{\frac{3}{26}}}{2} & -\frac{1}{6 \sqrt{2}} & \frac{1}{6} &
   -\frac{1}{2 \sqrt{15}} \\
 \frac{1}{\sqrt{6}} & \frac{1}{\sqrt{30}} & -2 \sqrt{\frac{2}{35}} &
   \frac{2}{\sqrt{105}} & -\sqrt{\frac{3}{65}} &
   \frac{\sqrt{\frac{3}{26}}}{2} & \frac{1}{6 \sqrt{2}} & -\frac{1}{6} &
   -\frac{1}{2 \sqrt{15}} \\
 -\frac{1}{\sqrt{6}} & -\frac{1}{\sqrt{30}} & -\frac{1}{\sqrt{70}} &
   -\sqrt{\frac{3}{35}} & -\sqrt{\frac{3}{65}} &
   \frac{\sqrt{\frac{3}{26}}}{2} & -\frac{1}{2 \sqrt{2}} & -\frac{1}{4} &
   -\frac{\sqrt{\frac{3}{5}}}{4} \\
 0 & 0 & 0 & 0 & 0 & 0 & 0 & 0 & \sqrt{\frac{3}{5}} \\
 0 & 0 & 0 & 0 & 0 & 0 & 0 & \frac{3}{4} & -\frac{\sqrt{\frac{3}{5}}}{4}
   \\
 0 & 0 & 0 & 0 & 0 & 0 & \frac{1}{\sqrt{2}} & -\frac{1}{4} &
   -\frac{\sqrt{\frac{3}{5}}}{4} \\
 0 & 0 & 0 & 0 & 0 & \frac{\sqrt{\frac{13}{6}}}{2} & -\frac{1}{6
   \sqrt{2}} & -\frac{1}{12} & \frac{\sqrt{\frac{3}{5}}}{4} \\
 0 & 0 & 0 & 0 & 2 \sqrt{\frac{5}{39}} & \frac{\sqrt{\frac{3}{26}}}{2} &
   \frac{1}{6 \sqrt{2}} & -\frac{1}{6} & \frac{1}{2 \sqrt{15}} \\
 0 & 0 & 0 & \sqrt{\frac{7}{15}} & -\sqrt{\frac{3}{65}} &
   \frac{\sqrt{\frac{3}{26}}}{2} & -\frac{1}{6 \sqrt{2}} & \frac{1}{6} &
   \frac{1}{2 \sqrt{15}} \\
 0 & 0 & \sqrt{\frac{5}{14}} & \frac{1}{\sqrt{105}} &
   -\frac{4}{\sqrt{195}} & \sqrt{\frac{2}{39}} & 0 & -\frac{1}{4} &
   -\frac{\sqrt{\frac{3}{5}}}{4} \\
 0 & \sqrt{\frac{3}{10}} & -\sqrt{\frac{2}{35}} &
   -\frac{\sqrt{\frac{5}{21}}}{2} & \frac{\sqrt{\frac{5}{39}}}{2} &
   \sqrt{\frac{2}{39}} & -\frac{1}{3 \sqrt{2}} & \frac{1}{12} &
   -\frac{\sqrt{\frac{3}{5}}}{4} \\
 \frac{1}{\sqrt{6}} & -\sqrt{\frac{2}{15}} & -\sqrt{\frac{2}{35}} &
   -\frac{\sqrt{\frac{5}{21}}}{2} & -\frac{\sqrt{\frac{5}{39}}}{2} &
   -\sqrt{\frac{2}{39}} & -\frac{1}{3 \sqrt{2}} & -\frac{1}{6} &
   \frac{1}{2 \sqrt{15}} \\
\end{array}
\right).
$$
\end{table}
%\end{itemize}
The other more complicated cases:\\
$$[3,1]\otimes[2]\rightarrow 5[5,1]+9[4,2]+10[4,1^2]+5[3^2]+16[3,2,1]$$
$$[3,1]\otimes[1^2]\rightarrow 9[4,2]+10[4,1^2]+16[3,2,1]+10[3,1^3]$$
$$[22]\otimes[2]\rightarrow 9[4,2]+16[3,2,1]+5[2^3]$$
$$[22]\otimes[1^2]\rightarrow 5[3^2]+16[3,2,1]+9[2^2,1^2]$$
$$[2,1^2]\otimes[2]\rightarrow 9[2^2,1^2]+10[3,1^3]+16[3,2,1]+10[4,1^2]$$
$$[2,1^2]\otimes[1^2]\rightarrow 5[2,1^4]+9[2^2,1^2]+10[3,1^3]+5[2^3]+16[3,2,1]$$
will appear in the appendix B.
	\section{applications}
	We will begin by considering  given symmetries $[f]$, $[f^{'}]$ etc of $n$ particles which can be distributed in $p$ single particle states $\phi_{1},\phi_2, \cdots,\phi_r$. Let us suppose that we want to evaluate the matrix elements of one-body and two-body operators, for which the one article matrix elements $m^{(1)}\left (\phi_{\alpha}(k),\phi_{\beta}(k)\right )$ for particle $k$ are known. Let us also suppose that the matrix elements $m^{(p),(q)}\left (\phi_{\alpha}(m),\phi_{\beta}(n),\phi_{\gamma}(m)\phi_{\delta}(n)\right )$, with $p$ and $q$ taking values $S$ (symmetric) and $A$ (antisymmetric) combinations of the particles $m$ and $n$, are also known.
	\begin{itemize}
	\item One-body operators.\\
	Let us indicate indicate the one-body expansion \cite{Hamermesh} coefficients by $C^{[f]}_{[f_i](n-1)j}$  running down the corresponding column of the table.
	Then the many-body matrix element is:
	\beq
	ME^{(1),\phi}([f],[f^{'}])=\sum_{\alpha,\beta,j}C^{[f]}_{[f_i](n-1)j}C^{[f^{'}]}_{[f_i](n-1)j}m^{(1)}\left (\phi_{\alpha}(j),\phi_{\beta}(j)\right ).
	\eeq
	Note that, since the  $ (n-1)$ particles are not  interacting the corresponding symmetry in the expansion must be the same. Clearly the number of single particle states must be at least equal to the length of the first column of the Young tableau \cite{Hamermesh} characterizing the symmetries $[f]$ and $[f^{'}]$, so that an antisymmetric combination is possible.
	\item Two-body operators.\\
	Now we get:
		\beq
	ME^{(2),\phi}([f],[f^{'}])=\sum_{\alpha,\beta,,\gamma,\delta,p[m,n],q[m,n]}C^{[f]}_{[f_i](n-2)p[m,n]}C^{[f^{'}]}_{[f_i](n-2)q[m,n]}
	m^{(p),(q)}\left (\phi_{\alpha}(m),\phi_{\beta}(n),\phi_{\gamma}(m)\phi_{\delta}(n)\right ),
	\eeq
	where the coefficients $C$ are just the columns of the corresponding 2-particle CFP's involving $[f]$ and $[f^{'}]$, which can be read off from the  tables.
	\end{itemize}
	 Sometimes one needs to consider a subgroup of the highest symmetry. Then the state is characterized by labels additional to $[f]$ $[f^{'}]$ etc and one needs to construct the 1-particle and 2-particle  CFP's associated with the symmetry involving these labels.  The two sets of CFP's  factor out.\\
	
	In practice one needs with symmetries involving more than one space of the $n$ particles involving let us say the single particle states $\phi_{1},\phi_2, \cdots,\phi_r$ and $\chi_{1},\chi_2, \cdots,\chi_s$ with given symmetries $[f_1]$ and $[f_2]$ to be combined to an overall symmetry $[f]$ of $S_n$, namely for  $[f_1] \times[f_2]\rightarrow [f]$. Thus one needs the standard (inner product) C-G coefficients\cite{Hamermesh}  of the symmetric group $S_n$. The simplest case is the one in which the full symmetry is symmetric or antisymmetric. In this case it terns out that the expansion coefficients are related to the dimension of the representation [f].
	If the overall symmetry is symmetric $[n]$ one finds:
	\beq
	ME^{(i)}=\sum_{[f],[f^{'}]}\frac{1}{\mbox{dim}[f] \mbox{dim}[f]^{'} }ME^{(i),\phi}([f],[f^{'}])\,ME^{(i),\chi}([f],[f^{'}]),\,i=1,2\,,
	\eeq
	where $\mbox{dim}[f]$ and  $\mbox{dim}[f^{'}]$ are the dimensions of $[f]$ and $[f^{'}$ respectively. 
		If the overall symmetry is antisymmetric $[1^n]$ one finds:
		\beq
	ME^{(i)}=\sum_{[f],[f^{'}]}\frac{1}{\mbox{dim}[f] \mbox{dim}[f]^{'} }ME^{(i),\phi}([f],[f^{'}])\,ME^{(i),\chi}([{\tilde f}],[{\tilde f}^{'}]),\,i=1,2\,,
	\eeq
	where ${\tilde [f]}$ corresponds to a Young table obtained from that of $[f]$ by interchanging rows and columns.  
	
	In the above formulas we assumed that the number of particles is up to six, but the number of single particle states is unrestricted. In practice often the number $r$  of these states is smaller, i.e. $r<n$. Then, if they form a basis for the fundamental representation  of the unitary group $U(r)$ one can use $U(r)$ in constructing states of a given symmetry, which is simpler than the one considered here 
with $S_n$. 

	If sub-symmetries are involved one must provide labels in addition to the symmetries $[f]$,  $[f^{'}]$ etc, which of course do not affect the overall symmetry. Thus, e.g., in atomic and nuclear spectroscopy one employs the chain as discussed, e.g., in the well known book  of Hamermesh \cite{Hamermesh}, chapter 10, i.e.  $SU(2n+1) \supset O(2n+1)$ or $SU(2n) \supset Sp(2n)$ (the latter is  symplectic group \cite{Flowers52} with the corresponding Lie algebra indicated by $C_n$).  Sometimes additional subgroups may be useful \cite{Pang69}, \cite{HechtPang69}. In most cases all of them must contain the orthogonal group O(3), rotation group or $SU(2)$, depending on the problem under consideration. It is a remarkable result of the structure of the symmetries involved that the needed CFP's factorize \cite{HechtPang69}, the most complicated being the one we considered here.
	%The corresponding expansion 
	%The 1-particle and 2-paricle CFP's are sufficient to determine the matrix elements of $\Omega_{1b}=\sum \omega_i$ (1-body) and 
	%$\Omega_{2b}=\sum{i>j} \omega_i \omega_j $ (2-body) operators.
	%In fact one finds that:
	%\beq
	%\prec[f]^{'}|\Omega_{1b}|[f]\succ=\sum [f]^{\alpha}_{1}, [1]_{\beta^{'}}\}[f]^{'} [f]^{\alpha}_{1}, [1]_{\beta}\} [f] \prec \phi_{\beta^{'}}|\omega_{1b}|\phi_{\beta}\succ
	%\eeq
%	where $\phi_{\beta}$ is the single paricle function occupied by the paricle $\alpha$. We assumed the most general case that there as many available  single particle as paricles. In pracice there are fewer states than paricles. In that case the above formula applies but the  by asigning which particle is in which state.
 
In the Example 1 considered in the second section
%, \ref{sec:InnerCG}, 
(structure of the nucleon), 
in the space involving three quarks, each quark can exist  in two spin states. So the group is $U(2)$. For three particles, however, the spin $s=1/2$ state is not unique. These can be distinguished by symmetry,  one has symmetry [3] and the other [2,1]. So the CFP can be found from  and the spin matrix elements can be computed  as outlined here.\\
In the Example 2 considered in the second  section
%, \ref{sec:InnerCG},
 we examined structures involving six quarks. In the orbital part we restricted the space to involve harmonic oscillator states $0s,0p,0d$ and $1s$ with energies 0,1 2 2  $ \hbar \omega$ respectively. Furthermore    we have restricted our configuration space to be such that the summed energies of all particles to be $\leq 2 \hbar \omega$. Thus the only possibilities are $0s^6$, $0s^5 1s$, $0s^50d$ and $0s^40p^2$. Thus symmetries with more than two rows are excluded, thus the only possibilities are $[f]_L=[6],[5,1]$ and [4,2]. Note that the symmetry must be supplemented by specifying the the allowed orbits that go with it. In our example the number of orbits is less than the number of particles, so the admissible representations are fewer than the ones obtained in the work. \\
We also note that in  this particular example we considered two quark flavors, u and d, which can be described as members of the (strong) isospin doublet $I=1/2$, i.e. that the symmetry involved is $U_I(2)$. This is isomoprhic to the spin group $U_s(2)$ so the study of the flavor structure can be accomplished by the known facts of the spin case.  Anyway  only symmetries with Young tableaux of at most two rows  can appear. Let us indicate them by $[f_1,f_2]$,with  dimension $f_1-f_2+1$. Let us now suppose that the total isospin projection for the six particles is $I_3$. Its maximum value can be $(1/2)(f_1-f_2)$.  So there must be one isospin with value $I=(1/2)(f_1-f_2)$. Since  there exist $(2I+1)=f_1-f_2+1$ components that go with it, this value of isospin is unique. So each representation is described by the isospin value, e.g. $[6]\leftrightarrow I=3$, $[5,1]\leftrightarrow I=2$,  $[4,2]\leftrightarrow I=1$and $[3,3]\leftrightarrow I=0$. In a world in which the isospin symmetry is exact, different representations $[f_1,f_2]$ do not mix. In short in this case the needed CFPs are nothing but the usual C-G coefficients for the rotation group and one does not need to use the tables produced here for the general case.\\
The above considerations, of course, put restrictions on the representations of the other  symmetries allowed by the requirement of total antisymmetry. Further restrictions are imposed by the requirement that the  allowed  6 particle states must contain a definite representation of a given subgroup, e.g. definite representations of the color group  $SU_c(3)$ must appear in the allowed representations of $SU_{sc}(6)$ group \cite{Strot79}. In particular the six quark states must  be a color singlet.\\
 We see that in this particular example only a small portion of the CFPs given in our tables may, in fact, be needed. For a different problem, however, another set of the CFP's of our tables will be needed.

It is obvious that each representation $[f]_i$ included in our tables must be specified by additional information, i.e.  the number of single particle states involved, before our tables maybe used. In our example $[6]0s^6$, $[6]0s^51s$, $[6]0s^51d$, $[6]0s^41p^2$ etc So for many single particle states the number of states of a given symmetry increases dramatically. The number of needed CFP's, however, does not change. This is the bonus contained in  the generality of our results, which compensates for the encountered complexity.

%They can be also be applied in the opposite situation for a number of particles and the number of orbits up to and including six. Then one orders the orbits $\phi_{1},\phi_2, \cdots,\phi_r$ and $\chi_{1},\chi_2, \cdots,\chi_n$ for the elements of $S_n$ and uses the above formulas after interchanging the roles of particle state labels, $k\rightarrow \phi_{k}$ and $\alpha->1,\beta->2$ etc.
\section{Conclusions}	
 In the present paper we presented results   C-G  for the outer product (coefficients of fractional parentage or CFP) of irreducible representations of the symmetry group $S_n$ up to $n=6$ in analytic form. These are adequate to calculate the matrix elements of 
one-body and two-body operators for up to  $n=6$ particles put in unrestricted number of particle states.
 %or up to $n=6$ particle states in which one can put an unlimited number of particles . 
Due to this generality combined with  the encountered degeneracies in the product of  representations of $S_n$  has caused our results to appear a bit cumbersome. It perhaps possible that a judicious choice of linear combinations of judicious choice of the degenerate states may have lead to simpler expressions. There is no general how to do this, and one cannot examine all possibilities. We believe the reader will find them easy to use in a variety of problems,  since no restriction was put on the available single particle states provided that the number of particles is $n$, $n\le6$. 
%or unlimited number of particles  restricting the number of available states to be $\le6$.
 \\If symmetries of different spaces need be combined, one needs the C-G  for the relevant inner product of $S_n$, for which some general formulas or tables already exist.

{\bf  Acknowledgments}: The author is indebted to prfessors George Voulgaris of USC and  George Akrivis of UoI for their invaluable assistance in making the tables tractable  and to professors F. Avignone and R. Creswick for their support and hospitality. This work was partly supported by a USC Provost’s Internal Visiting Distinguished Visitor Grant and IBS-Korea under system code IBS-R017-D1-2014-a00

\clearpage
\section{Appendix A:	Tables of  $S_5\subset S_6$ (1-particle CFP's).}
We begin with the reduction:
$$ [4,1]\otimes[1]\rightarrow 5[5,1]+9[4,2]+10[4,1,1].$$
For illustration purposes we will  express the above four [4,1] states in 20-dimensional space of the five particles (right set):
$$[4,1]_1=\left\{\frac{1}{2 \sqrt{2}},-\frac{1}{2
   \sqrt{2}},0,0,0,\frac{1}{2 \sqrt{2}},-\frac{1}{2
   \sqrt{2}},0,0,0,\frac{1}{2 \sqrt{2}},-\frac{1}{2
   \sqrt{2}},0,0,0,\frac{1}{2 \sqrt{2}},-\frac{1}{2
   \sqrt{2}},0,0,0\right\},$$
	$$[4,1]_2= \left\{0,0,\frac{1}{2 \sqrt{2}},-\frac{1}{2
   \sqrt{2}},0,0,0,\frac{1}{2 \sqrt{2}},-\frac{1}{2
   \sqrt{2}},0,0,0,\frac{1}{2 \sqrt{2}},-\frac{1}{2
   \sqrt{2}},0,0,0,\frac{1}{2 \sqrt{2}},-\frac{1}{2
   \sqrt{2}},0\right\},$$ 
	$$ [4,1]_3=\left\{\frac{1}{2 \sqrt{6}},\frac{1}{2
   \sqrt{6}},0,0,-\frac{1}{\sqrt{6}},\frac{1}{2
   \sqrt{6}},\frac{1}{2
   \sqrt{6}},0,0,-\frac{1}{\sqrt{6}},\frac{1}{2
   \sqrt{6}},\frac{1}{2
   \sqrt{6}},0,0,-\frac{1}{\sqrt{6}},\frac{1}{2
   \sqrt{6}},\frac{1}{2
   \sqrt{6}},0,0,-\frac{1}{\sqrt{6}}\right\},$$
	$$[4,1]_4=\left\{\frac{1}{\sqrt{30}},\frac{1}{\sqrt{30}},-\frac
   {1}{2}\sqrt{\frac{3}{10}},-\frac{1}{2}\sqrt{\frac{3}{10}
   },\frac{1}{\sqrt{30}},\frac{1}{\sqrt{30}},
	\frac{1}{\sqrt{30}},-\frac{1}{2}\sqrt{\frac{3}{10}},
	-\frac{1}{2}\sqrt{\frac{3}{10}},\frac{1}{\sqrt{30}},
	\frac{1}{\sqrt{30}},\frac{1}{\sqrt{30}},-\frac{1}{2}\sqrt
   {\frac{3}{10}}, -\frac{1}{2}\sqrt{\frac{3}{10}},\right .$$ $$\left .
   \frac{1}{\sqrt{30}},\frac{1}{\sqrt{30}},\frac{1}
	{\sqrt{30}},-\frac{1}{2}\sqrt{\frac{3}{10}},
	-\frac{1}{2}\sqrt{\frac{3}{10}},\frac{1}{\sqrt{30}}\right\}.$$
	The obtained reductions are given in the three tables below. The vertical axis is labeled by $[4,1]_i\otimes 6,\, i=1,2,3,4,\,[4,1]_i \otimes 5, \,i=1,2,3,4$, etc.

\begin{table}
\caption{The 1-particle CFP's (outer  C-G coefficients) involving the reduction$[4,1]\times[1]\rightarrow 5[5,1]+9[4,2]+10[4,1^2] $.}
\label{1cf6.51}
	$$
	\left(
\begin{array}{ccccc}
\hline
[5,1]_1&[5,1]_2&[5,1]_3&[5,1]_4&[5,1]_5\\
\hline
 0 & 0 & 0 & \frac{\sqrt{7}}{6} & -\frac{1}{6 \sqrt{2}} \\
 -\frac{3 \sqrt{\frac{5}{1186}}}{2} & -\frac{151
   \sqrt{\frac{15}{852734}}}{4} & -\frac{151
   \sqrt{\frac{5}{15099}}}{8} & \frac{5}{8 \sqrt{7}} & 0 \\
 \frac{3 \sqrt{\frac{5}{1186}}}{2} & \frac{151
   \sqrt{\frac{15}{852734}}}{4} & \frac{151
   \sqrt{\frac{5}{15099}}}{8} & \frac{13}{24 \sqrt{7}} & -\frac{1}{6
   \sqrt{2}} \\
 0 & 0 & 0 & \frac{\sqrt{7}}{6} & -\frac{1}{6 \sqrt{2}} \\
 0 & 0 & 0 & \frac{\sqrt{7}}{6} & -\frac{1}{6 \sqrt{2}} \\
 0 & 0 & 0 & \frac{\sqrt{7}}{6} & -\frac{1}{6 \sqrt{2}} \\
 0 & -\frac{\sqrt{\frac{2965}{4314}}}{2} &
   \frac{\sqrt{\frac{105}{719}}}{2} & 0 & 0 \\
\end{array}
\right),
	$$
	\end{table}
%We continue with  the C-G regarding the 5-dimensional representation [3,3].
\begin{table}
\caption{The 1-particle CFP's ( see table \ref{1cf6.51}) involving [4,2] of $S_6$}
\label{1cf6.51a}
	$$
	\left(
\begin{array}{ccccccccc}
\hline
[4,2]_1&[4,2]_2&[4,2]_3&[4,2]_4&[4,2]_5&[4,2]_6&[4,2]_7&[4,2]_8&[4,2]_9\\
\hline
 -3 \sqrt{\frac{10}{593}} & -\frac{\sqrt{\frac{5}{2}}}{4} &
   -\frac{15 \sqrt{\frac{5}{4382}}}{4} & -\frac{219}{\sqrt{6919178}}
   & -\frac{1467}{2 \sqrt{7083394}} & \frac{35
   \sqrt{\frac{5}{5979838}}}{2} & -110 \sqrt{\frac{2}{1003749}} &
   -\frac{25}{12 \sqrt{502}} & -\frac{5}{12 \sqrt{2}} \\
 -\frac{3 \sqrt{\frac{5}{1186}}}{2} & -\frac{1}{4 \sqrt{10}} &
   \frac{33}{4 \sqrt{21910}} & 92 \sqrt{\frac{2}{3459589}} &
   -\frac{1601}{2 \sqrt{7083394}} & \frac{925
   \sqrt{\frac{5}{5979838}}}{2} & \frac{89}{\sqrt{2007498}} &
   -\frac{7}{12 \sqrt{502}} & -\frac{5}{12 \sqrt{2}} \\
 -\frac{9 \sqrt{\frac{5}{1186}}}{2} & -\frac{1}{\sqrt{10}} &
   -\frac{27}{\sqrt{21910}} & -\frac{403}{\sqrt{6919178}} &
   \frac{67}{\sqrt{7083394}} & -445 \sqrt{\frac{5}{5979838}} & -103
   \sqrt{\frac{3}{669166}} & -\frac{3}{2 \sqrt{502}} & 0 \\
 -3 \sqrt{\frac{10}{593}} & \frac{3}{4 \sqrt{10}} & -\frac{19}{4
   \sqrt{21910}} & 279 \sqrt{\frac{2}{3459589}} & 318
   \sqrt{\frac{2}{3541697}} & -3 \sqrt{\frac{10}{2989919}} &
   -\frac{115}{\sqrt{2007498}} & -\frac{53}{12 \sqrt{502}} &
   -\frac{1}{12 \sqrt{2}} \\
 -3 \sqrt{\frac{10}{593}} & 0 & 4 \sqrt{\frac{10}{2191}} &
   \frac{29}{\sqrt{6919178}} & \frac{865}{\sqrt{7083394}} & 13
   \sqrt{\frac{5}{5979838}} & 9 \sqrt{\frac{3}{669166}} &
   \frac{43}{12 \sqrt{502}} & -\frac{1}{12 \sqrt{2}} \\
 -3 \sqrt{\frac{10}{593}} & \frac{1}{2 \sqrt{10}} & -\frac{33}{2
   \sqrt{21910}} & -184 \sqrt{\frac{2}{3459589}} & -\frac{1535}{2
   \sqrt{7083394}} & -\frac{49 \sqrt{\frac{5}{5979838}}}{2} & 154
   \sqrt{\frac{2}{1003749}} & \frac{35}{12 \sqrt{502}} & \frac{7}{12
   \sqrt{2}} \\
 0 & 0 & 6 \sqrt{\frac{10}{2191}} & -526 \sqrt{\frac{2}{3459589}} &
   -\frac{229}{4 \sqrt{7083394}} & -\frac{19
   \sqrt{\frac{5}{5979838}}}{4} & -\frac{71}{2 \sqrt{2007498}} &
   -\sqrt{\frac{2}{251}} & 0 \\
\end{array}
\right),
	$$
	\end{table}
%We continue with  the C-G regarding the 5-dimensional representation [3,3].
\begin{table}
\caption{The 1-particle CFP's ( see table \ref{1cf6.51}) involving $[4,1^2]$ of $S_6$}
\label{1cf6.51b}
	$$
	\left(
\begin{array}{cccccccccc}
\hline
[4,1^2]_1&[4,1^2]_2&[4,1^2]_3&[4,1^2]_4&[4,1^2]_5&[4,1^2]_6&[4,1^2]_7&[4,1^2]_8&[4,1^2]_9&[4,1^2]_{10}\\
\hline
 \frac{1}{4 \sqrt{6}} & -\frac{21}{4 \sqrt{310}} & -\frac{271}{4
   \sqrt{325965}} & \frac{1399}{4 \sqrt{5144639}} & -\frac{11663}{4
   \sqrt{227765865}} & \frac{47}{2 \sqrt{9093255}} & \frac{71}{4
   \sqrt{8790}} & -\frac{1}{4 \sqrt{30}} & 0 & -\frac{7}{4
   \sqrt{15}} \\
 -\frac{1}{4 \sqrt{6}} & \frac{21}{4 \sqrt{310}} & \frac{271}{4
   \sqrt{325965}} & \frac{1405}{4 \sqrt{5144639}} & \frac{8683}{4
   \sqrt{227765865}} & \frac{593}{2 \sqrt{9093255}} & \frac{473}{28
   \sqrt{8790}} & \frac{17}{28 \sqrt{30}} & 0 & \frac{7}{4
   \sqrt{15}} \\
 0 & 0 & 0 & 0 & 0 & 0 & 0 & 0 & 0 & \frac{\sqrt{\frac{5}{3}}}{2} \\
 -\frac{\sqrt{\frac{3}{2}}}{4} & -\frac{1}{4 \sqrt{310}} & \frac{163
   \sqrt{\frac{3}{108655}}}{4} & -\frac{699}{\sqrt{5144639}} &
   \frac{188}{\sqrt{227765865}} & -\frac{953}{2 \sqrt{9093255}} &
   -\frac{103}{28 \sqrt{8790}} & \frac{\sqrt{\frac{5}{6}}}{28} &
   \frac{\sqrt{\frac{3}{5}}}{2} & -\frac{1}{4 \sqrt{15}} \\
 0 & 2 \sqrt{\frac{2}{155}} & -32 \sqrt{\frac{3}{108655}} &
   -\frac{909}{\sqrt{5144639}} & -13 \sqrt{\frac{3}{75921955}} &
   \frac{55 \sqrt{\frac{15}{606217}}}{2} & -\frac{13}{4 \sqrt{8790}}
   & -\frac{1}{4 \sqrt{30}} & -\frac{\sqrt{\frac{3}{5}}}{2} &
   -\frac{1}{4 \sqrt{15}} \\
 \frac{1}{2 \sqrt{6}} & \frac{3}{2 \sqrt{310}} & \frac{83}{2
   \sqrt{325965}} & \frac{2229}{4 \sqrt{5144639}} & \frac{14047}{4
   \sqrt{227765865}} & -\frac{559}{2 \sqrt{9093255}} &
   -\frac{1273}{28 \sqrt{8790}} & -\frac{1}{28 \sqrt{30}} & 0 &
   -\frac{1}{4 \sqrt{15}} \\
 0 & 0 & \frac{\sqrt{\frac{465}{701}}}{2} & \frac{525}{8
   \sqrt{5144639}} & -\frac{10441}{8 \sqrt{227765865}} &
   \frac{413}{2 \sqrt{9093255}} & \frac{3 \sqrt{\frac{6}{1465}}}{7}
   & -\frac{3 \sqrt{\frac{6}{5}}}{7} & -\frac{1}{2 \sqrt{15}} & 0 \\
\end{array}
\right).
	$$
	\end{table}
We will next consider the reduction:
$[3,2]\otimes[1]\rightarrow 9[4,2]+5[3,3]+16[3,2,1]$.
We have already discussed some technical points in the main text. So we will present the obtained results in tables 
\ref{1cf6.42}-\ref{1cf6.321c}.
\begin{table}
\caption{The 1-particle CFP's (outer  C-G coefficients) involving the reduction$[3,2]\otimes[1]\rightarrow 9[4,2]+5[3,3]+16[3,2,1]$.}
\label{1cf6.42}
$$
\left(
\begin{array}{ccccccccc}
\hline
[4,2]_1&[4,2]_2&[4,2]_3&[4,2]_4&[4,2]_5&[4,2]_6&[4,2]_7&[4,2]_8&[4,2]_9\\
\hline
 \frac{2}{7} & \frac{6 \sqrt{\frac{15}{61}}}{7} & \frac{1}{4 \sqrt{305}} &
   -\frac{\sqrt{\frac{5}{11}}}{4} & -\frac{1}{5 \sqrt{77}} &
   \frac{\sqrt{\frac{3}{154}}}{20} & -\frac{1}{4 \sqrt{330}} & 0 & \frac{1}{2
   \sqrt{30}} \\
 \frac{2}{7} & \frac{6 \sqrt{\frac{15}{61}}}{7} & \frac{1}{4 \sqrt{305}} &
   -\frac{\sqrt{\frac{5}{11}}}{4} & -\frac{1}{5 \sqrt{77}} &
   \frac{\sqrt{\frac{3}{154}}}{20} & -\frac{1}{4 \sqrt{330}} & 0 & \frac{1}{2
   \sqrt{30}} \\
 \frac{1}{7} & \frac{3 \sqrt{\frac{15}{61}}}{7} & \frac{1}{8 \sqrt{305}} &
   -\frac{\sqrt{\frac{5}{11}}}{8} & -\frac{17}{5 \sqrt{77}} & \frac{17
   \sqrt{\frac{3}{154}}}{20} & -\frac{17}{4 \sqrt{330}} & 0 & -\frac{1}{2
   \sqrt{30}} \\
 \frac{1}{7} & \frac{3 \sqrt{\frac{15}{61}}}{7} & \frac{1}{8 \sqrt{305}} &
   -\frac{\sqrt{\frac{5}{11}}}{8} & -\frac{17}{5 \sqrt{77}} & -\frac{9
   \sqrt{\frac{3}{154}}}{10} & \frac{3 \sqrt{\frac{3}{110}}}{2} & 0 & 0 \\
 \frac{1}{7} & \frac{3 \sqrt{\frac{15}{61}}}{7} & \frac{1}{8 \sqrt{305}} &
   -\frac{\sqrt{\frac{5}{11}}}{8} & \frac{16}{5 \sqrt{77}} & -\frac{2
   \sqrt{\frac{6}{77}}}{5} & -\frac{\sqrt{\frac{3}{110}}}{2} & \frac{3}{4
   \sqrt{10}} & \frac{\sqrt{\frac{3}{10}}}{4} \\
 \frac{1}{7} & \frac{3 \sqrt{\frac{15}{61}}}{7} & \frac{1}{8 \sqrt{305}} &
   -\frac{\sqrt{\frac{5}{11}}}{8} & \frac{16}{5 \sqrt{77}} & \frac{19
   \sqrt{\frac{3}{154}}}{20} & \frac{\sqrt{\frac{3}{110}}}{4} & -\frac{3}{4
   \sqrt{10}} & \frac{\sqrt{\frac{3}{10}}}{4} \\
 -\frac{3}{7} & \frac{\sqrt{\frac{61}{15}}}{7} & 0 & 0 & 0 &
   -\frac{\sqrt{\frac{7}{66}}}{4} & -\frac{\sqrt{\frac{3}{110}}}{4} &
   \frac{1}{2 \sqrt{10}} & 0 \\
\end{array}
\right).
$$
\end{table}
%We continue with  the C-G regarding the 5-dimensional representation [3,3].
\begin{table}
\caption{The 1-particle CFP's ( see table \ref{1cf6.42}) involving [3,3] of $S_6$}
\label{1cf6.33}
$$
\left(
\begin{array}{ccccc}
\hline
[3,3]_1&[3,3]_2&[3,3]_3&[3,3]_4&[3,3]_5\\
\hline
 -\frac{1}{6} & -\frac{1}{6 \sqrt{5}} & \sqrt{\frac{2}{15}} & 0 & 0 \\
 \frac{1}{6} & \frac{1}{6 \sqrt{5}} & -\sqrt{\frac{2}{15}} & 0 & 0 \\
 0 & 0 & \frac{\sqrt{\frac{5}{6}}}{4} & \frac{1}{2 \sqrt{3}} & \frac{1}{4
   \sqrt{2}} \\
 -\frac{1}{6} & -\frac{1}{6 \sqrt{5}} & \frac{\sqrt{\frac{3}{10}}}{4} &
   \frac{1}{2 \sqrt{3}} & -\frac{1}{4 \sqrt{2}} \\
 -\frac{1}{6} & \frac{1}{3 \sqrt{5}} & \frac{1}{\sqrt{30}} & -\frac{1}{2
   \sqrt{3}} & 0 \\
 0 & -\frac{1}{2 \sqrt{5}} & \frac{1}{\sqrt{30}} & -\frac{1}{2 \sqrt{3}} & 0
   \\
 \frac{1}{6} & -\frac{\sqrt{5}}{6} & 0 & 0 & 0 \\
\end{array}
\right).
$$
\end{table}
%Finally we show the C-G regarding the 16-dimensional representation [3,2,1].
\begin{table}
\caption{The 1-particle CFP's  ( see table \ref{1cf6.42}) involving  the representation [3,2,1] of $S_6$}
\label{1cf6.321a}
$$
\left(
\begin{array}{ccccccc}
\hline
[3,2,1]_1&[3,2,1]_2&[3,2,1]_3&[3,2,1]_4&[3,2,1]_5&[3,2,1]_6&[3,2,1]_7\\
\hline
 -\frac{1}{\sqrt{31}} & \frac{3 \sqrt{\frac{3}{31}}}{2} & -\frac{3
   \sqrt{\frac{3}{494}}}{2} & -\frac{401}{2 \sqrt{6105346}} &
   -\frac{3807}{\sqrt{801147457}} & 774 \sqrt{\frac{3}{41600095427}} &
   -\frac{1161}{2 \sqrt{17940093295}} \\
 -\frac{2}{\sqrt{31}} & -\frac{17}{5 \sqrt{93}} & \frac{67}{5 \sqrt{1482}} &
   \frac{619}{\sqrt{6105346}} & -\frac{729}{\sqrt{801147457}} & -1231
   \sqrt{\frac{3}{41600095427}} & \frac{3693}{4 \sqrt{17940093295}} \\
 \frac{1}{\sqrt{31}} & -\frac{7}{5 \sqrt{93}} & -\frac{43}{5 \sqrt{1482}} &
   \frac{517}{\sqrt{6105346}} & \frac{4762}{\sqrt{801147457}} & -53225
   \sqrt{\frac{3}{41600095427}} & \frac{31935 \sqrt{\frac{5}{3588018659}}}{4}
   \\
 \frac{1}{\sqrt{31}} & -\frac{7}{5 \sqrt{93}} & \frac{16
   \sqrt{\frac{2}{741}}}{5} & -313 \sqrt{\frac{2}{3052673}} &
   \frac{5589}{\sqrt{801147457}} & 52653 \sqrt{\frac{3}{41600095427}} &
   -\frac{157959}{4 \sqrt{17940093295}} \\
 \frac{2}{\sqrt{31}} & -\frac{14}{5 \sqrt{93}} & -\frac{11}{5 \sqrt{1482}} &
   -\frac{109}{\sqrt{6105346}} & -\frac{2008}{\sqrt{801147457}} & 255
   \sqrt{\frac{3}{41600095427}} & -\frac{153 \sqrt{\frac{5}{3588018659}}}{4}
   \\
 0 & 0 & 0 & 0 & 0 & 0 & 0 \\
 \frac{1}{\sqrt{31}} & -\frac{7}{5 \sqrt{93}} & \frac{16
   \sqrt{\frac{2}{741}}}{5} & 181 \sqrt{\frac{2}{3052673}} &
   -\frac{5917}{\sqrt{801147457}} & 590 \sqrt{\frac{3}{41600095427}} &
   -\frac{177 \sqrt{\frac{5}{3588018659}}}{2} \\
\end{array}
\right),
$$
\end{table}
\begin{table}
\caption{The  \ref{1cf6.321a} continued}
\label{1cf6.321b}
$$
\left(
\begin{array}{ccccc}
\hline
[3,2,1]_{8}&[3,2,1]_{9}&[3,2,1]_{10}&[3,2,1]_{11}&[3,2,1]_{12}\\
\hline
 \frac{55275}{2 \sqrt{6738088697}} & -7370 \sqrt{\frac{30}{82453910639}} &
   256 \sqrt{\frac{2}{87992403623}} & \frac{36571}{4 \sqrt{236960995155}} &
   \frac{6157}{4 \sqrt{11371785645}} \\
 \frac{18345}{4 \sqrt{6738088697}} & -1223 \sqrt{\frac{30}{82453910639}} &
   10681 \sqrt{\frac{2}{87992403623}} & -\frac{35643
   \sqrt{\frac{3}{78986998385}}}{4} & -\frac{3381
   \sqrt{\frac{3}{3790595215}}}{4} \\
 -\frac{53121}{4 \sqrt{6738088697}} & 17707 \sqrt{\frac{6}{412269553195}} &
   -4345 \sqrt{\frac{2}{87992403623}} & 3956 \sqrt{\frac{3}{78986998385}} &
   398 \sqrt{\frac{3}{3790595215}} \\
 -\frac{52155}{4 \sqrt{6738088697}} & 3477 \sqrt{\frac{30}{82453910639}} &
   -4521 \sqrt{\frac{2}{87992403623}} & -\frac{19351
   \sqrt{\frac{3}{78986998385}}}{4} & -\frac{4201
   \sqrt{\frac{3}{3790595215}}}{4} \\
 -\frac{134169}{4 \sqrt{6738088697}} & 44723 \sqrt{\frac{6}{412269553195}} &
   -2327 \sqrt{\frac{2}{87992403623}} & \frac{11092}{\sqrt{236960995155}} &
   \frac{1414}{\sqrt{11371785645}} \\
 0 & 0 & 0 & 0 & 0 \\
 \frac{28959}{2 \sqrt{6738088697}} & -19306 \sqrt{\frac{6}{412269553195}} &
   -81944 \sqrt{\frac{2}{87992403623}} & \frac{50087
   \sqrt{\frac{3}{78986998385}}}{4} & -\frac{10591
   \sqrt{\frac{3}{3790595215}}}{4} \\
\end{array}
\right),
$$
\end{table}
\begin{table}
\caption{The  \ref{1cf6.321a} continued}
\label{1cf6.321c}
$$
\left(
\begin{array}{cccc}
\hline
[3,2,1]_{13}&[3,2,1]_{14}&[3,2,1]_{15}&[3,2,1]_{16}\\
\hline
 \frac{241 \sqrt{\frac{15}{30798779}}}{2} & -\frac{247}{6 \sqrt{850795}} &
   \frac{7}{12 \sqrt{3410}} & -\frac{7}{4 \sqrt{30}} \\
 -\frac{1185 \sqrt{\frac{15}{30798779}}}{4} & \frac{2051}{12 \sqrt{850795}} &
   \frac{1}{6 \sqrt{3410}} & -\frac{1}{2 \sqrt{30}} \\
 -\frac{1084}{\sqrt{461981685}} & -\frac{194}{3 \sqrt{850795}} & -\frac{7}{24
   \sqrt{3410}} & \frac{7}{8 \sqrt{30}} \\
 -\frac{86}{\sqrt{461981685}} & -\frac{194}{3 \sqrt{850795}} & -\frac{7}{24
   \sqrt{3410}} & \frac{7}{8 \sqrt{30}} \\
 \frac{533 \sqrt{\frac{15}{30798779}}}{4} & \frac{163}{4 \sqrt{850795}} &
   -\frac{3}{4 \sqrt{3410}} & -\frac{7}{4 \sqrt{30}} \\
 0 & 0 & 0 & 2 \sqrt{\frac{2}{15}} \\
 \frac{879 \sqrt{\frac{3}{153993895}}}{2} & -\frac{663}{2 \sqrt{850795}} &
   -\frac{53}{4 \sqrt{3410}} & -\frac{1}{4 \sqrt{30}} \\
\end{array}
\right).
$$
\end{table}

We next come to the reduction:
$$[2,1^3]\times[1]\rightarrow 5[2,1^4]+9[2^2,1^2]+10[3,1^3].$$
\begin{table}
\caption{The 1-particle CFP's (outer  C-G coefficients) involving the reduction
$[2,1^3\otimes[1]\rightarrow 5[2,1^4]+9[2^2,1^2]+10[3,1^3]$.}
\label{1cf6.21111}
$$
\left(
\begin{array}{ccccc}
\hline
[2,1^3]_1&[2,1^3]_2&[2,1^3]_3&[2,1^3]_4\\
\hline
 0 & 0 & \frac{\sqrt{\frac{115}{158}}}{2} & -\frac{5}{2 \sqrt{237}} & 0 \\
 0 & 0 & -\frac{\sqrt{\frac{115}{158}}}{2} & \frac{5}{2 \sqrt{237}} & 0 \\
 \sqrt{\frac{10}{509}} & \frac{295 \sqrt{\frac{5}{23414}}}{12} &
   -\frac{\sqrt{\frac{395}{46}}}{12} & 0 & 0 \\
 -\sqrt{\frac{10}{509}} & -\frac{295 \sqrt{\frac{5}{23414}}}{12} & -\frac{59
   \sqrt{\frac{5}{3634}}}{12} & \frac{5}{2 \sqrt{237}} & 0 \\
 0 & 0 & -\frac{\sqrt{\frac{115}{158}}}{2} & \frac{5}{2 \sqrt{237}} & 0 \\
 0 & 0 & -\frac{\sqrt{\frac{115}{158}}}{2} & \frac{5}{2 \sqrt{237}} & 0 \\
 3 \sqrt{\frac{10}{509}} & -\frac{133 \sqrt{\frac{5}{23414}}}{12} &
   \frac{\sqrt{\frac{5}{3634}}}{12} & \frac{1}{8 \sqrt{237}} & \frac{1}{8
   \sqrt{3}} \\
\end{array}
\right),
$$ 
\end{table}
\begin{table}
\caption{The 1-particle CFP's (see table \ref{1cf6.21111}) involving the representation 
$[2^2,1^2]$.}
\label{1cf6.21111a}
$$
\left(
\begin{array}{ccccccccc}
\hline
[2^2,1^2]_1&[2^2,1^2]_2&[2^2,1^2]_3&[2^2,1^2]_4&[2^2,1^2]_5&[2^2,1^2]_6&[2^2,1^2]_7&[2^2,1^2]_8&[2^2,1^2]_9\\
\hline
 \frac{1}{4 \sqrt{3}} & \frac{1}{4 \sqrt{33}} & -\frac{1}{\sqrt{6}} &
   \frac{21}{2 \sqrt{715}} & -\frac{1}{4 \sqrt{390}} & \frac{1}{10 \sqrt{3}}
   & \frac{\sqrt{3}}{40} & \frac{\sqrt{\frac{5}{3}}}{8} & 0 \\
 \frac{1}{2 \sqrt{3}} & \frac{1}{2 \sqrt{33}} & 0 & -\frac{1}{\sqrt{715}} &
   -\frac{\sqrt{\frac{3}{130}}}{4} & \frac{\sqrt{3}}{10} & \frac{3
   \sqrt{3}}{40} & \frac{\sqrt{15}}{8} & 0 \\
 \frac{1}{4 \sqrt{3}} & \frac{1}{4 \sqrt{33}} & \frac{1}{\sqrt{6}} &
   -\frac{1}{2 \sqrt{715}} & \frac{31}{4 \sqrt{390}} & -\frac{1}{10 \sqrt{3}}
   & -\frac{\sqrt{3}}{40} & -\frac{\sqrt{\frac{5}{3}}}{8} & 0 \\
 -\frac{1}{2 \sqrt{3}} & -\frac{1}{2 \sqrt{33}} & 0 & -2 \sqrt{\frac{5}{143}}
   & -\frac{\sqrt{\frac{15}{26}}}{2} & 0 & 0 & 0 & 0 \\
 0 & -\frac{\sqrt{\frac{3}{11}}}{2} & -\frac{1}{2 \sqrt{6}} & \frac{23}{4
   \sqrt{715}} & \frac{1}{4 \sqrt{390}} & -\frac{1}{4 \sqrt{3}} & \frac{1}{8
   \sqrt{3}} & -\frac{\sqrt{\frac{5}{3}}}{8} & \frac{1}{\sqrt{6}} \\
 -\frac{1}{4 \sqrt{3}} & \frac{5}{4 \sqrt{33}} & -\frac{1}{2 \sqrt{6}} &
   \frac{23}{4 \sqrt{715}} & \frac{1}{4 \sqrt{390}} & \frac{1}{20 \sqrt{3}} &
   -\frac{11}{40 \sqrt{3}} & -\frac{\sqrt{\frac{5}{3}}}{8} &
   -\frac{1}{\sqrt{6}} \\
 -\frac{1}{4 \sqrt{3}} & \frac{\sqrt{\frac{11}{3}}}{4} & 0 & 0 & 0 &
   -\frac{\sqrt{3}}{10} & \frac{2}{5 \sqrt{3}} & 0 & \frac{1}{2 \sqrt{6}} \\
\end{array}
\right),
$$
\end{table}
\begin{table}
\caption{The 1-particle CFP's (see table  \ref{1cf6.21111}) involving the representation 
$[3,1^3]$.}
\label{1cf6.21111b}
$$
\left(
\begin{array}{cccccccccc}
\hline
[3,1^3]_1&[3,1^3]_2&[3,1^3]_3&[3,1^3]_4&[3,1^3]_5&[3,1^3]_6&[3,1^3]_7&[3,1^3]_8&[3,1^3]_9&[3,1^3]_{10}\\
\hline
 \frac{1}{12} & \frac{1}{12 \sqrt{10}} & -\frac{1}{4 \sqrt{185}} &
   -\frac{151}{4 \sqrt{7918}} & -\frac{161}{12 \sqrt{15943}} & -\frac{3}{4
   \sqrt{745}} & -\frac{1}{4 \sqrt{6}} & -\frac{1}{4 \sqrt{30}} & 0 &
   \frac{7}{4 \sqrt{15}} \\
 \frac{1}{6} & \frac{1}{6 \sqrt{10}} & -\frac{1}{2 \sqrt{185}} & -\frac{3}{2
   \sqrt{7918}} & -\frac{121}{3 \sqrt{15943}} & -\frac{51}{14 \sqrt{745}} &
   -\frac{5}{4 \sqrt{6}} & \frac{1}{28 \sqrt{30}} & 0 & -\frac{1}{4
   \sqrt{15}} \\
 \frac{1}{12} & \frac{1}{12 \sqrt{10}} & -\frac{21}{4 \sqrt{185}} &
   -\frac{63}{4 \sqrt{7918}} & -\frac{53}{12 \sqrt{15943}} & \frac{19}{28
   \sqrt{745}} & \frac{1}{4 \sqrt{6}} & -\frac{17}{28 \sqrt{30}} & 0 &
   \frac{7}{4 \sqrt{15}} \\
 0 & 0 & 0 & 0 & 0 & 0 & 0 & 0 & 0 & \frac{\sqrt{\frac{5}{3}}}{2} \\
 \frac{1}{12} & -\frac{11}{12 \sqrt{10}} & \frac{11}{4 \sqrt{185}} &
   -\frac{115}{4 \sqrt{7918}} & \frac{175}{12 \sqrt{15943}} & -\frac{29}{28
   \sqrt{745}} & \frac{1}{4 \sqrt{6}} & -\frac{\sqrt{\frac{5}{6}}}{28} &
   -\frac{\sqrt{\frac{3}{5}}}{2} & -\frac{1}{4 \sqrt{15}} \\
 -\frac{1}{6} & \frac{\sqrt{\frac{5}{2}}}{6} & \frac{\sqrt{\frac{5}{37}}}{2}
   & -\frac{59}{2 \sqrt{7918}} & \frac{10}{3 \sqrt{15943}} &
   \frac{\sqrt{\frac{5}{149}}}{2} & \frac{1}{4 \sqrt{6}} & \frac{1}{4
   \sqrt{30}} & \frac{\sqrt{\frac{3}{5}}}{2} & -\frac{1}{4 \sqrt{15}} \\
 -\frac{1}{12} & -\frac{13}{12 \sqrt{10}} & -\frac{7}{4 \sqrt{185}} &
   -\frac{21}{4 \sqrt{7918}} & -\frac{49}{3 \sqrt{15943}} & \frac{26}{7
   \sqrt{745}} & 0 & \frac{3 \sqrt{\frac{6}{5}}}{7} & \frac{1}{2 \sqrt{15}} &
   0 \\
\end{array}
\right).
$$
\end{table}

Now we will consider the reduction:
$$[2,2,1]\times[1]\rightarrow 9[2,2,1,1]+5[2,2,2]+16[3,2,1].$$
\begin{table}
\caption{The 1-particle CFP's (outer  C-G coefficients) involving the reduction
$[2^2,1]\times[1]\rightarrow 9[2^2,1^2]+5[2^3]+16[3,2,1]$.}
\label{1cf6.2211}
$$
\left(
\begin{array}{ccccccccc}
\hline
[2^2,1^2]_1&[2^2,1^2]_2&[2^2,1^2]_3&[2^2,1^2]_4&[2^2,1^2]_5&[2^2,1^2]_6&[2^2,1^2]_7&[2^2,1^2]_8&[2^2,1^2]_9\\
\hline
 2 \sqrt{\frac{2}{255}} & -58 \sqrt{\frac{2}{45645}} & -\frac{49}{2
   \sqrt{67483}} & -\frac{367}{2 \sqrt{6245005}} & 529
   \sqrt{\frac{5}{14954882}} & \frac{217}{6 \sqrt{880230}} &
   -\frac{\sqrt{\frac{5}{39}}}{6} & 0 & \frac{1}{2 \sqrt{30}} \\
 -\sqrt{\frac{5}{102}} & \frac{43}{\sqrt{91290}} & \frac{33}{\sqrt{67483}} &
   -\frac{553}{\sqrt{6245005}} & -\frac{2311}{\sqrt{74774410}} &
   -\frac{577}{6 \sqrt{880230}} & \frac{23}{6 \sqrt{195}} & 0 & -\frac{1}{2
   \sqrt{30}} \\
 \frac{1}{\sqrt{510}} & \frac{73}{\sqrt{91290}} & -\frac{98}{\sqrt{67483}} &
   4 \sqrt{\frac{5}{1249001}} & -\frac{857}{\sqrt{74774410}} & -\frac{19
   \sqrt{\frac{3}{293410}}}{2} & -\frac{\sqrt{\frac{3}{65}}}{2} & 0 &
   -\frac{\sqrt{\frac{3}{10}}}{2} \\
 -2 \sqrt{\frac{2}{255}} & 58 \sqrt{\frac{2}{45645}} & \frac{49}{2
   \sqrt{67483}} & -\frac{1141}{2 \sqrt{6245005}} & -31
   \sqrt{\frac{37}{2020930}} & \frac{\sqrt{\frac{37}{23790}}}{3} & \frac{4}{3
   \sqrt{195}} & 0 & \frac{1}{\sqrt{30}} \\
 -\sqrt{\frac{2}{255}} & 29 \sqrt{\frac{2}{45645}} & \frac{57}{\sqrt{67483}}
   & \frac{827}{\sqrt{6245005}} & -181 \sqrt{\frac{10}{7477441}} & 19
   \sqrt{\frac{10}{88023}} & \frac{1}{4 \sqrt{195}} & -\frac{1}{4 \sqrt{15}}
   & -\frac{1}{2 \sqrt{30}} \\
 -\sqrt{\frac{2}{255}} & 29 \sqrt{\frac{2}{45645}} & \frac{57}{\sqrt{67483}}
   & \frac{827}{\sqrt{6245005}} & -181 \sqrt{\frac{10}{7477441}} &
   -\frac{1117}{6 \sqrt{880230}} & -\frac{17}{12 \sqrt{195}} & \frac{1}{4
   \sqrt{15}} & 0 \\
 \frac{3}{\sqrt{85}} & 3 \sqrt{\frac{5}{3043}} & 24 \sqrt{\frac{6}{67483}} &
   -128 \sqrt{\frac{6}{6245005}} & 184 \sqrt{\frac{3}{37387205}} &
   -\frac{389}{4 \sqrt{146705}} & -\frac{1}{\sqrt{130}} & \frac{1}{2
   \sqrt{10}} & -\frac{1}{4 \sqrt{5}} \\
\end{array}
\right),
$$
\end{table}
\begin{table}
\caption{The 1-particle CFP's (see table \ref{1cf6.2211} involving the representation $[2^3]$.}
\label{1cf6.2211a}
$$
\left(
\begin{array}{ccccc}
\hline
[2^3]_1&[2^3]_2&[2^3]_3&[2^3]_4&[2^3]_5\\
\hline
 \frac{1}{\sqrt{10}} & \frac{1}{\sqrt{15}} & 0 & 0 & 0 \\
 0 & 0 & \frac{1}{\sqrt{6}} & 0 & 0 \\
 -\frac{1}{\sqrt{10}} & -\frac{1}{\sqrt{15}} & 0 & 0 & 0 \\
 0 & 0 & 0 & -\frac{1}{\sqrt{6}} & 0 \\
 -\frac{1}{2 \sqrt{10}} & \frac{\sqrt{\frac{3}{5}}}{4} & -\frac{1}{2
   \sqrt{6}} & \frac{1}{2 \sqrt{6}} & -\frac{1}{4 \sqrt{3}} \\
 \frac{1}{2 \sqrt{10}} & -\frac{\sqrt{\frac{3}{5}}}{4} & -\frac{1}{2
   \sqrt{6}} & \frac{1}{2 \sqrt{6}} & \frac{1}{4 \sqrt{3}} \\
 -\frac{1}{2 \sqrt{15}} & \frac{1}{2 \sqrt{10}} & 0 & 0 & -\frac{1}{2
   \sqrt{2}} \\
\end{array}
\right),
$$
\end{table}
\begin{table}
\caption{The 1-particle CFP's (see table \ref{1cf6.2211} involving the representation $[3,2,1]$.}
\label{1cf6.2211b}
$$
\left(
\begin{array}{cccccccc}
\hline
[3,2,1]_1&[3,2,1]_2&[3,2,1]_3&[3,2,1]_4&[3,2,1]_5&[3,2,1]_6&[3,2,1]_7&[3,2,1]_8\\
\hline
 -\frac{3 \sqrt{\frac{3}{22}}}{2} & -\frac{3}{2 \sqrt{22}} & \frac{1}{4
   \sqrt{22}} & \frac{\sqrt{\frac{5}{66}}}{4} & 4 \sqrt{\frac{5}{861}} & 32
   \sqrt{\frac{2}{752227}} & -\frac{7 \sqrt{\frac{2}{2621}}}{75} &
   -\frac{\sqrt{\frac{2}{31}}}{75} \\
 0 & 0 & 0 & 0 & 2 \sqrt{\frac{15}{287}} & 48 \sqrt{\frac{2}{752227}} &
   -\frac{7}{25 \sqrt{5242}} & -\frac{1}{25 \sqrt{62}} \\
 -\sqrt{\frac{3}{22}} & -\frac{1}{\sqrt{22}} & \sqrt{\frac{2}{11}} &
   -\frac{1}{\sqrt{330}} & -\frac{13}{4 \sqrt{4305}} & -120
   \sqrt{\frac{2}{752227}} & \frac{7}{10 \sqrt{5242}} & \frac{1}{10
   \sqrt{62}} \\
 -\sqrt{\frac{3}{22}} & -\frac{1}{\sqrt{22}} & -\frac{3}{4 \sqrt{22}} &
   \frac{29}{4 \sqrt{330}} & -\frac{67}{4 \sqrt{4305}} & 88
   \sqrt{\frac{2}{752227}} & -\frac{77}{150 \sqrt{5242}} & -\frac{11}{150
   \sqrt{62}} \\
 -\sqrt{\frac{3}{22}} & \frac{1}{\sqrt{22}} & -\frac{1}{4 \sqrt{22}} &
   -\frac{3 \sqrt{\frac{3}{110}}}{4} & \frac{19 \sqrt{\frac{3}{1435}}}{4} &
   \frac{103}{\sqrt{1504454}} & -\frac{1847}{150 \sqrt{5242}} &
   \frac{379}{150 \sqrt{62}} \\
 0 & -\sqrt{\frac{2}{11}} & -\frac{1}{2 \sqrt{22}} &
   -\frac{\sqrt{\frac{3}{110}}}{2} & \frac{3 \sqrt{\frac{21}{205}}}{4} &
   -\sqrt{\frac{7}{214922}} & \frac{361}{30 \sqrt{5242}} & -\frac{77}{30
   \sqrt{62}} \\
 \frac{1}{4 \sqrt{11}} & -\frac{\sqrt{\frac{3}{11}}}{4} & \frac{5}{8
   \sqrt{33}} & -\frac{\sqrt{\frac{5}{11}}}{8} &
   -\frac{\sqrt{\frac{5}{574}}}{2} & -\frac{311}{2 \sqrt{2256681}} &
   -\frac{826}{75 \sqrt{7863}} & -\frac{118}{75 \sqrt{93}} \\
\end{array}
\right),
$$
\end{table}
\begin{table}
\caption{The 1-particle CFP's (see table \ref{1cf6.2211} involving the representation $[3,2,1]$ (continued).}
\label{1cf6.2211d}
$$
\left(
\begin{array}{cccccccc}
\hline
[3,2,1]_9&[3,2,1]_{10}&[3,2,1]_{11}&[3,2,1]_{12}&[3,2,1]_{13}&[3,2,1]_{14}&[3,2,1]_{15}&[3,2,1]_{16}\\
\hline
 \frac{1329 \sqrt{\frac{3}{7920035}}}{10} & \frac{1117
   \sqrt{\frac{6}{90084011}}}{5} & -\frac{1139 \sqrt{\frac{2}{14825067}}}{5}
   & 21 \sqrt{\frac{3}{5241610}} & -\frac{7 \sqrt{\frac{3}{24310}}}{4} &
   \frac{\sqrt{\frac{7}{130}}}{4} & 0 & 0 \\
 \frac{7834}{5 \sqrt{23760105}} & \frac{27683}{5 \sqrt{540504066}} &
   -\frac{1139 \sqrt{\frac{3}{9883378}}}{5} & -\frac{1307}{\sqrt{15724830}} &
   -\frac{21 \sqrt{\frac{3}{24310}}}{4} & \frac{3 \sqrt{\frac{7}{130}}}{4} &
   0 & 0 \\
 \frac{719 \sqrt{\frac{3}{7920035}}}{4} & \frac{7927
   \sqrt{\frac{3}{180168022}}}{10} & \frac{2981}{10 \sqrt{29650134}} &
   -\frac{21 \sqrt{\frac{15}{1048322}}}{2} & -19 \sqrt{\frac{3}{24310}} &
   -\sqrt{\frac{7}{130}} & 0 & 0 \\
 -\frac{169 \sqrt{\frac{111}{214055}}}{20} & -\frac{67
   \sqrt{\frac{111}{4869406}}}{2} & \frac{105 \sqrt{\frac{3}{9883378}}}{2} &
   \frac{63 \sqrt{\frac{3}{5241610}}}{2} & \frac{31}{2 \sqrt{72930}} &
   \frac{\sqrt{\frac{7}{130}}}{2} & 0 & 0 \\
 -\frac{6499}{20 \sqrt{23760105}} & \frac{7 \sqrt{\frac{82}{6591513}}}{5} &
   -\frac{31 \sqrt{\frac{41}{723174}}}{5} & \frac{1601}{2 \sqrt{15724830}} &
   \frac{19}{\sqrt{72930}} & \frac{\sqrt{\frac{7}{130}}}{3} &
   -\frac{\sqrt{\frac{7}{5}}}{6} & \frac{1}{2 \sqrt{5}} \\
 -\frac{61 \sqrt{\frac{3}{7920035}}}{4} & -\frac{2717
   \sqrt{\frac{3}{180168022}}}{5} & -\frac{1073 \sqrt{\frac{2}{14825067}}}{5}
   & \frac{1391}{2 \sqrt{15724830}} & \frac{37 \sqrt{\frac{3}{24310}}}{4} &
   -\frac{\sqrt{\frac{7}{130}}}{12} & \frac{\sqrt{\frac{7}{5}}}{6} &
   -\frac{1}{2 \sqrt{5}} \\
 \frac{3304 \sqrt{\frac{2}{7920035}}}{15} & -\frac{4130}{3 \sqrt{90084011}} &
   \frac{325}{\sqrt{4941689}} & 39 \sqrt{\frac{5}{524161}} &
   -\frac{\sqrt{\frac{65}{187}}}{4} & \frac{\sqrt{\frac{65}{21}}}{4} &
   \frac{13}{4 \sqrt{210}} & \frac{7}{4 \sqrt{30}} \\
\end{array}
\right).
$$
\end{table}
Finally we will compute the C-G series:
$$[3,1^2]\times[1]\rightarrow 10[4,1^2]+10[3,1^3]+16[3,2,1] $$
\begin{table}
\caption{The 1-particle CFP's (outer  C-G coefficients) involving the reduction$[3,1^2]\times[1]\rightarrow 10[4,1^2]+10[3,1^3]+16[3,2,1]  $.}
\label{1cf6.411}
$$
\left(
\begin{array}{ccccc}
\hline
[4,1^2]_1&[4,1^2]_2&[4,1^2]_3&[4,1^2]_4&[4,1^2]_5\\
\hline
 \frac{\sqrt{5}}{32} & \frac{83}{96 \sqrt{19}} & \frac{221}{24 \sqrt{418}} & -\frac{4681}{24
   \sqrt{2368894}} & -\frac{221 \sqrt{\frac{5}{9277941}}}{2} \\
 -\frac{\sqrt{5}}{96} & -\frac{83}{288 \sqrt{19}} & \frac{235}{72 \sqrt{418}} & -\frac{4127}{72
   \sqrt{2368894}} & -\frac{1391 \sqrt{\frac{5}{9277941}}}{6} \\
 -\frac{\sqrt{5}}{48} & -\frac{83}{144 \sqrt{19}} & \frac{235}{36 \sqrt{418}} & -\frac{4127}{36
   \sqrt{2368894}} & \frac{1307 \sqrt{\frac{5}{9277941}}}{6} \\
 \frac{\sqrt{5}}{16} & \frac{83}{48 \sqrt{19}} & -\frac{7}{12 \sqrt{418}} & -\frac{277}{12
   \sqrt{2368894}} & -\frac{193 \sqrt{\frac{5}{9277941}}}{2} \\
 \frac{\sqrt{5}}{32} & \frac{83}{96 \sqrt{19}} & \frac{221}{24 \sqrt{418}} & -\frac{4681}{24
   \sqrt{2368894}} & -\frac{221 \sqrt{\frac{5}{9277941}}}{2} \\
 \frac{\sqrt{5}}{32} & \frac{83}{96 \sqrt{19}} & \frac{221}{24 \sqrt{418}} & -\frac{4681}{24
   \sqrt{2368894}} & -\frac{221 \sqrt{\frac{5}{9277941}}}{2} \\
 0 & 0 & 0 & 0 & 0 \\
\end{array}
\right),
$$
\end{table}
\begin{table}
\caption{The 1-particle CFP's (see table \ref{1cf6.411}) involving the representation $[4,1^2]$.}
\label{1cf6.411a}
$$
\left(
\begin{array}{ccccc}
\hline
[4,1^2]_6&[4,1^2]_7&[4,1^2]_8&[4,1^2]_9&[4,1^2]_{10}\\
\hline
 \frac{493 \sqrt{\frac{17}{548328809}}}{2} & -\frac{\sqrt{\frac{17}{16142343}}}{2} & \frac{1}{2
   \sqrt{6330641}} & \frac{1}{9 \sqrt{7198}} & -\frac{1}{9 \sqrt{2}} \\
 \frac{3103 \sqrt{\frac{17}{548328809}}}{6} & -151 \sqrt{\frac{17}{16142343}} &
   \frac{151}{\sqrt{6330641}} & \frac{151 \sqrt{\frac{2}{3599}}}{9} & -\frac{\sqrt{2}}{9} \\
 -\frac{9199 \sqrt{\frac{17}{548328809}}}{6} & \frac{151 \sqrt{\frac{17}{16142343}}}{2} &
   -\frac{151}{2 \sqrt{6330641}} & -\frac{151}{9 \sqrt{7198}} & \frac{1}{9 \sqrt{2}} \\
 \frac{2525 \sqrt{\frac{17}{548328809}}}{2} & 25 \sqrt{\frac{51}{5380781}} &
   -\frac{75}{\sqrt{6330641}} & -\frac{25 \sqrt{\frac{2}{3599}}}{3} & 0 \\
 \frac{493 \sqrt{\frac{17}{548328809}}}{2} & -\frac{\sqrt{\frac{17}{16142343}}}{2} & \frac{1}{2
   \sqrt{6330641}} & \frac{1}{9 \sqrt{7198}} & -\frac{1}{9 \sqrt{2}} \\
 \frac{493 \sqrt{\frac{17}{548328809}}}{2} & -\frac{\sqrt{\frac{17}{16142343}}}{2} & \frac{1}{2
   \sqrt{6330641}} & \frac{1}{9 \sqrt{7198}} & -\frac{1}{9 \sqrt{2}} \\
 0 & -\frac{\sqrt{\frac{52003}{1759}}}{12} & \frac{3059}{4 \sqrt{18991923}} & -5
   \sqrt{\frac{6}{3599}} & 0 \\
\end{array}
\right),
$$
\end{table}
\begin{table}
\caption{The 1-particle CFP's (see table \ref{1cf6.411} involving the representation $[3,1^3]$ (continued).}
\label{1cf6.411b}
$$
\left(
\begin{array}{ccccc}
\hline
[3,1^3]_1&[3,1^3]_2&[3,1^3]_3&[3,1^3]_4&[3,1^3]_5\\
\hline
 -\sqrt{\frac{2}{65}} & -\frac{6}{\sqrt{3055}} & 17 \sqrt{\frac{2}{30597}} &
   -\frac{104}{\sqrt{5502903}} & -13 \sqrt{\frac{3}{121435798}} \\
 \frac{1}{\sqrt{130}} & \frac{3}{\sqrt{3055}} & \frac{43}{3 \sqrt{61194}} & \frac{328}{3
   \sqrt{5502903}} & \frac{41}{\sqrt{364307394}} \\
 \sqrt{\frac{2}{65}} & \frac{6}{\sqrt{3055}} & \frac{43 \sqrt{\frac{2}{30597}}}{3} & \frac{656}{3
   \sqrt{5502903}} & 41 \sqrt{\frac{2}{182153697}} \\
 0 & 0 & 0 & 0 & 0 \\
 0 & \frac{\sqrt{\frac{65}{47}}}{4} & -\frac{145}{2 \sqrt{61194}} & -\frac{83
   \sqrt{\frac{3}{1834301}}}{4} & 1313 \sqrt{\frac{3}{121435798}} \\
 \frac{3}{\sqrt{130}} & -\frac{29}{4 \sqrt{3055}} & -\frac{145}{2 \sqrt{61194}} & \frac{185}{4
   \sqrt{5502903}} & -1315 \sqrt{\frac{3}{121435798}} \\
 -\sqrt{\frac{3}{130}} & \frac{29}{4 \sqrt{9165}} & -\frac{43}{6 \sqrt{20398}} & \frac{1445}{4
   \sqrt{1834301}} & -\frac{5212 \sqrt{\frac{2}{60717899}}}{3} \\
\end{array}
\right),
$$
\end{table}
\begin{table}
\caption{The 1-particle CFP's (see table \ref{1cf6.411}) involving the representation $[3,1^3]$.}
\label{1cf6.411c}
$$
\left(
\begin{array}{ccccc}
\hline
[3,1^3]_6&[3,1^3]_7&[3,1^3]_8&[3,1^3]_9&[3,1^3]_{10}\\
\hline
 \frac{1079}{2 \sqrt{85829667}} & \frac{1417}{18 \sqrt{4441045}} & -\frac{13}{9 \sqrt{2230}} & 0 &
   -\frac{13}{9 \sqrt{10}} \\
 -\frac{5293}{3 \sqrt{85829667}} & -\frac{2381}{9 \sqrt{4441045}} & \frac{13}{9 \sqrt{2230}} & 0 &
   \frac{13}{9 \sqrt{10}} \\
 \frac{377}{6 \sqrt{85829667}} & \frac{293}{18 \sqrt{4441045}} & \frac{14 \sqrt{\frac{2}{1115}}}{9} &
   0 & \frac{13}{9 \sqrt{10}} \\
 0 & 0 & 0 & 0 & \frac{\sqrt{\frac{5}{2}}}{3} \\
 -\frac{617}{\sqrt{85829667}} & \frac{2069}{9 \sqrt{4441045}} & \frac{\sqrt{\frac{5}{446}}}{9} &
   \frac{\sqrt{\frac{2}{5}}}{3} & -\frac{1}{9 \sqrt{10}} \\
 100 \sqrt{\frac{7}{12261381}} & -\frac{56 \sqrt{\frac{35}{126887}}}{9} & -\frac{7}{9 \sqrt{2230}} &
   -\frac{\sqrt{\frac{2}{5}}}{3} & -\frac{1}{9 \sqrt{10}} \\
 -\frac{9817}{12 \sqrt{28609889}} & \frac{\sqrt{\frac{3345}{3983}}}{4} & 0 & 0 & 0 \\
\end{array}
\right),
$$
\end{table}
\begin{table}
\caption{The 1-particle CFP's (see table \ref{1cf6.411}) involving the representation $[3,2,1]$.}
\label{1cf6.411d}
$$
\left(
\begin{array}{cccccccc}
\hline
[3,2,1]_1&[3,2,1]_2&[3,2,1]_3&[3,2,1]_4&[3,2,1]_5&[3,2,1]_6&[3,2,1]_7&[3,2,1]_8\\
\hline
 \frac{1}{\sqrt{15}} & \frac{\sqrt{\frac{2}{5}}}{3} & -\sqrt{\frac{2}{21}} &
   \frac{20 \sqrt{\frac{2}{553}}}{3} & -40 \sqrt{\frac{5}{156183}} & 17
   \sqrt{\frac{6}{1918349}} & -17 \sqrt{\frac{19}{986829}} & 0 \\
 \frac{2}{3 \sqrt{15}} & \frac{2 \sqrt{\frac{2}{5}}}{9} &
   \frac{\sqrt{\frac{2}{21}}}{3} & \frac{\sqrt{\frac{2}{553}}}{9} & -\frac{160
   \sqrt{\frac{5}{156183}}}{3} & 68 \sqrt{\frac{2}{5755047}} & -\frac{68
   \sqrt{\frac{19}{986829}}}{3} & 0 \\
 \frac{4}{3 \sqrt{15}} & \frac{4 \sqrt{\frac{2}{5}}}{9} & -\frac{5}{3 \sqrt{42}}
   & -\frac{5}{9 \sqrt{1106}} & \frac{163 \sqrt{\frac{5}{156183}}}{3} & 79
   \sqrt{\frac{2}{5755047}} & -\frac{79 \sqrt{\frac{19}{986829}}}{3} & 0 \\
 -\frac{1}{\sqrt{15}} & -\frac{\sqrt{\frac{2}{5}}}{3} & -\frac{1}{\sqrt{42}} &
   \frac{41}{3 \sqrt{1106}} & -41 \sqrt{\frac{5}{156183}} & -32
   \sqrt{\frac{6}{1918349}} & 32 \sqrt{\frac{19}{986829}} & 0 \\
 0 & -\frac{\sqrt{\frac{5}{2}}}{6} & \frac{1}{\sqrt{42}} & -\frac{10
   \sqrt{\frac{2}{553}}}{3} & 20 \sqrt{\frac{5}{156183}} & \frac{625
   \sqrt{\frac{3}{3836698}}}{2} & \frac{670}{\sqrt{18749751}} & -34
   \sqrt{\frac{5}{191121}} \\
 -\frac{1}{\sqrt{15}} & \frac{1}{6 \sqrt{10}} & \frac{1}{\sqrt{42}} & -\frac{10
   \sqrt{\frac{2}{553}}}{3} & 20 \sqrt{\frac{5}{156183}} & -\frac{693
   \sqrt{\frac{3}{3836698}}}{2} & -\frac{347}{\sqrt{18749751}} & 34
   \sqrt{\frac{5}{191121}} \\
 \frac{1}{3 \sqrt{5}} & -\frac{1}{\sqrt{30}} & 0 & 0 & 0 &
   \frac{\sqrt{\frac{659}{5822}}}{5} & \frac{3 \sqrt{\frac{113}{55309}}}{5} &
   \frac{407}{3 \sqrt{318535}} \\
\end{array}
\right),
$$
\end{table}
\begin{table}
\caption{The 1-particle CFP's (see table \ref{1cf6.411}) involving the representation $[3,2,1]$ (continued).}
\label{1cf6.411e}
$$
\left(
\begin{array}{cccccccc}
\hline
[3,2,1]_9&[3,2,1]_{10}&[3,2,1]_{11}&[3,2,1]_{12}&[3,2,1]_{13}&[3,2,1]_{14}&[3,2,1]_{15}&[3,2,1]_{16}\\
\hline
 0 & 0 & 10 \sqrt{\frac{545}{36021801}} & 120 \sqrt{\frac{5}{55998493}} & \frac{4
   \sqrt{\frac{5}{527}}}{3} & 0 & 0 & 0 \\
 0 & 0 & -62 \sqrt{\frac{545}{36021801}} & -24 \sqrt{\frac{155}{1806403}} &
   \frac{\sqrt{\frac{155}{17}}}{9} & -\frac{\sqrt{5}}{9} & 0 & 0 \\
 0 & 0 & 31 \sqrt{\frac{545}{36021801}} & 12 \sqrt{\frac{155}{1806403}} &
   -\frac{\sqrt{\frac{155}{17}}}{18} & \frac{\sqrt{5}}{18} & 0 & 0 \\
 0 & 0 & 41 \sqrt{\frac{545}{36021801}} & 492 \sqrt{\frac{5}{55998493}} & -\frac{7
   \sqrt{\frac{5}{527}}}{18} & \frac{\sqrt{5}}{18} & 0 & 0 \\
 304 \sqrt{\frac{5}{4288487}} & -13 \sqrt{\frac{5}{139411}} & -884 \sqrt{\frac{5}{3926376309}} &
   -\frac{193 \sqrt{\frac{5}{55998493}}}{12} & -\frac{2 \sqrt{\frac{5}{527}}}{3} & 0 & -\frac{1}{6} &
   \frac{1}{4 \sqrt{3}} \\
 -304 \sqrt{\frac{5}{4288487}} & 13 \sqrt{\frac{5}{139411}} & -206 \sqrt{\frac{5}{3926376309}} &
   -\frac{1247 \sqrt{\frac{5}{55998493}}}{12} & -\frac{2 \sqrt{\frac{5}{527}}}{3} & 0 & \frac{1}{6} &
   -\frac{1}{4 \sqrt{3}} \\
 -215 \sqrt{\frac{5}{12865461}} & -\frac{91 \sqrt{\frac{5}{418233}}}{2} & 51
   \sqrt{\frac{565}{11582231}} & -\frac{17 \sqrt{\frac{2635}{318777}}}{4} & 0 & 0 & 0 & \frac{1}{12}
   \\
\end{array}
\right).
$$
\end{table}
\clearpage
\section{Appendix B:	Tables of  $S_4\subset S_6$ (2-particle CFP's).}
Will consider the cases $[f']=[2,2],[3,1],[2,1^2]$ of $S_4$ as follows:
\begin{itemize}
\item $ [2,2]\otimes[2]\rightarrow 9[4,2]+5[2^3]+16[3,2,1]$: tables \ref{2.2x2a}-\ref{2.2x2d}.
\item $ [2,2]\otimes[1^2]\rightarrow 9[2^2,1^2]+5[3^2]+16[3,2,1]$: tables \ref{2.2x11a}-\ref{2.2x11c}.
\item $ [3,1]\otimes[1^2]\rightarrow 9[2^2,1^2]+5[3^2]+16[3,2,1]$: tables \ref{2.2x11a}-\ref{2.2x11c}.
 \item $ [3,1]\otimes[2]\rightarrow 5[5,1]+9[4,2]+10[4,1^2]+5[3,3]+16[3,2,1]$: Tables \ref{t31x2.2a}-\ref{t31x2.2d}
\item $ [3,1]\otimes[1^2]\rightarrow 9[4,2]+10[4,1^2]+10[3,1^3]+16[3,2,1]$: Tables
\ref{t31x2.11a}-\ref{t31x2.11e}.
\item $ [2,1^2]\otimes[1^2]\rightarrow 5[2,1^4]+ 9[2^2,1^2]+10[3,1^3]+5[2^3] +16[3,2,1]$: Tables
\ref{t211x11.11a}-\ref{t211x11.11e}.
\item $[2,1^2]\otimes[2]\rightarrow 9[2^2,1^2]+10[3^3,1^3]+10[4,1^2]+16[3,2,1]$: Tables
\ref{t211x2.11a}-\ref{t211x2.11e}.
%See Table \ref{tab:22x2_42},\ref{tab:22x2_222}.
\end{itemize}

\begin{table}
\caption{The reduction  $ [2,2]\otimes[2]\rightarrow 9[4,2]+5[2^3]+16[3,2,1]$:}
\label{2.2x2a}
$$
\left(
\begin{array}{ccccccccc}
\hline
[4,2]_1&[4,2]_2&[4,2]_3&[4,2]_4&[4,2]_5&[4,2]_6&[4,2]_7&[4,2]_8&[4,2]_9\\
%&&&&[4_2]_i&&&&\\
%1&2&3&4&5*6&7&8&9\\
\hline
 -\frac{1}{2 \sqrt{6}} & \frac{7}{6 \sqrt{10}} & -\frac{7}{6 \sqrt{170}} &
   \frac{5}{6 \sqrt{102}} & \frac{14}{3 \sqrt{1185}} & \frac{2
   \sqrt{\frac{2}{79}}}{3} & \frac{2 \sqrt{\frac{2}{65}}}{3} &
   -\sqrt{\frac{2}{65}} & \frac{1}{\sqrt{30}} \\
 0 & 0 & 3 \sqrt{\frac{3}{170}} & -\frac{1}{5 \sqrt{34}} &
   \sqrt{\frac{5}{79}} & \frac{3 \sqrt{\frac{3}{158}}}{10} & \frac{3
   \sqrt{\frac{3}{130}}}{2} & \sqrt{\frac{3}{130}} & 0 \\
 0 & 0 & 0 & 0 & 0 & 0 & \frac{3}{\sqrt{130}} & -\frac{9}{2 \sqrt{130}} &
   \frac{\sqrt{\frac{3}{10}}}{2} \\
 0 & 0 & \frac{3 \sqrt{\frac{3}{170}}}{2} & \frac{13}{6 \sqrt{34}} &
   \frac{4}{3 \sqrt{395}} & -\frac{3 \sqrt{\frac{3}{158}}}{2} & \frac{3
   \sqrt{\frac{3}{130}}}{2} & \sqrt{\frac{3}{130}} & 0 \\
 0 & 0 & -\frac{3}{\sqrt{170}} & \frac{1}{5 \sqrt{102}} &
   \frac{13}{\sqrt{1185}} & -\frac{2 \sqrt{\frac{2}{79}}}{5} &
   \frac{1}{\sqrt{130}} & -\frac{3}{2 \sqrt{130}} &
   \frac{\sqrt{\frac{3}{10}}}{2} \\
 0 & 0 & \frac{3 \sqrt{\frac{3}{170}}}{2} & -\frac{71}{30 \sqrt{34}} &
   \frac{11}{3 \sqrt{395}} & \frac{9 \sqrt{\frac{3}{158}}}{5} & 0 & 0 & 0
   \\
 -\frac{1}{2 \sqrt{6}} & \frac{7}{6 \sqrt{10}} & \frac{1}{3 \sqrt{170}} &
   -\frac{23}{15 \sqrt{102}} & -\frac{2}{3 \sqrt{1185}} & \frac{203}{60
   \sqrt{158}} & -\frac{17}{12 \sqrt{130}} & \frac{1}{2 \sqrt{130}} &
   -\frac{1}{2 \sqrt{30}} \\
\end{array}
\right),
$$
\end{table}
\begin{table}
\caption{As in table \ref{2.2x2a}}
\label{2.2x2b}
$$
\left(
\begin{array}{ccccc}
\hline
[2^3]_1&[2^3]_2&[2^3]_3&[2^3]_4&[2^3]_5\\
\hline
 0 & \frac{1}{4 \sqrt{2}} & -\frac{\sqrt{\frac{3}{2}}}{4} &
   \frac{1}{4 \sqrt{6}} & \frac{1}{4 \sqrt{2}} \\
 0 & \frac{1}{4 \sqrt{6}} & \frac{1}{4 \sqrt{2}} & -\frac{1}{4
   \sqrt{2}} & \frac{\sqrt{\frac{3}{2}}}{4} \\
 \frac{1}{2 \sqrt{3}} & 0 & \frac{1}{4 \sqrt{6}} & -\frac{1}{2
   \sqrt{6}} & -\frac{1}{4 \sqrt{2}} \\
 \frac{1}{6} & 0 & \frac{1}{12 \sqrt{2}} & \frac{1}{2 \sqrt{2}}
   & -\frac{1}{4 \sqrt{6}} \\
 -\frac{1}{2 \sqrt{3}} & -\frac{1}{4 \sqrt{2}} & \frac{1}{2
   \sqrt{6}} & \frac{1}{4 \sqrt{6}} & 0 \\
 \frac{1}{6} & -\frac{\sqrt{\frac{3}{2}}}{4} & -\frac{1}{6
   \sqrt{2}} & \frac{1}{4 \sqrt{2}} & 0 \\
 0 & \frac{1}{4 \sqrt{2}} & \frac{\sqrt{\frac{3}{2}}}{4} &
   \frac{1}{4 \sqrt{6}} & -\frac{1}{4 \sqrt{2}} \\
\end{array}
\right),$$ 
\end{table}
\begin{table}
\caption{As in table \ref{2.2x2a}}
\label{2.2x2c}
$$
\left(
\begin{array}{cccccccc}
\hline
[3,2,1]_1&[3,2,1]_2&[3,2,1]_3&[3,2,1]_4&[3,2,1]_5&[3,2,1]_6&[3,2,1]_7&[3,2,1]_8\\
\hline
 0 & 0 & \frac{3}{\sqrt{35}} & -\frac{13 \sqrt{\frac{3}{770}}}{4} &
   -\frac{57 \sqrt{\frac{3}{5038}}}{4} & -\frac{21}{4 \sqrt{2290}} &
   -\frac{15}{4 \sqrt{482}} & -\frac{843 \sqrt{\frac{3}{9026173}}}{4} \\
 0 & 0 & -\sqrt{\frac{3}{35}} & \frac{13}{4 \sqrt{770}} & -\frac{119}{4
   \sqrt{5038}} & -2 \sqrt{\frac{10}{687}} & -5 \sqrt{\frac{2}{723}} &
   -\frac{321}{\sqrt{9026173}} \\
 0 & 0 & 0 & \frac{\sqrt{\frac{105}{22}}}{4} & -\frac{9
   \sqrt{\frac{3}{5038}}}{4} & \frac{69}{4 \sqrt{2290}} & -\frac{9}{4
   \sqrt{482}} & -\frac{313 \sqrt{\frac{3}{9026173}}}{4} \\
 0 & 0 & 0 & -\frac{\sqrt{\frac{35}{22}}}{4} & \frac{9}{4 \sqrt{5038}} & 4
   \sqrt{\frac{10}{687}} & -2 \sqrt{\frac{2}{723}} &
   -\frac{32}{\sqrt{9026173}} \\
 0 & 0 & 0 & 0 & 0 & 0 & 0 & 4 \sqrt{\frac{723}{37453}} \\
 0 & 0 & 0 & 0 & 0 & 0 & 0 & 4 \sqrt{\frac{241}{37453}} \\
 0 & -\frac{1}{2 \sqrt{3}} & -\frac{3}{4 \sqrt{35}} & \frac{1}{4
   \sqrt{2310}} & \frac{15 \sqrt{\frac{3}{5038}}}{4} & -\frac{1}{8
   \sqrt{2290}} & \frac{9}{8 \sqrt{482}} & -\frac{13
   \sqrt{\frac{39}{694321}}}{8} \\
\end{array}
\right),
$$
\end{table}
\begin{table}
\caption{As in table \ref{2.2x2c} (continued)}
\label{2.2x2d}
$$
\left(
\begin{array}{cccccccc}
\hline
[3,2,1]_9&[3,2,1]_{10}&[3,2,1]_{11}&[3,2,1]_{12}&[3,2,1]_{13}&[3,2,1]_{14}&[3,2,1]_{15}&[3,2,1]_{16}\\
\hline
 \frac{2495 \sqrt{\frac{15}{687150191}}}{2} & -\frac{3
   \sqrt{\frac{209}{91735}}}{2} & 0 & -\sqrt{\frac{6}{749}} & \frac{53}{12
   \sqrt{7490}} & \frac{127}{12 \sqrt{17710}} & 2 \sqrt{\frac{2}{3795}} &
   0 \\
 -\frac{809 \sqrt{\frac{35}{98164313}}}{2} & \frac{27
   \sqrt{\frac{105}{547789}}}{2} & \frac{\sqrt{\frac{10}{1463}}}{3} &
   -\frac{2 \sqrt{\frac{2}{749}}}{3} & -\frac{143}{12 \sqrt{22470}} &
   -\frac{109}{12 \sqrt{53130}} & \frac{9}{2 \sqrt{2530}} & \frac{1}{2
   \sqrt{30}} \\
 \frac{7653 \sqrt{\frac{3}{3435750955}}}{2} &
   -\frac{\sqrt{\frac{209}{91735}}}{2} & 0 & -2 \sqrt{\frac{6}{749}} &
   -\frac{9}{\sqrt{7490}} & 3 \sqrt{\frac{7}{2530}} & -4
   \sqrt{\frac{2}{3795}} & 0 \\
 -157 \sqrt{\frac{35}{98164313}} & 2 \sqrt{\frac{105}{547789}} &
   -\frac{\sqrt{\frac{14}{1045}}}{3} & -\frac{2 \sqrt{\frac{14}{107}}}{3}
   & -\sqrt{\frac{21}{1070}} & -3 \sqrt{\frac{21}{2530}} & 4
   \sqrt{\frac{2}{1265}} & 0 \\
 48 \sqrt{\frac{105}{98164313}} & 8 \sqrt{\frac{209}{91735}} & 0 & 0 &
   \frac{\sqrt{\frac{107}{70}}}{6} & \frac{1}{6 \sqrt{17710}} & 8
   \sqrt{\frac{2}{3795}} & 0 \\
 48 \sqrt{\frac{35}{98164313}} & -34 \sqrt{\frac{105}{547789}} &
   \frac{29}{3 \sqrt{14630}} & \frac{1}{3 \sqrt{1498}} & -\frac{263}{6
   \sqrt{22470}} & \frac{\sqrt{\frac{11}{4830}}}{6} & \frac{3
   \sqrt{\frac{11}{230}}}{2} & \frac{1}{2 \sqrt{30}} \\
 -\frac{621 \sqrt{\frac{39}{264288535}}}{4} &
   -\frac{\sqrt{\frac{209}{91735}}}{4} & -\frac{\sqrt{\frac{209}{210}}}{2}
   & -\frac{43}{2 \sqrt{4494}} & \frac{181}{12 \sqrt{7490}} & \frac{13
   \sqrt{\frac{11}{1610}}}{12} & \frac{\sqrt{\frac{11}{690}}}{2} &
   \frac{1}{2 \sqrt{10}} \\
\end{array}
\right).
$$
\end{table}
%\begin{itemize}
%\item $ [2,2]\otimes[1^2]\rightarrow 9[2^2,1^2]+5[3^2]+16[3,2,1]$:\\
%See Table \ref{tab:22x2_42},\ref{tab:22x2_222}.
%\end{itemize}
\begin{table}
\caption{The reduction  $ [2,2]\otimes[1^2]\rightarrow 9[4,2]+5[2^3]+16[3,2,1]$:}
\label{2.2x11a}
$$
\left(
\begin{array}{ccccccccc}
\hline
[2^2,1^2]_1&[2^2,1^2]_2&[2^2,1^2]_3&[2^2,1^2]_4&[2^2,1^2]_5&[2^2,1^2]_6&[2^2,1^2]_7&[2^2,1^2]_8&[2^2,1^2]_9\\
\hline
 0 & 0 & \frac{3 \sqrt{\frac{11}{5}}}{10} & \frac{7
   \sqrt{\frac{11}{195}}}{10} & -3 \sqrt{\frac{33}{9035}} &
   \frac{\sqrt{\frac{55}{834}}}{2} & -\frac{\sqrt{\frac{11}{6}}}{10} &
   \frac{1}{5 \sqrt{6}} & 0 \\
 \frac{1}{\sqrt{6}} & -\sqrt{\frac{3}{110}} & \frac{3
   \sqrt{\frac{3}{55}}}{20} & \frac{17 \sqrt{\frac{13}{55}}}{60} & 2
   \sqrt{\frac{13}{7645}} & -\frac{34 \sqrt{\frac{2}{7645}}}{3} & \frac{2
   \sqrt{\frac{2}{11}}}{15} & \frac{\sqrt{2}}{15} & \frac{1}{\sqrt{30}} \\
 0 & 0 & -\frac{3 \sqrt{\frac{11}{5}}}{20} & \frac{41
   \sqrt{\frac{3}{715}}}{20} & 8 \sqrt{\frac{3}{99385}} &
   \frac{\sqrt{\frac{33}{1390}}}{2} & \frac{\sqrt{\frac{33}{2}}}{10} &
   -\frac{\sqrt{\frac{3}{2}}}{5} & 0 \\
 0 & 4 \sqrt{\frac{2}{165}} & -\frac{1}{10 \sqrt{165}} & -\frac{23}{10
   \sqrt{715}} & -\frac{2}{\sqrt{99385}} & 8 \sqrt{\frac{2}{7645}} &
   \frac{1}{5 \sqrt{22}} & \frac{1}{10 \sqrt{2}} &
   -\frac{\sqrt{\frac{3}{10}}}{2} \\
 0 & 0 & -\frac{3 \sqrt{\frac{11}{5}}}{20} & -\frac{277}{20 \sqrt{2145}} &
   5 \sqrt{\frac{15}{19877}} & -2 \sqrt{\frac{22}{2085}} &
   -\frac{\sqrt{\frac{11}{6}}}{5} & \frac{\sqrt{\frac{2}{3}}}{5} & 0 \\
 0 & 0 & 0 & -\frac{4 \sqrt{\frac{5}{143}}}{3} & \frac{17}{\sqrt{99385}} &
   \frac{74 \sqrt{\frac{2}{7645}}}{3} & \frac{1}{15 \sqrt{22}} &
   \frac{1}{30 \sqrt{2}} & -\frac{\sqrt{\frac{3}{10}}}{2} \\
 0 & 2 \sqrt{\frac{2}{55}} & -\frac{17}{10 \sqrt{55}} & \frac{9
   \sqrt{\frac{3}{715}}}{10} & 7 \sqrt{\frac{3}{99385}} & -\frac{17
   \sqrt{\frac{15}{3058}}}{4} & \frac{\sqrt{\frac{3}{22}}}{4} & 0 & 0 \\
\end{array}
\right),
$$
\end{table}
\begin{table}
\caption{As in table \ref{2.2x11a}}
\label{2.2x11b}
$$
\left(
\begin{array}{ccccc}
\hline
[3^2]_1&[3^2]_2&[3^2]_3&[3^2]_4&[3^2]_5\\
\hline
 -\frac{1}{\sqrt{33}} & -\frac{1}{4 \sqrt{66}} & \frac{1}{4 \sqrt{6}} &
   \frac{\sqrt{\frac{3}{2}}}{4} & -\frac{1}{4 \sqrt{2}} \\
 \frac{1}{\sqrt{11}} & \frac{1}{4 \sqrt{22}} & \frac{1}{4 \sqrt{2}} &
   \frac{1}{4 \sqrt{2}} & \frac{1}{4 \sqrt{6}} \\
 -\frac{1}{2 \sqrt{33}} & -\frac{\sqrt{\frac{3}{22}}}{2} &
   \frac{\sqrt{\frac{3}{2}}}{4} & 0 & -\frac{1}{4 \sqrt{2}} \\
 \frac{1}{2 \sqrt{11}} & \frac{3}{2 \sqrt{22}} & \frac{1}{4 \sqrt{2}} & 0 &
   -\frac{1}{4 \sqrt{6}} \\
 \frac{1}{2 \sqrt{33}} & -\frac{5}{4 \sqrt{66}} & 0 & \frac{1}{4 \sqrt{6}}
   & -\frac{1}{2 \sqrt{2}} \\
 \frac{1}{2 \sqrt{11}} & -\frac{5}{4 \sqrt{22}} & 0 & \frac{1}{4 \sqrt{2}}
   & \frac{1}{2 \sqrt{6}} \\
 -\frac{1}{\sqrt{33}} & -\frac{1}{4 \sqrt{66}} & -\frac{1}{4 \sqrt{6}} &
   \frac{\sqrt{\frac{3}{2}}}{4} & \frac{1}{4 \sqrt{2}} \\
\end{array}
\right),
$$
\end{table}
\begin{table}
\caption{As in table \ref{2.2x11a}}
\label{2.2x11c}
$$
\left(
\begin{array}{cccccccc}
\hline
[3,2,1]_1&[3,2,1]_2&[3,2,1]_3&[3,2,1]_4&[3,2,1]_5&[3,2,1]_6&[3,2,1]_7&[3,2,1]_8\\
\hline
 0 & 0 & -\frac{1}{\sqrt{41}} & -\frac{19
   \sqrt{\frac{3}{902}}}{4} & \frac{137}{4 \sqrt{16302}} & -2
   \sqrt{\frac{10}{247}} & \frac{1}{\sqrt{157}} & -809
   \sqrt{\frac{2}{17640363}} \\
 0 & 0 & -\sqrt{\frac{3}{41}} & -\frac{57}{4 \sqrt{902}} &
   -\frac{3 \sqrt{\frac{13}{418}}}{4} & \frac{3
   \sqrt{\frac{39}{190}}}{4} & -\frac{3
   \sqrt{\frac{3}{157}}}{4} & \frac{2499}{4 \sqrt{11760242}}
   \\
 0 & 0 & -\frac{2}{\sqrt{41}} & \frac{3
   \sqrt{\frac{3}{902}}}{4} & \frac{\sqrt{\frac{247}{66}}}{4}
   & 0 & 0 & 2 \sqrt{\frac{314}{112359}} \\
 0 & 0 & -2 \sqrt{\frac{3}{41}} & \frac{9}{4 \sqrt{902}} &
   -\frac{105}{4 \sqrt{5434}} & -\frac{9
   \sqrt{\frac{3}{2470}}}{4} & \frac{\sqrt{\frac{3}{157}}}{4}
   & -\frac{1461}{4 \sqrt{11760242}} \\
 0 & 0 & 0 & 0 & 0 & 0 & \frac{4}{\sqrt{157}} & -32
   \sqrt{\frac{6}{5880121}} \\
 0 & 0 & 0 & 0 & 0 & 0 & -4 \sqrt{\frac{3}{157}} & 96
   \sqrt{\frac{2}{5880121}} \\
 0 & -\frac{1}{2 \sqrt{3}} & \frac{3}{4 \sqrt{41}} &
   -\frac{29}{4 \sqrt{2706}} & \frac{15
   \sqrt{\frac{3}{5434}}}{4} & \frac{43}{8 \sqrt{2470}} &
   -\frac{25}{8 \sqrt{157}} & -\frac{7607
   \sqrt{\frac{3}{11760242}}}{8} \\
\end{array}
\right),
$$
\end{table}
\begin{table}
\caption{As in table \ref{2.2x11a}}
\label{2.2x11d}
$$
\left(
\begin{array}{cccccccc}
\hline
[3,2,1]_9&[3,2,1]_{10}&[3,2,1]_{11}&[3,2,1]_{12}&[3,2,1]_{13}&[3,2,1]_{14}&[3,2,1]_{15}&[3,2,1]_{16}\\
\hline
 -\frac{321 \sqrt{\frac{105}{113744761}}}{2} & -\frac{71
   \sqrt{\frac{35}{634733}}}{2} &
   \frac{\sqrt{\frac{10}{1463}}}{3} & -\frac{8
   \sqrt{\frac{2}{2261}}}{3} & \frac{109}{12 \sqrt{67830}} &
   \frac{143}{60 \sqrt{714}} & \frac{7}{30 \sqrt{34}} &
   -\frac{1}{2 \sqrt{30}} \\
 -\frac{5961}{2 \sqrt{3981066635}} &
   -\frac{\sqrt{\frac{3135}{21259}}}{2} & 0 &
   -\sqrt{\frac{6}{2261}} & -\frac{23
   \sqrt{\frac{7}{3230}}}{12} & -\frac{43}{60 \sqrt{238}} &
   \frac{2 \sqrt{\frac{2}{51}}}{5} & 0 \\
 16 \sqrt{\frac{105}{113744761}} & 13 \sqrt{\frac{35}{634733}}
   & \frac{\sqrt{\frac{14}{1045}}}{3} & \frac{2
   \sqrt{\frac{14}{323}}}{3} & -3 \sqrt{\frac{21}{3230}} &
   -\frac{\sqrt{\frac{21}{34}}}{5} & -\frac{4
   \sqrt{\frac{2}{17}}}{5} & 0 \\
 \frac{1147}{2 \sqrt{3981066635}} & \frac{3
   \sqrt{\frac{627}{106295}}}{2} & 0 & -2
   \sqrt{\frac{6}{2261}} & \frac{27}{\sqrt{22610}} &
   -\frac{\sqrt{\frac{7}{34}}}{5} & -\frac{4
   \sqrt{\frac{2}{51}}}{5} & 0 \\
 -482 \sqrt{\frac{105}{113744761}} & -12
   \sqrt{\frac{35}{634733}} & -\frac{29}{3 \sqrt{14630}} &
   \frac{59}{3 \sqrt{4522}} & \frac{11}{6 \sqrt{67830}} &
   -\frac{263}{30 \sqrt{714}} & \frac{1}{30 \sqrt{34}} &
   \frac{1}{2 \sqrt{30}} \\
 -8 \sqrt{\frac{3037}{1310855}} & 0 & 0 & 4
   \sqrt{\frac{6}{2261}} & -\frac{1}{6 \sqrt{22610}} &
   \frac{\sqrt{\frac{17}{14}}}{6} & 0 & 0 \\
 \frac{13919 \sqrt{\frac{3}{3981066635}}}{4} & -\frac{59
   \sqrt{\frac{11}{2019605}}}{4} &
   -\frac{\sqrt{\frac{55}{266}}}{6} &
   \frac{\sqrt{\frac{7}{646}}}{6} & \frac{67}{12 \sqrt{67830}}
   & -\frac{5 \sqrt{\frac{7}{102}}}{12} & -\frac{1}{6
   \sqrt{34}} & \frac{\sqrt{\frac{3}{10}}}{2} \\
\end{array}
\right).
$$
\end{table}
%\begin{itemize}
%\item[i)] [3,1]x[2]
%\subsection{[3,1]x[2]}
%\subsection{[3,1]x[2]}
\clearpage
%\newpage
%\begin{itemize}
%\item $ [3,1]\otimes[2]\rightarrow 5[5,1]+9[4,2]+10[4,1^2]+5[3,3]+16[3,2,1]$:
%\end{itemize}
\begin{table}
\caption{The reduction $ [3,1]\otimes[2]\rightarrow 5[5,1]+9[4,2]+10[4,1^2]+5[3,3]+16[3,2,1]$}
\label{t31x2.2a}
$$
\left(
\begin{array}{ccccc}
\hline
[5,1]_1&[5,1]_2&[5,1]_3&[5,1]_4&[5,1]_5\\
\hline
 -\frac{1}{\sqrt{30}} & \frac{1}{2 \sqrt{5}} & \frac{1}{6} & 0 & 0 \\
 -\frac{1}{5 \sqrt{3}} & -\frac{\sqrt{2}}{15} & 0 & \frac{4}{15} &
   \frac{2}{15} \\
 \frac{2}{5 \sqrt{3}} & \frac{2 \sqrt{2}}{15} & 0 & \frac{2}{15} &
   \frac{1}{15} \\
 -\frac{1}{\sqrt{30}} & \frac{1}{2 \sqrt{5}} & 0 & \frac{1}{6
   \sqrt{10}} & \frac{1}{2 \sqrt{10}} \\
 -\frac{1}{5 \sqrt{3}} & -\frac{\sqrt{2}}{15} &
   \frac{\sqrt{\frac{2}{5}}}{3} & \frac{1}{5} & -\frac{1}{15} \\
 \frac{2}{5 \sqrt{3}} & \frac{2 \sqrt{2}}{15} & \frac{1}{3 \sqrt{10}}
   & \frac{1}{10} & -\frac{1}{30} \\
 -\frac{1}{\sqrt{30}} & \frac{1}{2 \sqrt{5}} & 0 & -\frac{1}{2
   \sqrt{10}} & \frac{1}{6 \sqrt{10}} \\
\end{array}
\right)
\left(
\begin{array}{ccccccccc}
\hline
[4,2]_1&[4,2]_2&[4,2]_3&[4,2]_4&[4,2]_5&[4,2]_6&[4,2]_7&[4,2]_8&[4,2]_9\\
\hline
 -\frac{1}{4 \sqrt{2}} & \frac{7}{4 \sqrt{230}} & \frac{9}{2
   \sqrt{1610}} & -\frac{9}{10 \sqrt{14}} &
   \frac{\sqrt{\frac{7}{2}}}{10} & 0 & \frac{1}{2 \sqrt{5}} & 0 & 0
   \\
 -\frac{1}{4 \sqrt{5}} & \frac{7}{20 \sqrt{23}} & \frac{9}{10
   \sqrt{161}} & \frac{1}{2 \sqrt{35}} & -\frac{1}{2 \sqrt{35}} &
   -\frac{1}{5 \sqrt{14}} & 0 & -\frac{4}{5 \sqrt{5}} & -\frac{2}{5
   \sqrt{5}} \\
 \frac{1}{2 \sqrt{5}} & -\frac{7}{10 \sqrt{23}} & -\frac{9}{5
   \sqrt{161}} & -\frac{1}{\sqrt{35}} & \frac{1}{\sqrt{35}} &
   \frac{\sqrt{\frac{2}{7}}}{5} & 0 & -\frac{2}{5 \sqrt{5}} &
   -\frac{1}{5 \sqrt{5}} \\
 0 & -\sqrt{\frac{2}{115}} & 4 \sqrt{\frac{2}{805}} & -\frac{1}{5
   \sqrt{14}} & \frac{3}{5 \sqrt{14}} & -\frac{3}{2 \sqrt{35}} & 0 &
   \frac{1}{10 \sqrt{2}} & \frac{3}{10 \sqrt{2}} \\
 0 & -\frac{2}{5 \sqrt{23}} & \frac{8}{5 \sqrt{161}} & -\frac{1}{5
   \sqrt{35}} & -\frac{2}{5 \sqrt{35}} & \frac{\sqrt{\frac{2}{7}}}{5}
   & -\frac{\sqrt{2}}{5} & -\frac{3}{5 \sqrt{5}} & \frac{1}{5
   \sqrt{5}} \\
 0 & \frac{4}{5 \sqrt{23}} & -\frac{16}{5 \sqrt{161}} & \frac{2}{5
   \sqrt{35}} & \frac{4}{5 \sqrt{35}} & -\frac{2
   \sqrt{\frac{2}{7}}}{5} & -\frac{1}{5 \sqrt{2}} & -\frac{3}{10
   \sqrt{5}} & \frac{1}{10 \sqrt{5}} \\
 -\frac{1}{4 \sqrt{2}} & -\frac{9}{4 \sqrt{230}} &
   -\frac{\sqrt{\frac{5}{322}}}{2} & -\frac{9}{10 \sqrt{14}} &
   -\frac{3}{10 \sqrt{14}} & -\frac{1}{\sqrt{35}} & 0 & -\frac{3}{10
   \sqrt{2}} & \frac{1}{10 \sqrt{2}} \\
\end{array}
\right)
$$
\end{table}
\begin{table}
\caption{As in table \ref{t31x2.2a}}
\label{t31x2.2b}
$$
\left(
\begin{array}{cccccccc}
\hline
[4,1^2]_{1}&4,1^2]_{2}&[4,1^2]_{3}&[4,1^2]_{4}&[4,1^2]_{5}&[4,1^2]_{6}&[4,1^2]_{7}&[4,1^2]_{8}\\
\hline
 \frac{1}{3 \sqrt{17}} & \frac{4 }{3}\sqrt{\frac{2}{19397}} &
   -\frac{458}{13 \sqrt{12551}} & -\frac{102}{13}
   \sqrt{\frac{3}{35365}} & 738 \sqrt{\frac{3}{23247665}}
   & -\frac{1088}{\sqrt{219554853}} & -\frac{862}{3
   \sqrt{27660693}} & -\frac{1796}{3 \sqrt{37422969}} \\
 0 & 3 \sqrt{\frac{17}{5705}} & \frac{167}{13 \sqrt{125510}} &
   \frac{943}{130} \sqrt{\frac{3}{14146}} & -\frac{291}{5}
   \sqrt{\frac{3}{9299066}} & -438
   \sqrt{\frac{6}{365924755}} & -\frac{4779}{2}
   \sqrt{\frac{3}{92202310}} & 331
   \sqrt{\frac{3}{124743230}} \\
 0 & 9 \sqrt{\frac{17}{5705}} & \frac{501}{13 \sqrt{125510}} &
   -\frac{202 \sqrt{\frac{2}{21219}}}{65} & \frac{407}{5}
   \sqrt{\frac{3}{9299066}} & 107
   \sqrt{\frac{3}{731849510}} & 1261 \sqrt{\frac{2}{138303465}}
   & -\frac{1841}{\sqrt{374229690}} \\
 -\frac{1}{\sqrt{17}} & -4 \sqrt{\frac{2}{19397}} &
   \frac{233}{13 \sqrt{12551}} & -\frac{216}{13}
   \sqrt{\frac{3}{35365}} & 239 \sqrt{\frac{3}{23247665}}
   & -26 \sqrt{\frac{3}{73184951}} &
   -\frac{802}{\sqrt{27660693}} & -\frac{314}{\sqrt{37422969}}
   \\
 0 & 0 & 0 & 0 & 0 & -\sqrt{\frac{21693}{101210}} & 196
   \sqrt{\frac{2}{138303465}} & -505 \sqrt{\frac{5}{74845938}}
   \\
 0 & 0 & 0 & 0 & 0 & \sqrt{\frac{7231}{303630}} & -\frac{196}{3}
   \sqrt{\frac{2}{138303465}} & -\frac{4961}{3}
   \sqrt{\frac{5}{74845938}} \\
 -\frac{1}{3 \sqrt{17}} & -\frac{4}{3} \sqrt{\frac{2}{19397}} &
   \frac{233}{39 \sqrt{12551}} & -\frac{72}{13}
   \sqrt{\frac{3}{35365}} & \frac{239}{\sqrt{69742995}} &
   -\frac{26}{\sqrt{219554853}} & -\frac{802}{3
   \sqrt{27660693}} & -\frac{314}{3 \sqrt{37422969}} \\
\end{array}
\right)
$$
\end{table}
\begin{table}
\caption{As in table \ref{t31x2.2a}}
\label{t31x2.2b2}
$$
\left(
\begin{array}{cc}
\hline
[4,1^2]_{9}&[4,1^2]_{10}\\
\hline
 -\frac{409}{3 \sqrt{5271805}} & -\frac{83}{3 \sqrt{28105}} \\
 31 \sqrt{\frac{2}{1054361}} & \frac{4
   \sqrt{\frac{2}{5621}}}{3} \\
 -2 \sqrt{\frac{154}{13693}} & 0 \\
 -\frac{53 \sqrt{\frac{5}{1054361}}}{2} & -\frac{59
   \sqrt{\frac{5}{5621}}}{6} \\
 \frac{439}{5 \sqrt{2108722}} & \frac{79}{15 \sqrt{11242}} \\
 \frac{757 \sqrt{\frac{11}{191702}}}{30} & \frac{3
   \sqrt{\frac{11}{1022}}}{10} \\
 -\frac{53 \sqrt{\frac{5}{1054361}}}{6} & \frac{83
   \sqrt{\frac{5}{5621}}}{6} \\
\end{array}
\right)
\left(
\begin{array}{ccccc}
\hline
[3,3]_{1}&[3,3]_{2}&[3,3]_{3}&[3,3]_{4}&[3,3]_{5}\\
\hline
 0 & 0 & 0 & 0 & 0 \\
 0 & 0 & 0 & 0 & 0 \\
 0 & 0 & 0 & 0 & 0 \\
 0 & 0 & 0 & 0 & 0 \\
 0 & 0 & 0 & 0 & 0 \\
 0 & 0 & 0 & 0 & 0 \\
 0 & 0 & 0 & 0 & 0 \\
\end{array}
\right)
$$
\end{table}
\begin{table}
\caption{As in table \ref{t31x2.2a}}
\label{t31x2.2c}
$$
\left(
\begin{array}{cccccccc}
\hline
[3,2,1]_1&[3,2,1]_2&[3,2,1]_3&[3,2,1]_4&[3,2,1]_5&[3,2,1]_6&[3,2,1]_7&[3,2,1]_8\\
\hline
 0 & 0 & 0 & 2 \sqrt{\frac{2}{589}} & \frac{5}{2} \sqrt{\frac{21}{1178}} &
   \frac{1}{4 \sqrt{7}} & \frac{9}{4 \sqrt{91}} & \frac{1}{2 \sqrt{273}}
   \\
 0 & 0 & 0 & \frac{2}{\sqrt{2945}} & \frac{\sqrt{\frac{105}{589}}}{4} &
   \frac{1}{4 \sqrt{70}} & \frac{41}{4 \sqrt{910}} &
   -\frac{1}{2}\sqrt{\frac{7}{390}} \\
 0 & 0 & 0 & \frac{6}{\sqrt{2945}} & \frac{3 \sqrt{\frac{105}{589}}}{4} &
   \frac{3}{4 \sqrt{70}} & -\frac{37}{4 \sqrt{910}} & \frac{19}{2
   \sqrt{2730}} \\
 \frac{1}{2 \sqrt{3}} & \frac{7}{4 \sqrt{78}} & -\frac{1}{4 \sqrt{26}} &
   -\frac{23}{4 \sqrt{1178}} & -\frac{27 \sqrt{\frac{3}{8246}}}{4} & 0 &
   -\frac{5}{4 \sqrt{91}} & -\frac{1}{\sqrt{273}} \\
 \frac{1}{2 \sqrt{30}} & \frac{5}{8} \sqrt{\frac{5}{39}} & -\frac{11}{8
   \sqrt{65}} & -\frac{23}{8 \sqrt{2945}} & -\frac{27
   \sqrt{\frac{3}{20615}}}{8} & \frac{1}{2 \sqrt{70}} & -\frac{23}{4
   \sqrt{910}} & -\sqrt{\frac{2}{1365}} \\
 \frac{\sqrt{\frac{3}{10}}}{2} & -\frac{\sqrt{\frac{15}{13}}}{8} &
   \frac{17}{8 \sqrt{65}} & -\frac{69}{8 \sqrt{2945}} & -\frac{81
   \sqrt{\frac{3}{20615}}}{8} & -\frac{1}{\sqrt{70}} & \frac{3
   \sqrt{\frac{7}{130}}}{4} & -\frac{1}{\sqrt{2730}} \\
 -\frac{1}{2 \sqrt{3}} & -\frac{7}{4 \sqrt{78}} & \frac{1}{4 \sqrt{26}} &
   \frac{7}{4 \sqrt{1178}} & -\frac{43 \sqrt{\frac{3}{8246}}}{4} &
   -\frac{1}{4 \sqrt{7}} & -\frac{1}{\sqrt{91}} & \frac{1}{2 \sqrt{273}}
   \\
\end{array}
\right)
$$
\end{table}
\begin{table}
\caption{As in table \ref{t31x2.2a} }
\label{t31x2.2d}
$$
\left(
\begin{array}{cccccccc}
\hline
[3,2,1]_9&[3,2,1]_{10}&[3,2,1]_{11}&[3,2,1]_{12}&[3,2,1]_{13}&[3,2,1]_{14}&[3,2,1]_{15}&[3,2,1]_{16}\\
\hline
 0 & 2 \sqrt{\frac{2}{589}} & \frac{1}{4 \sqrt{7}} & \frac{1}{2
   \sqrt{273}} & -\frac{4}{7 \sqrt{165}} & -\frac{333
   \sqrt{\frac{6}{3318007}}}{5} & \frac{2}{3 \sqrt{5}} & 0 \\
 0 & \frac{2}{\sqrt{2945}} & \frac{1}{4 \sqrt{70}} &
   -\frac{1}{2}\sqrt{\frac{7}{390}} & \frac{1}{35}\sqrt{\frac{22}{3}} & 89
   \sqrt{\frac{11}{4524555}} & 0 & \frac{2}{15 \sqrt{5}} \\
 0 & \frac{6}{\sqrt{2945}} & \frac{3}{4 \sqrt{70}} & \frac{19}{2
   \sqrt{2730}} & \frac{1}{35}\sqrt{\frac{6}{11}} & \frac{1079}{2
   \sqrt{49770105}} & 0 & \frac{1}{15 \sqrt{5}} \\
 \frac{7}{4 \sqrt{78}} & -\frac{23}{4 \sqrt{1178}} & 0 &
   -\frac{1}{\sqrt{273}} & -\frac{16}{7 \sqrt{165}} & -\frac{2}{5}
   \sqrt{\frac{462}{43091}} & 0 & \frac{\sqrt{2}}{5} \\
 \frac{5}{8} \sqrt{\frac{5}{39}} & -\frac{23}{8 \sqrt{2945}} & \frac{1}{2
   \sqrt{70}} & -\sqrt{\frac{2}{1365}} & \frac{4}{35} \sqrt{\frac{22}{3}}
   & \frac{38}{5} \sqrt{\frac{77}{646365}} & \frac{\sqrt{2}}{15} &
   -\frac{1}{15 \sqrt{5}} \\
 -\frac{1}{8}\sqrt{\frac{15}{13}} & -\frac{69}{8 \sqrt{2945}} &
   -\frac{1}{\sqrt{70}} & -\frac{1}{\sqrt{2730}} & \frac{4}{35}
   \sqrt{\frac{6}{11}} & -\frac{1}{5}\sqrt{\frac{77}{646365}} &
   \frac{1}{15 \sqrt{2}} & -\frac{1}{30 \sqrt{5}} \\
 -\frac{7}{4 \sqrt{78}} & \frac{7}{4 \sqrt{1178}} & -\frac{1}{4 \sqrt{7}}
   & \frac{1}{2 \sqrt{273}} & \frac{4}{7} \sqrt{\frac{5}{33}} & \frac{487}{5}
   \sqrt{\frac{6}{3318007}} & 0 & \frac{\sqrt{2}}{15} \\
\end{array}
\right)
$$
\end{table}
%\clearpage
%\newpage
%\begin{itemize}
%\item [ii)] [3,1]x[1,1]
%\end{itemize}
%\subsection{[3,1]x[1,1]}
\begin{table}
\caption{The reduction:$ [3,1]\otimes[1^2]\rightarrow 9[4,2]+10[4,1^2]+10[3,1^3]+16[3,2,1]$: }
\label{t31x2.11a}
$$
\left(
\begin{array}{ccccccccc}
\hline
[4,2]_1&[4,2]_2&[4,2]_3&[4,2]_4&[4,2]_5&[4,2]_6&[4,2]_7&[4,2]_8&[4,2]_9\\
\hline
 0 & -\sqrt{\frac{2}{165}} & -\sqrt{\frac{3}{55}} & \frac{1}{\sqrt{15}} & \frac{1}{5}\sqrt{\frac{3}{2}} & -\frac{1}{5 \sqrt{6}} & 0 &
   0 & 0 \\
 0 & \frac{4}{5 \sqrt{33}} & \frac{2}{5} \sqrt{\frac{6}{11}} & \frac{1}{5}\sqrt{\frac{2}{3}} & \frac{1}{5}\sqrt{\frac{3}{5}} &
   -\frac{1}{5 \sqrt{15}} & 0 & \frac{1}{5 \sqrt{5}} & -\frac{2}{5 \sqrt{5}} \\
 0 & \frac{2}{5 \sqrt{33}} & \frac{1}{5}\sqrt{\frac{6}{11}} & \frac{1}{5 \sqrt{6}} & \frac{\sqrt{\frac{3}{5}}}{10} & -\frac{1}{10
   \sqrt{15}} & 0 & -\frac{2}{5 \sqrt{5}} & \frac{4}{5 \sqrt{5}} \\
 -\frac{1}{2 \sqrt{6}} & \frac{1}{2 \sqrt{330}} & -\frac{2}{\sqrt{165}} & 0 & \frac{2 }{5}\sqrt{\frac{2}{3}} &
   \frac{1}{5}\sqrt{\frac{2}{3}} & 0 & 0 & 0 \\
 \frac{1}{\sqrt{15}} & -\frac{1}{5 \sqrt{33}} & \frac{4}{5}\sqrt{\frac{2}{33}} & 0 & \frac{4}{5 \sqrt{15}} & \frac{2}{5 \sqrt{15}}
   & -\frac{1}{5 \sqrt{2}} & \frac{3}{10 \sqrt{5}} & -\frac{1}{10 \sqrt{5}} \\
 \frac{1}{2 \sqrt{15}} & -\frac{1}{10 \sqrt{33}} & \frac{2}{5} \sqrt{\frac{2}{33}} & 0 & \frac{2}{5 \sqrt{15}} & \frac{1}{5
   \sqrt{15}} & \frac{\sqrt{2}}{5} & -\frac{3}{5 \sqrt{5}} & \frac{1}{5 \sqrt{5}} \\
 -\frac{1}{2 \sqrt{6}} & -\frac{1}{2}\sqrt{\frac{11}{30}} & 0 & 0 & \frac{1}{5}\sqrt{\frac{3}{2}} & -\frac{1}{5 \sqrt{6}} & 0 & 0 & 0
   \\
\end{array}
\right)
$$
\end{table}
\begin{table}
\caption{As in table \ref{t31x2.11a} }
\label{t31x2.11b}
$$
\left(
%
\right)
$$
\end{table}
\begin{table}
\caption{As in table \ref{t31x2.11a} }
\label{t31x2.11c}
$$
\left(
%
\right)
$$
\end{table}
\begin{table}
\caption{As in table \ref{t31x2.11a} }
\label{t31x2.11d}
$$
\left(
%
\right)
$$
\end{table}
\begin{table}
\caption{As in table \ref{t31x2.11a} }
\label{t31x2.11e}
$$
\left(
%
\right)
$$
\end{table}
%\clearpage
%\begin{itemize}
%\item [iii)] [2,1,1]x[1,1]
%\end{itemize}
%\subsection{The case [f^']=[2,1,1]}
%\begin{itemize}
%\item
 %$ [2,1^2]\otimes[1^2]\rightarrow 5[2,1^3]+ 9[2^2,1^2]+15[3,1^3]+16[3,2,1]$:
%\end{itemize}
%\newpage
%\clearpage
\begin{table}
 \caption{The reduction:$ [2,1^2]\otimes[1^2]\rightarrow 5[2,1^4]+ 9[2^2,1^2]+10[3,1^3]+5[2^3] +16[3,2,1]$ }
\label{t211x11.11a}
$$
\left(
\begin{array}{ccccc}
\hline
[2,1^4]_1&[2,1^4]_2&[2,1^4]_3&[2,1^4]_4&[2,1^4]_5\\
\hline
 -\frac{2}{5 \sqrt{3}} & \frac{2 \sqrt{2}}{15} & \frac{1}{15} & \frac{2}{15} &
   0 \\
 \frac{1}{5 \sqrt{3}} & -\frac{\sqrt{2}}{15} & \frac{2}{15} & \frac{4}{15} & 0
   \\
 \frac{1}{\sqrt{30}} & \frac{1}{2 \sqrt{5}} & 0 & 0 & \frac{1}{6} \\
 \frac{2}{5 \sqrt{3}} & -\frac{2 \sqrt{2}}{15} & \frac{1}{30} & -\frac{1}{10} &
   -\frac{1}{3 \sqrt{10}} \\
 -\frac{1}{5 \sqrt{3}} & \frac{\sqrt{2}}{15} & \frac{1}{15} & -\frac{1}{5} &
   -\frac{\sqrt{\frac{2}{5}}}{3} \\
 -\frac{1}{\sqrt{30}} & -\frac{1}{2 \sqrt{5}} & -\frac{1}{2 \sqrt{10}} &
   -\frac{1}{6 \sqrt{10}} & 0 \\
 \frac{2}{5 \sqrt{3}} & -\frac{2 \sqrt{2}}{15} & -\frac{1}{10} & -\frac{1}{30}
   & \frac{1}{3 \sqrt{10}} \\
\end{array}%
\right)
$$
\end{table}
\begin{table}
\caption{As in table \ref{t211x11.11a}}
\label{t211x11.11b}
$$
\left (
%
\right)
$$
\end{table}

\begin{table}
\caption{As in table \ref{t211x11.11a}}
\label{t211x11.11c}
$$
\left(
%
\right)
$$
\end{table}
\begin{table}
\caption{As in table \ref{t211x11.11a}}
\label{t211x11.11d}
$$
\left(
%
\right)
$$
\end{table}

\begin{table}
\caption{As in table \ref{t211x11.11a}}
\label{t211x11.11e}
$$
\left(
%
\right)
$$
\end{table}
%\clearpage
%\begin{itemize}
%\item [iv)] [2,1,1]x[2]
%\end{itemize}
%\subsection{[2,1,1]x[2]}
%$[2,1^2]\otimes[2]\rightarrow 9[2^2,1^2]+10[3^3,1^3]+10[4,1^2]+16[3,2,1]$
%\newpage
\begin{table}[h!]
\caption{The reduction $[2,1^2]\otimes[2]\rightarrow 9[2^2,1^2]+10[3^3,1^3]+10[4,1^2]+16[3,2,1]$}
\label{t211x2.11a}
$$ 
\left(
%
\right)
$$
\end{table}
\begin{table}[h!]
\caption{As in  table \ref{t211x2.11a}}
\label{t211x2.11b}
$$
\left(
%
\right)
$$
\end{table}
%\clearpage
\begin{table}[h!]
\caption{As in  table \ref{t211x2.11a}}
\label{t211x2.11c}
$$
\left(
%
\right)
$$
\end{table}
%\end{document}
%\newpage
%\section{new}
%\clearpage. 
\begin{table}[h!]
\caption{As in table \ref{t211x2.11a}}
\label{t211x2.11d}
$$
\left(
%
\right)
$$
\end{table}
%\clearpage
%\begin{sidewaystable}[h!]
%\end{document}
\begin{table}[h!]
\caption{As in \ref{t211x2.11a}}
\label{t211x2.11e}
$$
\left(
%
\right)
$$
\end{table}
%\clearpage
%\end{sidewaystable}
\end{document}